\documentclass[11pt]{article}
\usepackage{mathrsfs}
\usepackage{natbib,graphicx,setspace,lscape,longtable}
\usepackage{mathrsfs,amsmath,amsthm,amssymb,color}
\usepackage{natbib,epsfig,graphicx,pdfpages}
\usepackage{rotating}
\usepackage{subfigure}
\usepackage{graphicx}
\usepackage{algorithmic}
\usepackage{algorithm}
\newcommand\sbullet[1][.5]{\mathbin{\vcenter{\hbox{\scalebox{#1}{$\bullet$}}}}}
\usepackage{url}
\usepackage{hyperref}
\usepackage[table]{xcolor}
\usepackage{colortbl}
\bibpunct{(}{)}{;}{a}{,}{,}

\newtheorem{theorem}{Theorem}
\newtheorem{lemma}{Lemma}
\newtheorem{proposition}{Proposition}
\newtheorem{remark}{Remark}
\usepackage{amsthm}

\newtheorem{assumption}{{Assumption}}
\def\beq{\begin{equation}}
\def\eeq{\end{equation}}
\def\beqr{\begin{eqnarray}}
\def\eeqr{\end{eqnarray}}
\def\beqrs{\begin{eqnarray*}}
\def\eeqrs{\end{eqnarray*}}
\def\bet{\begin{theorem}}
\def\eet{\end{theorem}}
\def\bel{\begin{lemma}}
\def\eel{\end{lemma}}
\def\bep{\begin{proposition}}
\def\eep{\end{proposition}}
\def\bg{\begin{figure}[tbph]\begin{center}}
\def\eg{\end{center}\end{figure}}

\def\bc{\begin{center}}
\def\ec{\end{center}}

\def\wh{\widehat}

\def\bJ{\mathbf J}

\def\cov{\mbox{Cov}}

\def\diag{\mbox{diag}}

\numberwithin{equation}{section}

\newcommand{\Var}{\textnormal{Var}}
\newcommand{\Cov}{\textnormal{Cov}}

\newcommand{\vc}{\textnormal{vec}}

\newcommand{\bA}{{\mathbf A}}
\newcommand{\bB}{{\mathbf B}}
\newcommand{\bF}{{\mathbf F}}
\newcommand{\bE}{{\mathbf E}}
\newcommand{\bG}{{\mathbf G}}
\newcommand{\bH}{{\mathbf H}}
\newcommand{\bI}{{\mathbf I}}

\newcommand{\bL}{{\mathbf L}}
\newcommand{\bM}{{\mathbf M}}
\newcommand{\bN}{{\mathbf N}}
\newcommand{\bQ}{{\mathbf Q}}
\newcommand{\bP}{{\mathbf P}}
\newcommand{\bR}{{\mathbf R}}
\newcommand{\bS}{{\mathbf S}}
\newcommand{\bT}{{\mathbf T}}
\newcommand{\bU}{{\mathbf U}}
\newcommand{\bV}{{\mathbf V}}
\newcommand{\bW}{{\mathbf W}}
\newcommand{\bX}{{\mathbf X}}
\newcommand{\bY}{{\mathbf Y}}
\newcommand{\bZ}{{\mathbf Z}}
\newcommand{\ba}{{\mathbf a}}
\newcommand{\bb}{{\mathbf b}}

\newcommand{\bff}{{\mathbf f}}
 
\newcommand{\bh}{{\mathbf h}}

\newcommand{\bp}{{\mathbf p}}
\newcommand{\bq}{{\mathbf q}}
\newcommand{\br}{{\mathbf r}}

\newcommand{\bu}{{\mathbf u}}
\newcommand{\bv}{{\mathbf v}}

\newcommand{\bx}{{\mathbf x}}
\newcommand{\by}{{\mathbf y}}
\newcommand{\bz}{{\mathbf z}}

\newcommand{\bfeta}  {\boldsymbol{\eta}}

\newcommand{\blambda}{\boldsymbol{\lambda}}

\newcommand{\bOmega}{\boldsymbol{\Omega}}

\newcommand{\bSigma}{\boldsymbol{\Sigma}}
\newcommand{\bDelta}{\boldsymbol{\Delta}}

\newcommand{\bPhi} {\boldsymbol{\Phi}}
\newcommand{\bPsi} {\boldsymbol{\Psi}}

\newcommand{\bxi} {\boldsymbol{\xi}}
\newcommand{\bXi} {\boldsymbol{\Xi}}

\newcommand{\bGamma} {\boldsymbol{\Gamma}}
\newcommand{\bLambda} {\boldsymbol{\Lambda}}

\newcommand{\bD}{{\mathbf D}}
\newcommand{\bzero}{{\mathbf 0}}

\newcommand{\ve}{{\varepsilon}}
\renewcommand{\epsilon}{{\ve}}
\renewcommand{\hat}{\widehat}

\def\JRSSB{{\sl Journal of the Royal Statistical Society}, {\bf B}}

\def\BKA{{\sl Biometrika}}
\def\JASA{{\sl Journal of the American Statistical Association}}

\begin{document}

\title{\bf A Two-Way Transformed Factor Model for Matrix-Variate Time Series}

\author{
Zhaoxing Gao$^1$ and Ruey S. Tsay$^2$ \\
$^1$Department of Mathematics, Lehigh University\\
$^2$Booth School of Business, University of Chicago
}

 \date{}

\maketitle

\begin{abstract}
We propose a new framework for modeling high-dimensional matrix-variate time series by a 
two-way transformation, where the transformed data consist of a matrix-variate factor process, which is dynamically dependent, and three other blocks of white noises. Specifically, for a given $p_1\times p_2$ matrix-variate time series, we  seek common nonsingular transformations to project the rows and columns onto  another $p_1$ and $p_2$  directions according to the strength of the dynamic dependence of the series on the past values. Consequently, we treat the data as nonsingular  linear row and column transformations of dynamically dependent common factors and white noise idiosyncratic components. We propose a common orthonormal projection method to estimate the front and back loading matrices of the matrix-variate factors. Under the setting that the largest eigenvalues of the covariance of the vectorized idiosyncratic term diverge for large $p_1$ and $p_2$, 
we  introduce a two-way projected Principal Component Analysis (PCA) to estimate the associated loading matrices of the idiosyncratic terms to  mitigate such diverging noise effects. A diagonal-path white noise testing procedure is proposed to estimate the order of the factor matrix. 
%under the assumption that the idiosyncratic term is a matrix-variate white noise process.  
Asymptotic properties of the proposed method are established for both fixed and diverging dimensions as the sample size increases to infinity. We use simulated and real examples to assess the performance of the proposed method. We also compare our method with some existing ones in the literature and find that the proposed approach not only provides interpretable results but also performs well in out-of-sample forecasting.
\end{abstract}


\newpage

\section{Introduction}
%In the era of big data, the volume, scale and structure of contemporary data pose new and exciting challenges that cannot be tackled by the traditional statistical methods, and 
Modern scientific studies often  collect data under combinations of multiple factors. For example, neuroimaging experiments record brain activities at multiple spatial locations and multiple time points under a variety of experimental stimuli. Studies of social networks record social links for a variety of settings from multiple initiators of social activity to multiple receivers of the activity.  Data such as these are naturally represented not as lists or tables of numbers, but as multi-indexed arrays, or tensors. As many types of such data are collected over time, it is natural to view them as tensor-valued time series. 
The matrix-variate time series is a sequence of second-order random tensors. For example, financial and economic studies often collect data from different countries with a number of economic indicators (e.g., growth rate of the gross domestic product, unemployment rate, etc.) every quarter. Therefore, it is important and interesting to develop appropriate statistical methods to analyze such data. 
The most commonly used approach to modeling such data is to stack the matrix into a long vector and to apply the standard multivariate methods. However, such an approach  ignores the matrix structure of the data and often overlooks some important patterns embedded in the data. For example, \cite{werner2008} pointed out that after vectorizing the matrices the resulting vectors have a Kronecker structure, and ignoring this structure means that a much larger number of parameters need to be estimated. Furthermore, the dimension of a matrix-variate time series itself can become 
large in the current era of big data.  
Therefore, it is important to make use of the matrix structure and  to find an effective way to reduce the number of parameters, especially when the dimension is high. There are some works on tensor 
time series, e.g., \cite{rogers2013} and \cite{surana2016}, but these articles focus on data processing 
rather than on statistical properties or the high dimensional case. 

In modeling vector time series, the available methods to reduce the number of parameters  
can be classified in two categories: regularization and dimension reduction. The former imposes some conditions on the structure of a vector autoregressive moving-average (VARMA) model, and the latter assumes there is a lower dimensional representation for the high-dimensional process. For the regularization methods, some special structures are often imposed on the VARMA model. For example, Chapter 4 of \cite{Tsay_2014} and the references therein discussed two different canonical structures.  \cite{Davis2012} studied the VAR model with sparse coefficient matrices based on partial spectral coherence. The Lasso regularization has also been applied to VAR models, see \cite{ShojaieMichailidis_2010}, \cite{SongBickel_2011}, and \cite{HanTsay2020}, among others.  
For dimension reduction, popular methods include the canonical correlation analysis (CCA) of \cite{BoxTiao_1977}, the principle component analysis (PCA) of \cite{StockWatson_2002}, the scalar component analysis of \cite{TiaoTsay_1989}. The factor model approach can be found in \cite{BaiNg_Econometrica_2002}, \cite{StockWatson_2005}, \cite{forni2000,forni2005}, \cite{panyao2008}, \cite{LamYaoBathia_Biometrika_2011}, \cite{lamyao2012}, \cite{gaotsay2018a,gaotsay2020a,gaotsay2018b}, among others. However, none of the methods mentioned above can directly be used to model matrix-variate time series without vectorization. 
The matrix-variate time series has not been well studied in the literature; \cite{walden2002} handled this type of data in signal and image processing, \cite{wang2018} proposed a factor model for matrix-variate time series, which maintains and utilizes the matrix structure to achieve the dimension reduction, and  \cite{chentsaychen2018} studied the constrained matrix-variate factor models by incorporating domain or prior knowledge in the model through linear constraints. However, the mechanism of the proposed matrix factor model deserves a further study and the bounded eigenvalue assumption of the covariance matrix of the vectorized idiosyncratic term is often violated in the high-dimensional setting, especially for the notable case of low signal-to-noise ratio commonly seen in finance and economics. See, for example, \cite{black1986}.

The  goal of this paper is to study the common dynamic dependence of matrix-variate time series   from a new perspective. We first illustrate our primitive idea below and propose 
our approach in Section 2. Let $\bY_t\in\mathbb{R}^{p_1\times p_2}$ be an observable 
matrix-variate time series. For simplicity, we assume that $\bY_t$ is weakly stationary 
with $E(\bY_t)$ = $\bzero$.
 We postulate that there exist two full-rank matrices $\bT_1\in R^{p_1\times p_1}$ and $\bT_2\in R^{p_2\times p_2}$ such that $\bT_1\bY_t\bT_2'$ is of the form
\begin{equation}\label{int:m}
\bT_1\bY_t\bT_2'=\left[\begin{array}{cc}
\bF_t&\bZ_{12,t}\\
\bZ_{21,t}&\bZ_{22,t}
\end{array}\right],
\end{equation}
where $\bF_t\in\mathbb{R}^{r_1\times r_2}$ is a matrix-variate factor that captures the dynamic dependence of $\bY_t$,  and $\bZ_{12,t}$, $\bZ_{21,t}$ and $\bZ_{22,t}$ are matrix-variate  idiosyncratic components, which are white noise processes. Equivalently, model (\ref{int:m}) is to seek two nonsingular transformation matrices $\bL=(\bL_1,\bL_2):=\bT_1^{-1}$ and $\bR=(\bR_1,\bR_2):=\bT_2^{-1}$ with $\bL_1\in \mathbb{R}^{p_1\times r_1}$ and $\bR_1\in \mathbb{R}^{p_2\times r_2}$ such that $\bL_1$ and $\bR_1$ are the front and back loading matrices associated with the common factors.
To see the rationale of model (\ref{int:m}), let $\vc(\cdot)$ be the conventional vectorization operator that converts a matrix to a vector by stacking columns of the matrix on top of each other.  By the basic properties of Kronecker product, we rewrite the model in the following vector form:
\begin{equation}\label{vec:m}
\by_t:=\vc(\bY_t)=\bA\left[\begin{array}{c}
\bff_t\\
\bz_t
\end{array}\right],
\end{equation}
where $\bA=[\bR_1\otimes\bL_1,\bR_1\otimes\bL_2,\bR_2\otimes\bL_1,\bR_2\otimes\bL_2]\in\mathbb{R}^{p_1p_2\times p_1p_2}$, $\bff_t=\vc(\bF_t)\in\mathbb{R}^{r_1r_2}$ and $\bz_t=[\vc(\bZ_{21,t})',\vc(\bZ_{12,t})',\vc(\bZ_{22,t})']'\in\mathbb{R}^{p_1p_2-r_1r_2}$. 
For identifiability, we assume that both $\bff_t$ and $\bz_t$ have zero mean and identity  
covariance matrices. This is a special case of the model considered in \cite{gaotsay2018b} for vector time series by assuming that the covariance of the vectorized data has a Kronecker structure. That is, we expect there exists a transformation matrix $\bA^{-1}$ with a Kronecker structure such that $\bA^{-1}\by_t=(\bff_t',\bz_t')'$, and this can be done via canonical correlation analysis between $\by_t$ and its past lagged variables, and the resulting vector $(\bff_t',\bz_t')'$ are contemporaneously uncorrelated with an identity covariance matrix. See the discussions in \cite{gaotsay2018b} and \cite{TiaoTsay_1989}. The structure of $\bA$ is different from that in \cite{gaotsay2018b} in order to preserve the structure of the matrix-valued data. Consequently,  the main task of the proposed 
method is to estimate $\bL_1$, $\bR_1$ and to recover the matrix factor $\bF_t$.

To summarize, we propose in this paper a new framework for statistical modeling of 
matrix-variate time series based on the aforementioned motivation and the concepts of factor models.  We reparametrize the model by compressing the strengths of the linear transformation matrices to the corresponding factors and idiosyncratic components, and the resulting front and back loading matrices associated with the common factor  and the idiosyncratic terms are all half-orthonormal. Our first step is to find common orthonormal projections for the row and column vectors respectively based on an eigen-analysis of certain matrices, and the top few projected coordinates form a matrix-variate  common factor process. The rest of the projected coordinates form a matrix-variate white noise process. When recovering the factor matrix, we introduce a two-way projected principal component analysis (PCA) to estimate the loading matrices associated with the idiosyncratic matrix; see Section 2 for details. In the presence of diverging noise components, the projected PCA helps to mitigate the effect of the idiosyncratic component in estimating the common factor matrix. Furthermore, we propose a diagonal-path selection method to estimate the order (dimension) of the factor matrix based on a white noise testing procedure. The testing procedure is more reasonable and statistically interpretable than the ratio-based method in \cite{wang2018}, which essentially follows the method in \cite{LamYaoBathia_Biometrika_2011}. Consequently, the extracted matrix-variate factors capture most of the dynamic dependence of the data and  is useful if one is interested in out-of-sample forecasting of matrix-variate time series. 
An autoregressive type of model can be used to model the low-dimensional common factor process.  See, for example, the model in \cite{chenxiaoyang2020}.
Asymptotic properties of the proposed method are established for both fixed and diverging dimensions as the sample size $n$ tends to infinity. We use simulated and real examples to assess the performance of the proposed method.

The rest of the paper is organized as follows. We introduce the proposed model and 
estimation methodology in Section 2.  
 In Section 3, we study the theoretical properties of the proposed model and its associated 
 estimates. 
Numerical illustrations with both simulated and real data sets are
reported in Section 4. Section 5 provides concluding remarks. 
All technical proofs are given in an Appendix. Throughout the article,
 we use the following notation. For a $p\times 1$ vector
$\bu=(u_1,..., u_p)',$  $||\bu||_2 =\|\bu'\|_2= (\sum_{i=1}^{p} u_i^2)^{1/2} $
is the Euclidean norm, and $\bI_p$ denotes a $p\times p$ identity matrix. For a matrix $\bH=(h_{ij})$, $|\bH|_\infty=\max_{i,j}|h_{ij}|$,  $\|\bH
\|_F=\sqrt{\sum_{i,j}h_{ij}^2}$ is the Frobenius norm, $\|\bH
\|_2=\sqrt{\lambda_{\max} (\bH' \bH ) }$ is the operator norm, where
$\lambda_{\max} (\cdot) $ denotes for the largest eigenvalue of a matrix, and $\|\bH\|_{\min}$ is the square root of the minimum non-zero eigenvalue of $\bH'\bH$. The superscript ${'}$ denotes 
the transpose of a vector or matrix. We also use the notation $a\asymp b$ to denote $a=O(b)$ and $b=O(a)$.

\section{Models and Methodology}

\subsection{Setting}
Let $\bY_t= [y_{ij,t}] = (\by_{1,t},...,\by_{p_2, t})$ be an observable $p_1\times p_2$ matrix-variate  time series with $\by_{jt}=(y_{1j,t},...,y_{p_1 j, t})'\in \mathbb{R}^{p_1}$ and $E(\by_{jt})={\bf 0}$ for $1\leq j\leq p_2$. We assume $\bY_t$ admits a latent structure:
\begin{equation}\label{m-factor}
\bY_t=\bL\left[\begin{array}{cc}
\bF_t&\bZ_{12,t}\\
\bZ_{21,t}&\bZ_{22,t}
\end{array}\right]\bR'=\bL_1\bF_t\bR_1'+\bL_2\bZ_{21,t}\bR_1'+\bL_1\bZ_{12,t}\bR_2'+\bL_2\bZ_{22,t}\bR_2',
\end{equation}
where $\bF_t\in \mathbb{R}^{r_1\times r_2}$ is a matrix-variate common factor process, $\bZ_{12,t}\in \mathbb{R}^{r_1\times v_2}$, $\bZ_{21,t}\in\mathbb{R}^{v_1\times r_2}$, and $\bZ_{22,t}\in \mathbb{R}^{v_1\times v_2}$ are matrix-variate idiosyncratic noise processes with $r_1+v_1=p_1$ and $r_2+v_2=p_2$.
$\bL=(\bL_1,\bL_2)\in \mathbb{R}^{p_1\times p_1}$ is the front loading matrix with $\bL_1\in \mathbb{R}^{p_1\times r_1}$ and $\bL_2\in \mathbb{R}^{p_1\times v_1}$, and $\bR=(\bR_1,\bR_2)\in \mathbb{R}^{p_2\times p_2}$ is the back loading matrix with $\bR_1\in \mathbb{R}^{p_2\times r_2}$ and $\bR_2\in \mathbb{R}^{p_2\times v_2}$. We assume $\bL$ and $\bR$ are full-rank so that $\bF_t$, $\bZ_{12,t}$, $\bZ_{21,t}$, and $\bZ_{22,t}$ can be viewed as transformed processes by applying the inverses of $\bL$ and $\bR'$, respectively, to the left and right of the data matrix $\bY_t$ as discussed in Section 1. Furthermore, letting $\bff_t$ and $\bz_t$ be the vectorized factor and idiosyncratic terms, we assume that $\Cov(\bff_t)=\bI_{r_1r_2}$ and $\Cov(\bz_t)=\bI_{p_1p_2-r_1r_2}$. This assumption holds because one can adjust the 
scales of $\bL$ and $\bR$ accordingly.  Therefore, the three noise terms are uncorrelated with each other and individually identified.  Model (\ref{m-factor})  is general if one allows $r_1$, $r_2$, $v_1$ and $v_2$ to be zero, but for effective dimension reduction, $r_1$ and $r_2$ should be small and  fixed positive integers. In addition, we assume $\bff_t$ and $\bz_s$ are uncorrelated for any $t$ and $s$. This is only for the simplicity in illustration, and it can be relaxed by imposing some dynamic dependence between $\bff_t$ and $\bz_s$. See \cite{gaotsay2018b} for details. 
We do not pursue it here. Note that $\bL$ and $\bR$ are not uniquely identified because 
$c\bL$ and $\bR/c$, where $c\neq 0$, also holds for Equation (\ref{m-factor}).

To proceed, we further decompose $\bL$ and $\bR$ as follows:
\[\bL_1=\bA_1\bW_1,\quad\bL_2=\bA_2\bW_2,\quad\bR_1=\bP_1\bG_1,\,\, \text{and}\,\,\bR_2=\bP_2\bG_2,\]
where $\bA_i$ and $\bP_i$ ($i=1,2$) are half orthonormal matrices, i.e., $\bA_i'\bA_i=\bI_{r_i}$ and $\bP_i'\bP_i=\bI_{v_i}$. This can be done via QR or singular value decomposition. Furthermore, let $\bX_t=\bW_1\bF_t\bG_1'$, $\bE_{21,t}=\bW_2\bZ_{21,t}\bG_1'$, $\bE_{12,t}=\bW_1\bZ_{12,t}\bG_2'$, and $\bE_{22,t}=\bW_2\bZ_{22,t}\bG_2'$, then model (\ref{m-factor}) can be rewritten as
\begin{equation}\label{mf}
\bY_t=\bA_1\bX_t\bP_1'+\bA_2\bE_{21,t}\bP_1'+\bA_1\bE_{12,t}\bP_2'+\bA_2\bE_{22,t}\bP_2'.
\end{equation}
Even though $\bL$ and $\bR$ are full rank, $\bA_1$ ($\bP_1$) is not orthogonal to $\bA_2$ ($\bP_2$) in general. Note that model (\ref{mf}) is still not identified since we can replace the triplets ($\bA_1$,$\bX_t$,$\bP_1$) by ($\bA_1\bH_1$,$\bH_1'\bX_t\bH_2$, $\bP_1\bH_2$) for any orthonormal matrices $\bH_1\in\mathbb{R}^{r_1\times r_1}$ and $\bH_2\in\mathbb{R}^{r_2\times r_2}$ without altering the data generating process. The same issue exists for the idiosyncratic terms. Nevertheless the linear spaces spanned by the columns of $\bA_i$ and $\bP_i$, denoted respectively by $\mathcal{M}(\bA_i)$ and $\mathcal{M}(\bP_i)$, are uniquely defined and $\mathcal{M}(\bA_i)=\mathcal{M}(\bL_i)$ and $\mathcal{M}(\bP_i)=\mathcal{M}(\bR_i)$.

\subsection{Common Orthonormal Projections}
To illustrate our estimation method, we first introduce some notation. For $i=1,2$, let $\bB_i$ and $\bQ_i$ be the orthonormal complements of $\bA_i$ and $\bP_i$, respectively, i.e., $\bB_i\in\mathbb{R}^{p_i\times r_i}$ and $\bQ_i=\mathbb{R}^{p_i\times v_i}$ are half orthonormal matrices with $\bB_i'\bA_i=\bf 0$ and $\bQ_i'\bP_i=\bf{0}$. Furthermore, denote $\boldsymbol{\ell}_{i,j}$, $\br_{i,j}$, $\ba_{i,j}$, $\bb_{i,j}$, $\bp_{i,j}$ and $\bq_{i,j}$  the $j$-th columns of $\bL_i$, $\bR_i$, $\bA_i$, $\bB_i$, $\bP_i$ and $\bQ_i$, respectively, where the range of $j$ depends on the dimension of the corresponding matrix. 

Let $\bfeta_{t}=[\vc(\bY_{t-1})',...,\vc(\bY_{t-k_0})']'$ be the vector of past $k_0$ lagged values of $\bY_t$, where $\vc(\bY_t)=(\by_{1,t}',...,\by_{p_2,t}')'$ and $k_0$ is a prescribed positive integer. 
 Define  $\bSigma_{y,ij}(k)=\cov(\by_{i,t},\by_{j,t-k})$. We seek the direction $\ba\in \mathbb{R}^{p_1}$ that solves the following optimization problem:
\begin{equation}\label{opm}
\max_{\ba\in\mathbb{R}^{p_1}}\sum_{i=1}^{p_2}\|\cov(\ba'\by_{i,t},\bfeta_t)\|_2^2,\quad\text{subject to}\quad \ba'\ba=1.
\end{equation}
That is, we look for a common direction $\ba$ with $\ba'\ba=1$ such that it maximizes the sum of the covariance between $\ba'\by_{i,t}$ and the past lagged variables, which characterize the dynamic dependence of the columns.
Note that
\[\sum_{i=1}^{p_2}\|\cov(\ba'\by_{i,t},\bfeta_t)\|_2^2=\ba'\left[\sum_{k=1}^{k_0}\sum_{i=1}^{p_2}\sum_{j=1}^{p_2}\bSigma_{y,ij}(k)\bSigma_{y,ij}(k)'\right]\ba.\]
Then, $\ba$ is an eigenvector of the matrix
\begin{equation}\label{m1}
\bM_1=\sum_{k=1}^{k_0}\sum_{i=1}^{p_2}\sum_{j=1}^{p_2}\bSigma_{y,ij}(k)\bSigma_{y,ij}(k)'.
\end{equation}
On the other hand, under model (\ref{mf}), let $\bp_{1,i\sbullet}$  be the $i$-th  row vectors of $\bP_1$ and  define $\bSigma_{xp,ij}(k)=\cov(\bX_t\bp_{1,i\sbullet}',\bX_{t-k}\bp_{1,j\sbullet}')$. Then
\begin{equation}\label{sigma:k}
\bSigma_{y,ij}(k)=\bA_1\bSigma_{xp,ij}(k)\bA_1',
\end{equation}
where we assume $\bff_t$ and $\bz_s$ are uncorrelated for any $t$ and $s$.  Therefore,
\begin{equation}\label{M1:x}
\bM_1=\bA_1\left\{\sum_{k=1}^{k_0}\sum_{i=1}^{p_2}\sum_{j=1}^{p_2}[\bSigma_{xp,ij}(k)\bSigma_{xp,ij}(k)']\right\}\bA_1'.
\end{equation}
We observe that $\bM_1\bB_1=\bf 0$, that is, the columns of $\bB_1$ are the eigenvectors associated with the zero eigenvalues of $\bM_1$, and the front factor loading space $\mathcal{M}(\bA_1)$ is spanned by the eigenvectors corresponding to the $r_1$ non-zero eigenvalues of $\bM_1$. Equivalently, the space spanned by the first $r_1$ solutions to the problem (\ref{opm}) are just the front factor loading space $\mathcal{M}(\bA_1)$.

The $r_2$ orthonormal directions of the columns of $\bP_1$ can be obtained by performing the same procedure on the the transpose of $\bY_t$'s. We can similarly construct $\bM_2$ as $\bM_1$ in (\ref{m1}) such that $\bM_2\bQ_1=\bf 0$, and therefore, $\mathcal{M}(\bP_1)$ is the space spanned by the first $r_2$ non-zero eigenvectors of $\bM_2$. We omit the details here.

\subsection{Two-Way Projected Principal Component Analysis}
In this section, we introduce the idea of a 2-way projected PCA in order to recover the true factor matrix $\bX_t$ and, hence, $\bF_t$. Using the notation in Section 2.2,  it follows from model (\ref{mf}) that
\begin{equation}\label{byq}
\bB_1'\bY_t\bQ_1=\bB_1'\bA_2\bE_{22,t}\bP_2'\bQ_1,
\end{equation}
which implies that $\bB_1'\bY_t\bQ_1$ is a matrix-variate white noise process, and hence $\{\bb_{1,i}'\bY_t\bq_{1,j}|t=0,\pm 1,...\}$ is a univariate white noise process for all $1\leq i\leq v_1$ and $1\leq j\leq v_2$. Furthermore,
\begin{equation}\label{b2q2}
\bB_2'\bY_t=\bB_2'\bA_1\bX_t\bP_1'+\bB_2\bA_1\bE_{12,t}\bP_2'\,\,\text{and}\,\, \bY_t\bQ_2=\bA_1\bX_t\bP_1'\bQ_2+\bA_2\bE_{21,t}\bP_1'\bQ_2.
\end{equation}
Therefore, $\bB_2'\bY_t$ and $\bY_t\bQ_2$ are uncorrelated with $\bB_1'\bY_t\bQ_1$ defined in (\ref{byq}). Let $\bOmega_{y_i}=\cov(\by_{i,t},\vc(\bY_t))$ and $\bOmega_{e_{22,i}p}=\cov(\bE_{22,t}\bp_{2,i\sbullet}',\vc(\bE_{22,t}))$, where  $\bp_{2,i\sbullet}$ is the $i$-th row vector 
of $\bP_2$.
It follows from (\ref{mf}) and (\ref{byq}) that
\begin{equation}\label{omgy}
\cov(\by_{i,t},\vc(\bB_1'\bY_t\bQ_1))=\bOmega_{y_i}(\bQ_1\otimes\bB_1)=\bA_2\bOmega_{e_{22,i}p}(\bP_2'\bQ_1\otimes \bA_2'\bB_1).
\end{equation}
For each $1\leq i\leq p_2$, $\bB_2'\by_{i,t}$ is uncorrelated with $\vc(\bB_1'\bY_t\bQ_1)$ and we define
\begin{equation}\label{s1}
\bS_1:=\sum_{i=1}^{p_2}[\bOmega_{y_i}(\bQ_1\otimes\bB_1)][\bOmega_{y_i}(\bQ_1\otimes\bB_1)]',
\end{equation}
from which we can see, via (\ref{omgy}), that $\bS_1\bB_2=\bzero$. In addition, the rank of $\bS_1\in \mathbb{R}^{p_1\times p_1}$ is $v_1$, therefore, $\bB_2$ contains all the eigenvectors corresponding to the zero eigenvalues of $\bS_1$. From the form of $\bS_1$, we can see that the middle component contains the information of the noise, and we seek the direction $\bb_{2,j}\in \mathbb{R}^{p_2}$ such that $\bb_{2,j}'\by_{i,t}$ minimizes the covariance between the projected direction and the noise, and therefore, $\bb_{2,j}'\by_{i,t}$ contains the information of the signal $\bF_t$. 

Similarly, we can construct $\bS_2$ such that $\bS_2\bQ_2=0$, and $\bQ_2$ contains all the eigenvectors associated with the zero eigenvalues of $\bS_2$ at the population level. Furthermore, if $\bA_1$, $\bP_1$, $\bB_2$ and $\bQ_2$ are known, it follows from (\ref{mf}) that
\begin{equation}\label{r:ft}
\bB_2'\bY_t\bQ_2=\bB_2'\bA_1\bX_t\bP_1'\bQ_2,
\end{equation}
and consequently, 
\begin{equation}\label{ft:e}
\bX_t=(\bB_2'\bA_1)^{-1}\bB_2'\bY_t\bQ_2(\bP_1'\bQ_2)^{-1},
\end{equation}
where $\bB_2'\bA_1\in\mathbb{R}^{r_1\times r_1}$ and $\bP_1'\bQ_2\in\mathbb{R}^{r_2\times r_2}$ are two invertible matrices. To see this, note that $\bL$ is a matrix of full rank, thus there exist matrices $\bH_1\in \mathbb{R}^{r_1\times r_1}$ and $\bH_2\in\mathbb{R}^{v_1\times r_1}$ such that 
\[\bB_2=\bL_1\bH_1+\bL_2\bH_2=\bA_1\bW_1\bH_1+\bA_2\bW_2\bH_2.\]
Then,
\[\bI_{r_1}=\bB_2'\bB_2=\bB_2'\bA_1\bW_1\bH_1,\]
implying that rank$(\bB_2'\bA_1)=r_1$ which is of full rank. The invertibility of $\bP_1'\bQ_2$ follows from a similar argument.

\subsection{Estimation}
In practice, given a sample $\{\bY_t:t=1,...,n\}$, the goal is to estimate $\bA_1$ and $\bP_1$ or equivalently $\mathcal{M}(\bA_1)$ and $\mathcal{M}(\bP_1)$, the dimension $(r_1,r_2)$ of the matrix factor, and to recover the latent factor matrix process $\bX_t$. To illustrate the main idea, we first assume $(r_1,r_2)$ is known, but propose a way to estimate them in the next subsection.

For the estimation of $\bA_1$ and $\bP_1$, we construct the sample version of $\bM_1$ defined in (\ref{m1}) as follows:
\begin{equation}\label{m1hat}
\wh\bM_1=\sum_{k=1}^{k_0}\sum_{i=1}^{p_2}\sum_{j=1}^{p_2}\wh\bSigma_{y,ij}(k)\wh\bSigma_{y,ij}(k)',
\end{equation}
where
\begin{equation}\label{yij:hat}
\wh\bSigma_{y,ij}(k)=\frac{1}{n}\sum_{t=k+1}^{n}(\by_{i,t}-\bar{\by}_i)(\by_{j,t-k},-\bar{\by}_j)',
\end{equation}
and $\bar{\by}_i=n^{-1}\sum_{t=1}^n\by_{i,t}$ which is essentially ${\bf 0}$ if the data are centered. Then, $\mathcal{M}(\bA_1)$ can be estimated by $\mathcal{M}(\wh\bA_1)$, where $\wh\bA_1=(\wh\ba_{1,1},...,\wh\ba_{1,r_1})$ with $\wh\ba_{1,1},...,\wh\ba_{1,r_1}$ being the eigenvectors corresponding to the $r_1$ largest eigenvalues of $\wh\bM_1$. Consequently, the orthogonal space $\mathcal{M}(\wh\bB_1)$ can be similarly obtained by $\wh\bB_1=(\wh\bb_{1,1},...,\wh\bb_{1,v_1})$, where $\wh\bb_{1,1},...,\wh\bb_{1,v_1}$ are the eigenvectors corresponding to the $v_1$ smallest eigenvalues of $\wh\bM_1$.

By a similar procedure on $\{\bY_t',t=1,...,n\}$, we can construct $\wh\bM_2$, and the estimator $\wh\bP_1$ for $\bP_1$ is then obtained. Once we have the estimators $\wh\bA_1$ and $\wh\bP_1$,  we consider methods for obtaining the estimators of $\bB_2$ and $\bQ_2$. The choices of the estimators $\wh\bB_2$ and $\wh\bQ_2$ are different for small and large dimensions. 
We only discuss the cases when $p_1$ and $p_2$ are both small or large, and the case when  one of them is small can be solved by applying both methods jointly. Let 
\begin{equation}\label{s1:hat}
\wh\bS_1=\sum_{i=1}^{p_2}[\wh\bOmega_{y_i}(\wh\bQ_1\otimes\wh\bB_1)][\wh\bOmega_{y_i}(\wh\bQ_1\otimes\wh\bB_1)]',
\end{equation}
where $\wh\bOmega_{y_i}$ is the sample estimator of $\bOmega_{y_i}$ defined in Section 2.3. 
When $p_1$ and $p_2$ are small, we perform an eigen-analysis on $\wh\bS_1$, and $\wh\bB_2=(\wh\bb_{2,1},...,\wh\bb_{2,r_1})$, where $\wh\bb_{2,1},...,\wh\bb_{2,r_1}$ are the eigenvectors of $\bS_1$ corresponding to its $r_1$ smallest eigenvalues. We can similarly obtain $\wh\bQ_1$ based on the eigen-analysis on $\wh\bS_2$, which is calculated based on the transposed data $\bY_t'$.

When the dimensions $p_1$ and $p_2$ are relatively large, the choices of $\wh\bB_2$ and $\wh\bQ_2$ by selecting the eigenvectors associated with the smallest $r_1$ and $r_2$ eigenvalues 
of $\wh\bS_1$ and $\wh\bS_2$ respectively may not fare well because the linear spaces spanned by the chosen vectors are not consistent to the true ones in the high-dimensional case. Suppose that the elements $z_{ij,t}$ of $\bZ_{22, t}$ are independent of each other for $1\leq i\leq v_1$ and $1\leq j\leq v_2$. A reasonable assumption is that the top eigenvalues of the covariance matrix of the idiosyncratic component $\vc(\bL_2\bZ_{22,t}\bR_2')$ or equivalently $\vc(\bA_2\bE_{22,t}\bP_2')$ are diverging. Thus, we assume the top singular values of $\bL_2$ and $\bR_2$ are diverging. See also Assumption 4 in Section 3. We can partition the singular vectors $\bA_2=(\bA_{21},\bA_{22})$ and $\bP_2=(\bP_{21},\bP_{22})$ with $\bA_{21}\in \mathbb{R}^{p_1\times k_1}$ and $\bP_{21}\in \mathbb{R}^{p_2\times k_2}$ which correspond to the $k_1$ and $k_2$ diverging singular values of $\bL_2$ and $\bR_2$, respectively. Let  $\bB_2^*=(\bA_{22},\bB_{2})\in \mathbb{R}^{p_1\times (p_1-k_1)}$ and $\bQ_2^*=(\bP_{22},\bQ_2)\in \mathbb{R}^{p_2\times(p_2-k_2)}$. Under the assumption that the top $k_1$ singular values of $\bL_2$ and $k_2$ of $\bR_2$ are diverging, we can consistently estimate the spaces $\mathcal{M}(\bA_{21})$ and $\mathcal{M}(\bQ_{21})$ and hence their orthogonal parts $\mathcal{M}(\bB_2^*)$ and $\mathcal{M}(\bQ_2^*)$. From the above discussion, $\mathcal{M}(\bB_2)$ and $\mathcal{M}(\bQ_2)$ are subspaces of $\mathcal{M}(\bB_2^*)$ and $\mathcal{M}(\bQ_2^*)$, respectively. Once we have the consistent estimators for $\bB_2^*$ and $\bQ_2^*$, denoted by $\wh{\bB}_2^*$ and $\wh{\bQ}_2^*$, respectively, there are half orthonormal matrices $\bXi_1\in\mathbb{R}^{(p_1-k_1)\times r_1 }$ and $\bXi_2\in\mathbb{R}^{(p_2-k_2)\times r_2}$ such that $\wh\bB_2=\wh{\bB}_2^*\bXi_1$ and $\wh\bQ_2=\wh{\bQ}_2^*\bXi_2$. In practice, it is not easy to find $\bXi_1$ and $\bXi_2$ such that  $\wh\bB_2$ and $\wh\bQ_2$ are consistent to $\bB_2$ and $\bQ_2$. Nevertheless, any choices of $\bXi_1$ and $\bXi_2$ can mitigate the diverging effect of the top eigenvalues since they are all orthogonal to $\wh\bA_{21}$ and $\wh\bP_{21}$, respectively. Thus, we only need to guarantee the invertiblities of the the matrices $\wh\bB_2'\wh\bA_1$ and $\wh\bP_1'\wh\bQ_2$ in order to recover the latent factors.

In practice, with the estimators $\wh{\bB}_2^*$ and $\wh{\bQ}_2^*$,  the  columns of $\bXi_1$ are chosen as the $r_1$ eigenvectors of $\wh{\bB}_2^*{'}\wh\bA_1\wh\bA_1'\wh{\bB}_2^*$ 
corresponding the $r_1$ largest eigenvalues, and the columns of $\bXi_2$ are the $r_2$ eigenvectors of $\wh{\bQ}_2^*{'}\wh\bP_1\wh\bP_1'\wh{\bQ}_2^*$ corresponding to the largest $r_2$ eigenvalues. These choices guarantee that both $\wh\bB_2'\wh\bA_1$ and $\wh\bP_1'\wh\bQ_2$ behave well in practical calculations. Finally, we recover the latent factor matrix as
\begin{equation}\label{rec:ft}
\wh\bX_t=(\wh\bB_2'\wh\bA_1)^{-1}\wh\bB_2'\bY_t\wh\bQ_2(\wh\bP_1'\wh\bQ_2)^{-1}.
\end{equation}

With $\wh\bA_1$, $\wh\bP_1$ and the estimated factor process $\wh\bX_t$, we can make an $h$-step ahead prediction for the $\bY_t$ series using the formula $\wh\bY_{n+h}=\wh\bA_1\wh\bX_{n+h}\wh\bP_1'$, where $\wh\bX_{t+h}$ is an $h$-step ahead forecast for $\bX_t$ based on the estimated past values $\wh\bX_1,...,\wh\bX_n$. This can be done, for example, by fitting a matrix-autoregressive model to $\{\wh\bX_1,...,\wh\bX_n\}$ as, for example, the one introduced in \cite{chenxiaoyang2020}.

\subsection{Diagonal-Path Selections of the Order of Factor Matrix}
The estimation of $\bA_1$, $\bP_1$, and $\bX_t$ of the prior sections are based on given $r_1$ and $r_2$, which are unknown in practice. To the best of our knowledge, there is no efficient method available to estimate them in the literature. The most relevant one is the ratio-based method of  \cite{wang2018}, but it can be shown that the method is not appropriate when the top eigenvalues of the covariance of the idiosyncratic term are diverging. See the simulation results in Section 4. For the vector factor models, there are  some methods available. See, for example, the information criterion in Bai and Ng (2002) and Bai (2003), the random matrix theory method in \cite{onatski2010}, the ratio-based method in Lam and Yao (2012), the canonical correlation analysis in \cite{gaotsay2018a}, and the white noise testing approach in   \cite{gaotsay2018b}, among others. However, those methods
 cannot apply to the matrix-factor models directly. 

In this section, we propose a diagonal-path method to search the dimension $(r_1,r_2)$ by modifying the approach of \cite{gaotsay2018b}.  The idea of our method follows from equation (\ref{byq}) that $\bB_1'\bY_t\bQ_1$ is a matrix-variate white noise process. Let $\wh \bGamma_1$ and $\wh\bGamma_2$ be the matrices of eigenvectors (in the decreasing order of corresponding eigenvalues) of the sample matrix $\wh\bM_1$ in (\ref{m1hat}) and $\wh \bM_2$, respectively.  Define  $\wh\bW_t=\wh\bGamma_1{'}\bY_t\wh\bGamma_2$ and let $\wh\bW_t(i,j)\in\mathbb{R}^{(p_1-i+1)\times (p_2-j+1)}$ be the lower-right submatrix consisting of the $i$-th to the $p_1$-th rows and the $j$-th to the $p_2$-th columns of $\wh\bW_t$, and $\wh\bW_t^*(i,j)\in \mathbb{R}^{(i-1)\times (j-1)}$ be the upper-left  submatrix of $\wh\bW_t$. Our test procedure searches  
the  order $(i,j)$ such that $\wh\bW_t^*(i,j)$ consists of all the factors and the remaining 
elements of $\wh\bW_t$ are white noises. The estimate of $(r_1,r_2)$ is then $(i-1,j-1)$. The testing procedure is discussed  below, and the test statistic used depends on the dimension $p_1p_2$.

If the dimension $p_1p_2$ is small, implying that $\bY_t$ is a low dimensional matrix, we recommend using the well-known Ljung-Box statistic $Q_s(m)$ for multivariate time series, where $s$ and $m$ denote the dimension of the vector and the number of lags used. See, for example, \cite{hosking1980} and \cite{Tsay_2014}. Specifically, we first search the minimum of $r_1$ and $r_2$ along the diagonal of $\wh\bW_t$. Consider the null hypothesis
\[H_0(l): \vc(\wh\bW_t(l,l))\,\, \text{is a vector white noise}, \]
with type-I error $\alpha$. $H_0(l)$ is rejected if $Q_{d_l}(m)\geq \chi_{d_l^2 m,1-\alpha}^2$, where  $d_l=(p_1-l+1)(p_2-l+1)$ is the dimension of $\vc(\wh\bW_t(l,l))$ and $\chi_{d_l^2 m,1-\alpha}^2$ is the $(1-\alpha)$-th quantile of a chi-squared distribution with $d_l^2m$ degrees of freedom. We start with $l=1$. If $H_0(1)$ is rejected, we increase $l$ by 1 and repeat the test until we  cannot reject $H_0(l)$, and denote the resulting  
order as $l^*$. Two situations can happen. If $l^*=\min(p_1,p_2)$ and we still  reject $H_0(l^*)$, we fix one dimension (say $p_1$ when $p_1=l^*$), and test whether $\vc(\wh\bW_t(p_1,p_1+j))$ is white noise or not by starting with $j=1$ until we cannot reject $H_0$. If $l^*<\min(p_1,p_2)$, then we perform a back testing to determine the maximum order of the factor matrix. That is, we first test whether $\vc(\wh\bW_t(l^*-1+i,l^*-1))$ is a vector white noise starting with $i=1$. Increase $i$ by 1 
and repeat the test until we cannot reject $H_0$ at $i=i^*$. Second, we test whether $\vc(\wh\bW_t(l^*+i^*-2,l^*-1+j))$ is a vector white noise starting with $j=1$. Increase $j$ by 1 
and repeat the test until we reject $H_0$ at $j=j^*$. Then, we have $\wh r_1=l^*+i^*-2$ and $\wh r_2=l^*+j^*-2$. Finally, $\wh \bGamma_1=[\wh\bA_1,\wh\bB_1]$ and $\wh\bGamma_2=[\wh\bP_1,\wh\bQ_1]$, where $\wh\bA_1\in\mathbb{R}^{p_1\times \wh r_1}$ and $\wh\bP_1\in\mathbb{R}^{p_2\times \wh r_2}$.

For large $p_1$ and/or $p_2$, we use the same testing procedure, but the multivariate white noise test statistics are no longer adequate.  Instead, some methods  have been developed in recent years to test high-dimensional white noise series. 
We consider two such methods in this paper. The first method is introduced by \cite{changyaozhou2017} and makes use of the maximum absolute auto- and cross-correlations of the component series. The second method of high-dimensional white noise test is by \cite{Tsay_2018} and uses rank correlations and the extreme value theory. The test 
 is simple and easy to use with a close-form limiting distribution under 
some weak assumptions. Details of the two test statistics can be found in \cite{changyaozhou2017} and \cite{Tsay_2018}, respectively. See also the formulation and 
a brief discussion of the two test statistics $T_n$ and $T(m)$ in Section 2.3 of \cite{gaotsay2018b}.

\section{Theoretical Properties}
In this section, we first present the asymptotic theory for the estimation method described in Section 2 assuming $r_1$ and $r_2$ are fixed. The consistency of the white noise test to determine 
$r_1$ and $r_2$ of the matrix factor is shown thereafter. The conventional asymptotic properties are established under the setting that the sample size $n$ tends to $\infty$ and everything else is fixed. Modern time series analysis encounters the situation that the number of time series $p_1p_2$ is as large as, or even larger than, the sample size $n$. We deal with these two settings separately in Sections 3.1 and 3.2 below.

\subsection{Asymptotics When $n\rightarrow\infty$ But $p_1$ and $p_2$ Are Fixed}
We first consider asymptotic properties under the assumption that $n\rightarrow\infty$ with $p_1$ and $p_2$ being fixed. These properties reflect the behavior of our estimation method when $n$ is large and the dimensions $p_1$ and $p_2$ are relatively small. We begin with some assumptions.

%%%%%%%%%%%%%%%%%%%%
\begin{assumption}
The process $\{\vc(\bY_t),\vc(\bF_t)\}$ is $\alpha$-mixing with the mixing coefficient satisfying the condition $\sum_{k=1}^\infty\alpha_p(k)^{1-2/\gamma}<\infty$ for some $\gamma>2$, where
\[\alpha_p(k)=\sup_{i}\sup_{A\in\mathcal{F}_{-\infty}^i,B\in \mathcal{F}_{i+k}^\infty}|P(A\cap B)-P(A)P(B)|,
\]
and $\mathcal{F}_i^j$ is the $\sigma$-field generated by $\{(\vc(\bY_t),\vc(\bF_t)):i\leq t\leq j\}$.
\end{assumption}
%%%%%%%%%%%%%%%%%%%%%%%%%%%%%%%%%%%%%%%%
\begin{assumption}
For any $i=1,...,r_1 r_2$ and $1\leq j\leq p_1p_2-r_1 r_2$, $E|f_{i,t}|^{2\gamma}<C_1$ and $E|z_{j,t}|^{2\gamma}<C_2$, where $f_{i,t}$ and $z_{j,t}$ are the $i$-th and $j$-th element of $\bff_t$ and $\bz_t$, respectively, $C_1$ and $C_2>0$ are constants, and $\gamma$ is given in Assumption 1.
\end{assumption}
%%%%%%%%%%%%%%%%%%%%%%%%%%%%%%

%%%%%%%%%%%%%%%%%%%%%%%%%%%%%
Assumption 1 is standard for dependent random processes. See \cite{gaoetal2017} for a theoretical justification for VAR models. The conditions in Assumption 2 imply that $E|y_{ij,t}|^{2\gamma}<C$ under the setting that $p_1$ and $p_2$ are fixed. 
To this end, we adopt the discrepancy measure used by
\cite{panyao2008}: for two $p\times r$ half orthogonal
matrices ${\bf H}_1$ and ${\bf H}_2$ satisfying the condition ${\bf
H}_1'{\bf H}_1={\bf H}_2'{\bf H}_2=\bI_{r}$, the difference
between the two linear spaces $\mathcal{M}({\bf H}_1)$ and
$\mathcal{M}({\bf H}_2)$ is measured by
\begin{equation}
D(\mathcal{M}({\bf H}_1),\mathcal{M}{\bf
H}_2)=\sqrt{1-\frac{1}{r}\textrm{tr}({\bf H}_1{\bf H}_1'{\bf
H}_2{\bf H}_2')}.\label{eq:D}
\end{equation}
Note that $D(\mathcal{M}({\bf H}_1),\mathcal{M}{\bf
H}_2) \in [0,1].$ 
It is equal to $0$ if and only if
$\mathcal{M}({\bf H}_1)=\mathcal{M}({\bf H}_2)$, and to $1$ if and
only if $\mathcal{M}({\bf H}_1)\perp \mathcal{M}({\bf H}_2)$.  The following theorem establishes the consistency of the estimated loading matrices $\wh\bA_1$ and $\wh\bP_1$, their orthonormal complements $\wh\bB_1$ and $\wh\bQ_1$, the matrices $\wh\bB_2$ and $\wh\bQ_2$, and the extracted common factor  $\wh\bA_1\wh\bX_t\wh\bP_1'$.

%%%%%%%%%%%%%%

\begin{theorem}
Suppose Assumptions 1-2 hold and $(r_1,r_2)$ are known and fixed. Then, for fixed $p_1$ and $p_2$, 
%As $n\rightarrow\infty$ (but $p$ fixed),
\[D(\mathcal{M}(\wh\bA_1),\mathcal{M}(\bA_1))=O_p(n^{-1/2}),\quad D(\mathcal{M}(\wh\bB_1),\mathcal{M}(\bB_1))=O_p(n^{-1/2}),\]
\[D(\mathcal{M}(\wh\bP_1),\mathcal{M}(\bP_1))=O_p(n^{-1/2}),\quad D(\mathcal{M}(\wh\bQ_1),\mathcal{M}(\bQ_1))=O_p(n^{-1/2}) \]
and
\[\quad D(\mathcal{M}(\wh\bB_2),\mathcal{M}(\bB_2))=O_p(n^{-1/2}),\quad D(\mathcal{M}(\wh\bQ_2),\mathcal{M}(\bQ_2))=O_p(n^{-1/2}), \]
as $n\rightarrow\infty$. 
Furthermore,
\[\|\wh\bA_1\wh\bX_t\wh\bP_1'-\bA_1\bX_t\bP_1'\|_2=O_p(n^{-1/2}).\]
\end{theorem}

From Theorem 1 and as expected, the convergence rates of all estimates 
are standard at $\sqrt{n}$, which is commonly seen in the traditional statistical theory. If the 
largest $r_1$ and $r_2$ eigenvalues of $\bM_1$ and $\bM_2$ are distinct, then $\bA_1$ and $\bP_1$ are uniquely defined up to a change of signs in columns. 
In fact, 
the consistency of the linear spaces of $\mathcal{M}(\bB_1)$ and $\mathcal{M}(\bB_2)$ is more meaningful since their columns correspond to the zero eigenvalues of $\bM_1$ and $\bS_1$, respectively, and they cannot be uniquely characterized.

\subsection{Asymptotics When $n\rightarrow\infty$ and $p_1,p_2\rightarrow\infty$}

Turn to the case of high-dimensional matrices. For vectorized variables, it is well known that if the dimension $p_1p_2$ 
diverges faster than $n^{1/2}$, the sample covariance matrix is no longer a consistent 
estimate of the population covariance matrix. 
When $p_1p_2=o(n^{1/2})$, it is still possible to consistently estimate the factor loading matrix  and the number of common factors. See \cite{gaotsay2018a}  for details. 
%when we deal with the residual processes.  
Therefore, without any additional assumptions on the underlying structure of 
time series, $p_1p_2$ can only be as large as $o(n^{1/2})$. 
To deal with the case of large $p_1p_2$, we impose some conditions on the transformation matrices $\bL$ and $\bR$ of Equation (\ref{m-factor}) and the cross dependence of $\bY_t$. 

%%%%%%%%%%%%%%
\begin{assumption}
(i) $\bL_1=(\boldsymbol{\ell}_1,...,\boldsymbol{\ell}_{r_1})$ and $\bR_1=(\br_1,...,\br_{r_2})$ such that $\|\boldsymbol\ell_i\|_2^2\asymp p_1^{1-\delta_1}$, $\|\br_j\|_2^2\asymp p_2^{1-\delta_1}$, $i=1,...,r_1$, $j=1,...,r_2$, and $\delta_1\in[0,1)$; (ii) For each $i=1,...,r_1$, $j=1,...,r_2$ and $\delta_1$ given in (i), $\min_{\theta_k\in \mathbb{R},k\neq i}\|\boldsymbol\ell_i-\sum_{1\leq k\leq r_1,k\neq i}\theta_k\boldsymbol\ell_k\|_2^2\asymp p_1^{1-\delta_1}$ and $\min_{\theta_k\in \mathbb{R},k\neq j}\|\br_j-\sum_{1\leq k\leq r_2,k\neq j}\theta_k\br_k\|_2^2\asymp p_2^{1-\delta_1}$ .
\end{assumption}

%%%%%%%%%%%%%%%%%
\begin{assumption}
(i) $\bL_2$ and $\bR_2$ admit a singular value decomposition $\bL_2=\bA_2\bD_2\bU_2'$ and $\bR_2=\bP_2\bLambda_2\bV_2'$, where $\bA_2\in \mathbb{R}^{p_1\times v_1}$ and $\bP_2\in \mathbb{R}^{p_2\times v_2}$ are given in Equation (\ref{mf}), $\bD_2=\diag(d_1,...,d_{v_1})$ and $\bU_2\in \mathbb{R}^{v_1\times v_1}$ satisfying $\bU_2'\bU_2=\bI_{v_1}$, $\bLambda_2=\diag(\gamma_1,...,\gamma_{v_2})$, $\bV_2\in \mathbb{R}^{v_2\times v_2}$ satisfying $\bV_2'\bV_2=\bI_{v_2}$;  (ii) There exist finite integers $1\leq k_1< v_1$ and $1\leq k_2<v_2$ such that $d_1\asymp...\asymp d_{k_1}\asymp p_1^{(1-\delta_2)/2}$ and $\gamma_1\asymp...\asymp\gamma_{k_2}\asymp p_2^{(1-\delta_2)/2}$ for some $\delta_2\in[0,1)$ and $d_{k_1+1}\asymp...\asymp d_{v_1}\asymp 1\asymp\gamma_{k_2+1}\asymp...\asymp\gamma_{v_2}$.
\end{assumption}

%%%%%%%%%%%%%%%%%%%%%%%%%%%%%%
\begin{assumption}
(i) For any $1\leq l_1\leq v_1$, $1\leq l_2\leq v_2$, $\bh\in \mathbb{R}^{l_1\times l_2}$, $\bU\in\mathbb{R}^{v_1\times l_1 }$ and $\bV\in \mathbb{R}^{v_2\times l_2}$ with $\|\bh\|_2=c<\infty$, $\bU'\bU=\bI_{l_1}$ and $\bV'\bV=\bI_{l_2}$, we assume $E|\bh'\vc(\bU'\bZ_{22,t}\bV)|^{2\gamma}<\infty$; (ii)  $\sigma_{\min}(\bXi_1'\bB_2^{*}{'}\bA_1)\geq C_3$ and $\sigma_{\min}(\bXi_2'\bQ_2^{*}{'}\bP_1)\geq C_4$ for some constants $C_3, C_4>0$ and some half orthogonal matrices $\bXi_1\in \mathbb{R}^{(p_1-v_1)\times r_1}$ and $\bXi_2\in \mathbb{R}^{(p_2-v_2)\times r_2}$ satisfying $\bXi_1'\bXi_1=\bI_{r_1}$ and $\bXi_2'\bXi_2=\bI_{r_2}$, where $\sigma_{\min}$ denotes the minimum non-zero singular value of a matrix. 
\end{assumption}
%%%%%%%%%%%%%%%%%

The quantity $\delta_1$ of Assumption 3 is used to quantify the strength of the factors. 
If $\delta_1=0$, the corresponding factors are called strong factors, since it includes the case where each element of $\boldsymbol{\ell}_i$ and $\br_{j}$ is $O(1)$. If $\delta_1>0$, the corresponding 
factors are weak factors and the smaller the $\delta_1$ is, the stronger the factors are. One advantage of using index $\delta_1$ is to link the convergence rates of the estimated factors explicitly to the strength of the factors. This assumption is slightly different from Condition 4 in \cite{wang2018}, which actually impose two different strengths $\varsigma_1$ and $\varsigma_2$ on the front and back loading matrices, respectively. Due to the non-uniqueness of the loading matrices, we can always choose $\delta_1$ such that $(p_1p_2)^{(1-\delta_1)/2}\asymp p_1^{(1-\varsigma_1)/2}p_2^{(1-\varsigma_2)/2}$. Hence Assumption 4 ensures that all common factor components in $\bF_t$ are of equal strength $\delta_1$.  
There are many sufficient conditions  for Assumption 4 to hold. See the discussion of Assumption 5 in \cite{gaotsay2018b}.  Assumption 5(i) is mild and includes the  standard normal distribution as a special case. Assumption 5(ii) is reasonable since $\bB_2$ is a subspace of $\bB_2^*$, $\wh\bQ_2$ is a subspace of $\wh\bQ_2^*$, and the discussion in Section 2.3 implies that  that $\bXi_1'\bB_2^{*}{'}\bA_1$ and $\bXi_2'\bQ_2^{*}{'}\bP_1$ are invertible. The choices of $\wh\bXi_1$ and $\wh\bXi_2$, and hence $\wh\bB_2=\wh{\bB}_2^*\wh\bXi_1$ and $\wh\bQ_2=\wh{\bQ}_2^*\wh\bXi_2$ will be discussed later.
%%%%%%%%%%%%%%%%%%%

If $p_1$ and $p_2$ are large, it is not possible to consistently estimate $\bB_2$ (also $\bQ_2$) or even $\mathcal{M}(\bB_2)$ (also $\mathcal{M}(\bQ_2)$). Instead, we will estimate $\bB_2^*=(\bA_{22},\bB_2)$ or equivalently $\mathcal{M}(\bB_2^*)$, which is the subspace spanned by the eigenvectors associated with the $p_1-k_1$ smallest eigenvalues of $\bS_1$. Assume $\wh{\bB}_2^*$ consists of the eigenvectors corresponding to the smallest $p-k_1$ eigenvalues of $\wh\bS_1$. Under some conditions, we can show that $\mathcal{M}(\wh{\bB}_2^*)$ is consistent to $\mathcal{M}(\bB_2^*)$. This is also the case in the literature on high-dimensional PCA with i.i.d. data. See, for example, \cite{shenetal2016} and the references therein. Therefore, the choice of $\wh\bB_2$ should be a subspace of $\wh{\bB}_2^*$, and we discuss it before Theorem 3 below. 

%%%%%%%%%%%%%%%%%%%%
\begin{theorem}
Suppose Assumptions 1-5 hold and $r_1$ and $r_2$ are known and fixed. As $n\rightarrow\infty$, if $p_1^{\delta_1}p_2^{\delta_1}n^{-1/2}=o(1)$, then 
\[\|D(\mathcal{M}({\wh\bA}_1),\mathcal{M}{\bA}_1)\|_2=O_p(p_1^{\delta_1}p_2^{\delta_1}n^{-1/2})\,\,\text{and}\,\,\|D(\mathcal{M}({\wh\bP}_1),\mathcal{M}{\bP}_1)\|_2=O_p(p_1^{\delta_1}p_2^{\delta_1}n^{-1/2}),\]
and the above results also hold for $\|D(\mathcal{M}({\wh\bB}_1),\mathcal{M}{\bB}_1)\|_2$ and $\|D(\mathcal{M}({\wh\bQ}_1),\mathcal{M}{\bQ}_1)\|_2$.
Furthermore,
\begin{equation*}
\|D(\mathcal{M}(\wh\bB_2^*),\mathcal{M}(\bB_2^*))\|_2
=O_p(p_1^{\delta_2}p_2^{3\delta_2/2}n^{-1/2}+p_1^{\delta_1}p_2^{\delta_1+\delta_2}n^{-1/2}),
\end{equation*}
and
\[\|D(\mathcal{M}(\wh\bQ_2^*),\mathcal{M}(\bQ_2^*))\|_2=O_p(p_1^{3\delta_2/2}p_2^{\delta_2}n^{-1/2}+p_1^{\delta_1+\delta_2}p_2^{\delta_1}n^{-1/2}).\]

\end{theorem}

%%%%%%%%%%%%%%%%%%%%%%%%%%
\begin{remark}
(i) For the consistencies of $\wh\bA_1$ and $\wh\bP_1$, we require $p_1p_2=o(n^{\frac{1}{2\delta_1}})$. When the strength $\delta_1\in[0,1/2]$, the range of the total dimensions $p_1p_2$ can be greater than $\sqrt{n}$. \\
(ii) The conditions for the consistencies of $\wh\bB_2^*$ and $\wh\bQ_2^*$ are slightly stronger since they depend on the estimation error in the first step. Specifically, we require $p_1^{\delta_2}p_2^{3\delta_2/2}n^{-1/2}=o(1)$, $p_1^{\delta_1}p_2^{\delta_1+\delta_2}n^{-1/2}=o(1)$,  $p_1^{3\delta_2/2}p_2^{\delta_2}n^{-1/2}=o(1)$ and $p_1^{\delta_1+\delta_2}p_2^{\delta_1}n^{-1/2}=o(1)$. To give a better illustration, we assume $p_1\asymp p_2\asymp p$, then we have $p^2=o(n^{\frac{1}{2\delta_1}})$ for the consistency of $\wh\bA_1$ (also $\wh\bP_1$), and $p^2=o(\min\{n^{\frac{2}{5\delta_2}},n^{\frac{1}{2\delta_1+\delta_2}}\} )$ for that of $\wh\bB_2^*$ and $\wh\bQ_2^*$, which is slightly stronger than the former.
\end{remark}
%%%%%%%%%%%%%%%%%%%%%%%%%%%%%%%%%%%

 Once we have $\wh{\bB}_2^*$ and $\wh\bQ_2^*$, we suggest to choose $\wh\bB_2$ and $\wh\bQ_2$ as $\wh\bB_2=\wh{\bB}_2^*\wh\bXi_1$ and $\wh\bQ_2=\wh{\bQ}_2^*\wh\bXi_2$, 
where $\wh\bXi_1=(\wh\bxi_{1,1},..,\wh\bxi_{1,r_1})\in \mathbb{R}^{(p_1-k_1)\times r_1}$ and $\wh\bXi_2=(\wh\bxi_{2,1},..,\wh\bxi_{2,r_2})\in \mathbb{R}^{(p_2-k_2)\times r_2}$, where $\wh\bxi_{1,i}$ is the vector associated with the $i$-th largest eigenvalues of $\wh{\bB}_2^*{'}\wh\bA_1\wh\bA_1'\wh{\bB}_2^*$ and $\wh\bxi_{2,j}$ is the vector associated with the $j$-th largest eigenvalues of $\wh{\bQ}_2^*{'}\wh\bP_1\wh\bP_1'\wh{\bQ}_2^*$. These choices can guarantee that the matrices $(\wh\bB_2'\wh\bA_1)^{-1}$ and $(\wh\bQ_2'\wh\bP_1)^{-1}$ behave well when recovering the factor $\wh\bX_t$. On the other hand, they  could still eliminate the diverging part of the noise covariance matrix and give prominent convergence rate, as shown in Theorem 3.  There are many ways to choose the numbers of components $k_1$ and $k_2$ in Assumption 4 so long as $p_1-k_1>r_1$ and $p_2-k_2>r_2$. We discuss the choices of $k_1$ and $k_2$ in Remark 2 below. The following theorem states the convergence rate of the extracted common factors. 
%%%%%%%%%%%%%%%%%%%%%%%%%%%%%%

\begin{theorem}
Under the conditions in Theorem 2, we have
\begin{align*}
(p_1p_2)^{-1/2}\|\wh\bA_1\wh\bX_t\wh\bP_1'-\bA_1\bX_t\bP_1'\|_2=&O_p\left(p_1^{-\delta_1/2}p_2^{-\delta_1/2}(\|D(\mathcal{M}({\wh\bA}_1),\mathcal{M}{\bA}_1)\|_2\right. \notag\\
&\left.+\|D(\mathcal{M}({\wh\bP}_1),\mathcal{M}({\bP}_1))\|_2)+p_1^{-\delta_2/2}\|D(\mathcal{M}(\wh\bB_2^*),\mathcal{M}(\bB_2^*))\|_2\right.\\
&\left.+p_2^{-\delta_2/2}\|D(\mathcal{M}(\wh\bQ_2^*),\mathcal{M}(\bQ_2))\|_2+p_1^{-1/2}p_2^{-1/2}\right).
\end{align*}
\end{theorem}

\begin{remark}
(i) A similar result is given in Theorem 3 of \cite{LamYaoBathia_Biometrika_2011} and Theorem 5 of \cite{gaotsay2018b}, which deal with the approximate factor model and a structured factor model, respectively. When  $\delta_1=\delta_2=0$, i.e. the factors and the noise terms are all strong, the convergence rate in Theorem 3 is $O_p((p_1p_2)^{-1/2}+n^{-1/2})$, which is the optimal rate specified in Theorem 3 of \cite{Bai_Econometrica_2003} when dealing with the traditional approximate factor models.\\
(ii) It is a common issue to select the number of principle components in the literature and there are many possible approaches available. 
Since it is impossible to eliminate all the noise effects in recovering the factors 
and we only need to guarantee that the diverging part of the noises 
is removed for large $p_1$, we may select $k_1$ in a range of possible values. 
In practice, let $\wh\mu_{1,1}\geq...\geq  \wh\mu_{1,p_1}$ be the sample eigenvalues of $\wh\bS_1$ and define $\wh k_{1,L}$ as
\begin{equation}\label{kl}
 \wh k_{1,L}=\arg\min_{1\leq j\leq \wh k_{1,U}}\{\wh\mu_{1,j+1}/\wh\mu_{1,j}\},
\end{equation} 
and $\wh k_{1,U}$ is a pre-specified integer. We suggest $\wh k_{1,U}=\min\{\sqrt{p_1},\sqrt{n},p_1-\wh r_1,5\}$. Then the estimator $\wh k_1$ for $k_1$ can assume some value between $\wh k_{1,L}$ and $\wh k_{1,U}$.  We can select $\wh k_2$ in a similar manner.
\end{remark}
%%%%%%%%%%%%%%%%%%%%%%%
Next, we study the consistency of the white noise tests described in Section 2. In fact, the consistency conditions depend on which test statistic we use. We only consider the two test statistics $T_n$ and $T(m)$ discussed in Section 2.3 of \cite{gaotsay2018b}
and  present the consistency when $p_1$ and $p_2$ are large since the case of small $p_1$ and $p_2$ is trivial. For any random vector  $\bx_t$ to be sub-Gaussian we mean there exists a constant $C>0$ such that $P(|\bv'(\bx_t-E\bx_t)|>x)\leq C\exp(-Cx^2)$ for any constant vector $\|\bv\|_2=1$.
We need an additional assumption.
\begin{assumption}
 $\vc({\bF_t})$, $\vc(\bZ_{12,t})$, $\vc(\bZ_{21,t})$,  and $\vc(\bZ_{22,t})$ are sub-Gaussian random vectors.
\end{assumption}

\begin{theorem}
Assume Assumptions 1-6 hold.\\
(i) If $p_1p_2=o\left\{\min\left(n^{\frac{2}{1+3\delta_1}},n^{\frac{1}{1+2\delta_1-\delta_2}}\right)\right\}$, then the test statistic $T_n$  can consistently estimate $r_1$ and $r_2$, i.e. $P(\wh r_1=r_1,\wh r_2=r_2)\rightarrow 1$ as $n\rightarrow\infty$.\\
(ii) If $p_1^{1+\delta_1-\delta_2/2}p_2^{1+\delta_1-\delta_2/2}n^{-1/2}\sqrt{\log(np_1p_2)}=o(1)$, then 
the test statistic $T(m)$ can consistently estimate $r_1$ and $r_2$.
\end{theorem}

With the estimator $\wh r_1$, we may define the estimator for $\bA_1$ as $\wh\bA_1=(\wh\ba_1,...,\wh\ba_{\wh r_1})$, where $\wh\ba_1,...,\wh\ba_{\wh r_1}$ are the orthonormal eigenvectors of $\wh \bM_1$, defined in (\ref{m1hat}), corresponding to the $\wh r_1$ largest eigenvalues. In addition, we may also replace $r_1$ by $\wh r_1$ in the whole methodology described in Section 2. We can define $\wh\bP_1$ in a similar way.

\section{Numerical Properties}

\subsection{Simulation}
In this section, we illustrate the finite-sample properties of the proposed methodology under different choices of $p_1$ and $p_2$. Because the actual dimension is $p_1p_2$ which can easily go to hundreds for even relatively small $p_1$ and $p_2$, we focus on the case of high dimension, which is of more interest. As the dimensions of $\wh\bA_1$  and $\bA_1$ are not necessarily the same,
and  $\bL_1$ is not an orthogonal matrix in general, we first extend the discrepancy measure
in Equation (\ref{eq:D}) to a more general form below. Let $\bH_i$ be a
$p\times h_i$ matrix with rank$(\bH_i) = h_i$, and $\bP_i =
\bH_i(\bH_i'\bH_i)^{-1} \bH_i'$, $i=1,2$. Define
\begin{equation}\label{dmeasure}
\bar{D}(\mathcal{M}(\bH_1),\mathcal{M}(\bH_2))=\sqrt{1-
\frac{1}{\max{(h_1,h_2)}}\textrm{tr}(\bP_1\bP_2)}.
\end{equation}
Then $\bar{D} \in [0,1]$. Furthermore,
$\bar{D}(\mathcal{M}(\bH_1),\mathcal{M}(\bH_2))=0$ if and only if
either $\mathcal{M}(\bH_1)\subset \mathcal{M}(\bH_2)$ or
$\mathcal{M}(\bH_2)\subset \mathcal{M}(\bH_1)$, and  it is 1 if and only if
$\mathcal{M}(\bH_1) \perp \mathcal{M}(\bH_2)$.
When $h_1 = h_2=h$ and $\bH_i'\bH_i= \bI_r$,
$\bar{D}(\mathcal{M}(\bH_1),\mathcal{M}(\bH_2))
$ reduces to that in Equation (\ref{eq:D}). We only present the simulation results for $k_0=2$ in Equation (\ref{m1hat}) to save space since other choices of $k_0$ produce similar patterns.\\

{\noindent \bf Example 1.} Consider model (\ref{m-factor}) with common factors satisfying 
\[\bF_t=\bPhi\bF_{t-1}\bPsi'+\bN_t,\]
where $\bN_t$ is a matrix-variate white noise process with independent entries, $\bPhi\in \mathbb{R}^{r_1\times r_1}$ and $\bPsi\in \mathbb{R}^{r_2\times r_2}$ are two diagonal coefficient matrices.  We set the true dimension of the matrix factors $(r_1,r_2)=(2,3)$, the orders of the diverging noise components $(k_1,k_2)=(1,2)$  as defined in Assumption 4, the dimensions $(p_1,p_2)=(7,7)$, $(10,15)$, $(20,20)$ and $(20,30)$, and the sample sizes are $n=300$, $500$, $1000$, $1500$, $3000$. We consider three scenarios for $\delta_1$ and $\delta_2$: $(\delta_1,\delta_2)=(0,0.9)$, $(0.2,0.8)$ and $(0.5,0.5)$. We can also obtain similar results for other settings but omit the details to save space.  For each scenario mentioned above, the elements of $\bL$ and $\bR$ are drawn independently from $U(-2,2)$, and then we divide $\bL_1$ (also $\bR_1$) by $p_1^{\delta_1/2}$ (also $p_2^{\delta_1/2}$),  the first $k_1$ (also $k_2$) columns of $\bL_2$ (also $\bR_2$) by $p_1^{\delta_2/2}$ (also $p_2^{\delta_2/2}$) and the rest $v_1-k_1$ (also $v_2-k_2$) columns by $p_1$ (also $p_2$) to satisfy Assumptions 3 and 4. $\bPhi$ and $\bPsi$ are diagonal matrices with their diagonal elements drawn independently from $U(0.5,0.9)$, $\vc(\bZ_{12,t})\sim N(0,\bI_{r_1v_2})$, $\vc(\bZ_{21,t})\sim N(0,\bI_{v_1r_2})$, $\vc(\bZ_{22,t})\sim N(0,\bI_{v_1v_2})$, $\vc(\bN_t)\sim N(0,\bI_{r_1r_2})$. We use $500$ replications in each experiment.

We first study the performance of estimating the dimension of the matrix-variate factors. For simplicity, we only report the results of the test statistic $T(m)$ with $m=10$ defined in \cite{gaotsay2018b}, and the results for the other test are similar. When $p_1p_2>n$, we only keep the upper $\epsilon\sqrt{n}$ row- and column-transformed series of $\wh\bGamma_1'\bY_t\wh\bGamma_2$ with $\epsilon=0.9$ in the testing. Similar results can be obtained for other choices of $\epsilon$ and we do not report them here.  The testing results are given in Table~\ref{Table1}. From the table, we see that for each setting of $(\delta_1,\delta_2)$ and fixed $(p_1,p_2)$, the performance of the white noise test improves as the sample size increases. The performance is also quite satisfactory for moderately large $p_1p_2$ when the factor strength is stronger than that of the noises. When $(\delta_1,\delta_2)=(0.5,0.5)$, we see that the test does not perform well for small sample sizes, which is understandable since the factors and the noises have the same level of strength but the diverging noise effect is much more prominent by Equation (\ref{m-factor}), yet the performance improves significantly when the sample size increases.

%%%%%%%%%%%%%%%%%%%%%%%%%%%%%%%%%%%%%%%%%%%%%%%%%%%
\begin{table}
 \caption{Empirical probabilities $P(\hat{r}_2=r_1, \hat{r}_2=r_2)$ for Example 1 with $(r_1,r_2)=(2,3)$ and $(k_1,k_2)=(1,2)$, where $(p_1,p_2)$ and $n$ are the dimension and the sample size, respectively. $\delta_1$ and $\delta_2$ are the strength parameters of the factors and the errors, respectively. $500$ iterations are used.}
          \label{Table1}
\begin{center}
 \setlength{\abovecaptionskip}{0pt}
\setlength{\belowcaptionskip}{3pt}

\begin{tabular}{c|cc|ccccc}
\hline
 & &&\multicolumn{5}{c}{$n$}\\
$(\delta_1, \delta_2)$ &$(p_1,p_2)$&$p_1p_2$&$300$&$500$&$1000$&$1500$&$3000$\\
\hline
 (0,0.9) &$(7,7)$&49&0.956&0.982&0.984&0.980&0.976\\
 &$(10,15)$&150&0.930&0.988&0.964&0.984&0.978\\
&$(20,20)$&400&0.818&0.976&0.962&0.970&0.968\\
&$(20,30)$&600&0.834&0.986&0.976&0.964&0.972\\
\hline
 (0.2,0.8) &$(7,7)$&49&0.848&0.992&0.988&0.972&0.986\\
 &$(10,15)$&150&0.882&0.982&0.974&0.978&0.984\\
&$(20,20)$&400&0.742&0.964&0.972&0.982&0.972\\
&$(20,30)$&600&0.816&0.994&0.976&0.968&0.966\\
   \hline 
 (0.5,0.5) &$(7,7)$&49&0.104&0.438&0.950&0.974&0.972\\
 &$(10,15)$&150&0.304&0.710&0.946&0.974&0.980\\
&$(20,20)$&400&0.028&0.074&0.334&0.696&0.980\\
&$(20,30)$&600&0.020&0.080&0.296&0.636&0.938\\
\hline
\end{tabular}
          \end{center}
\end{table}
%%%%%%%%%%%%%%%%%%%%%%%%%%%%%%%%%%%%%%%%%%%%%%%%%%

\begin{figure}
\begin{center}
\subfigure[]{\includegraphics[width=0.49\textwidth]{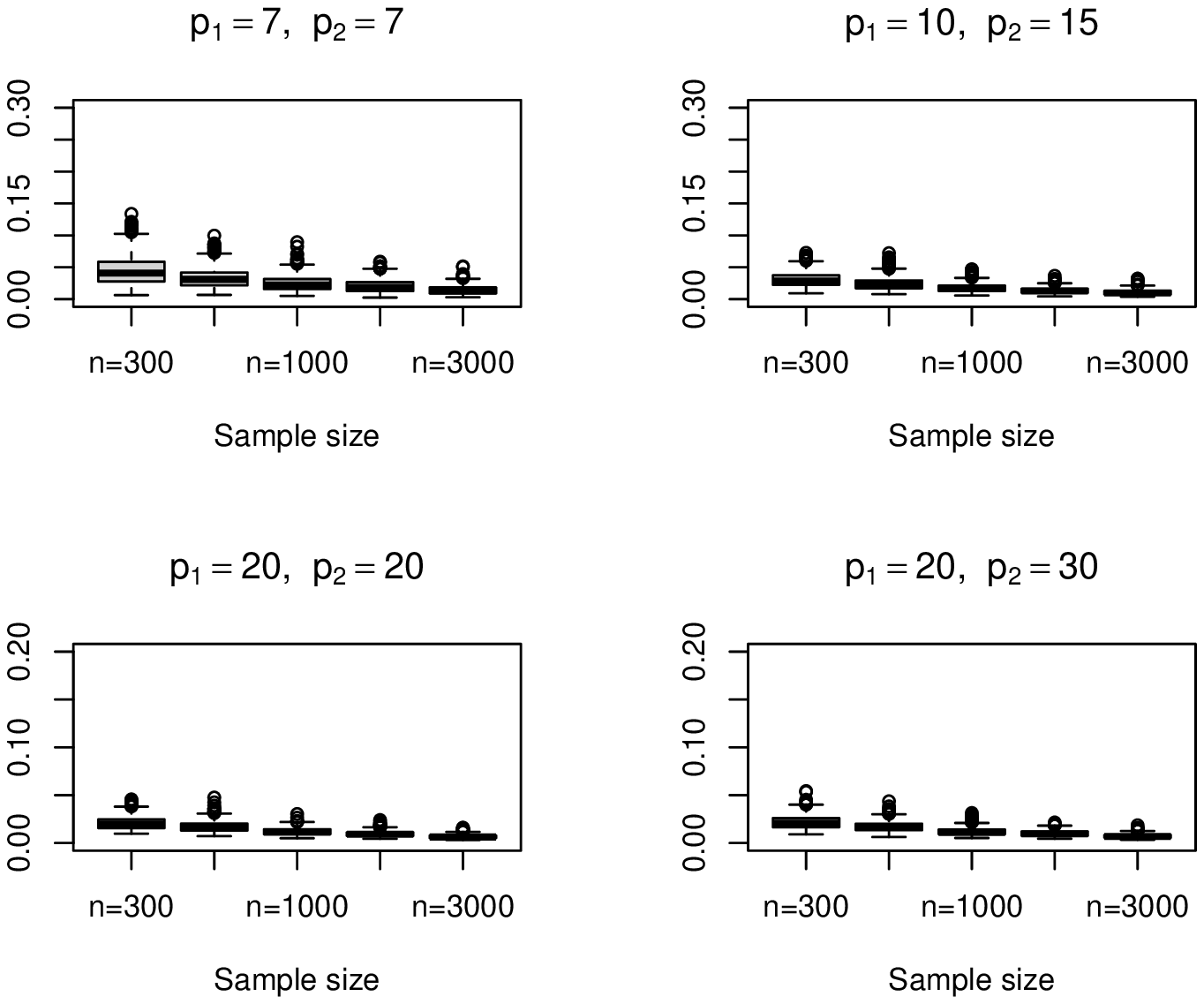}}
\subfigure[]{\includegraphics[width=0.49\textwidth]{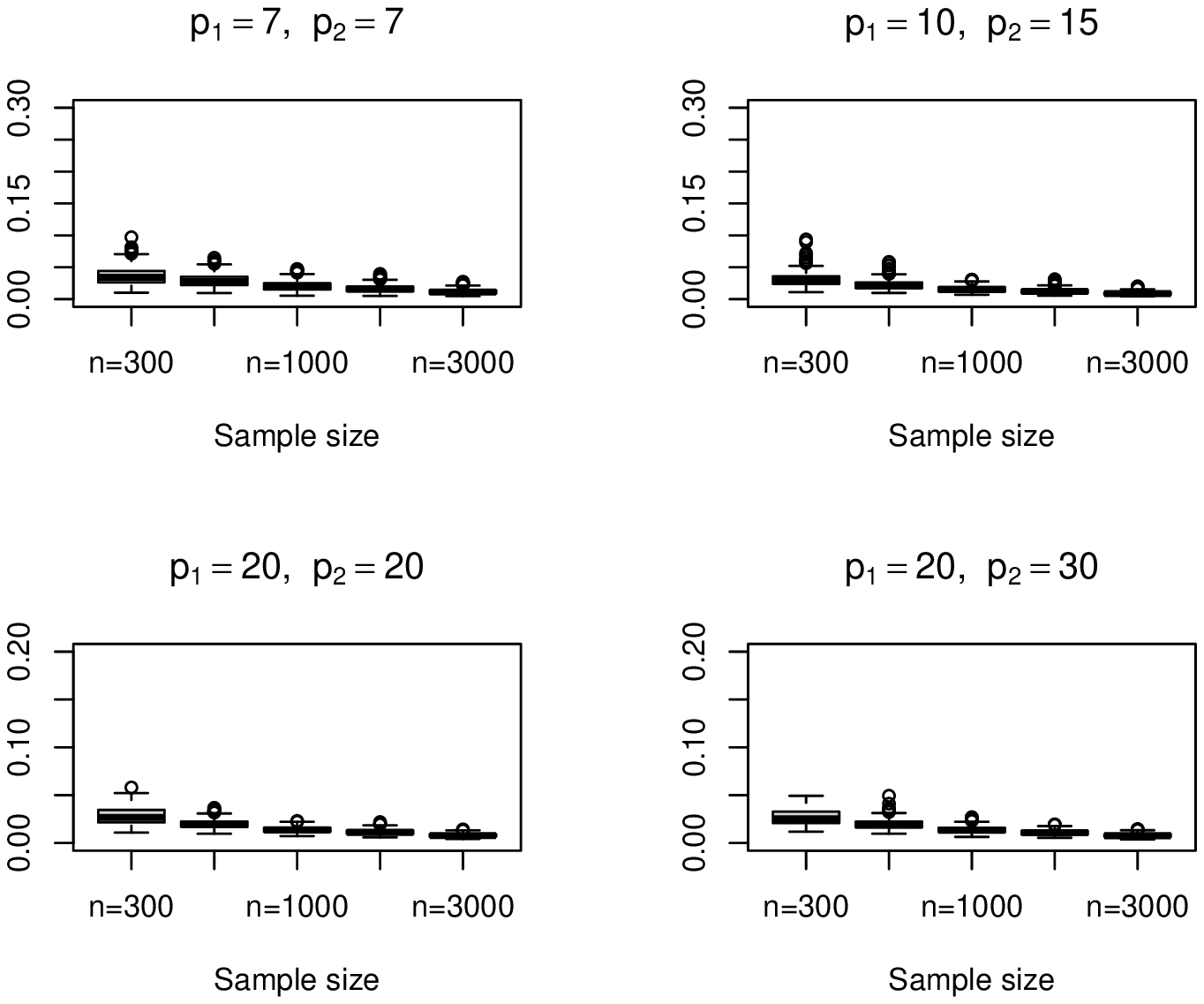}}
\caption{(a) Boxplots of $\bar{D}(\mathcal{M}(\wh\bA_1),\mathcal{M}(\bL_1))$; (b) Boxplots of $\bar{D}(\mathcal{M}(\wh\bP_1),\mathcal{M}(\bR_1))$.  We set $(r_1,r_2)=(2,3)$, $(k_1,k_2)=(1,2)$, and $(\delta_1,\delta_2)=(0,0.9)$ in Example 1. The sample sizes are $300, 500, 1000, 1500, 3000$, respectively. $500$ iterations are used.}\label{fig1}
\end{center}
\end{figure}

\begin{figure}
\begin{center}
{\includegraphics[width=0.7\textwidth]{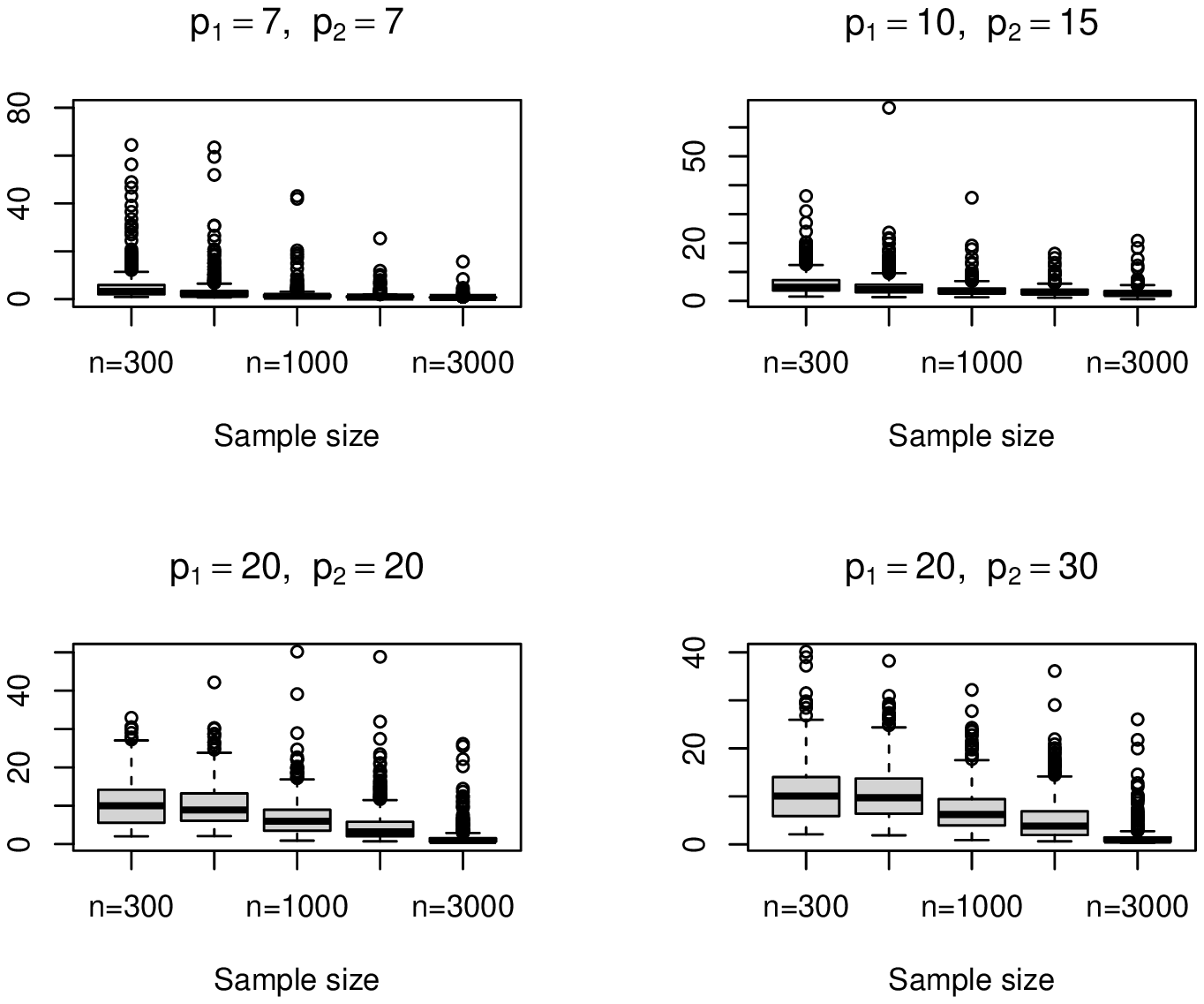}}
%{\includegraphics[width=0.9\textwidth]{p3q6-ycorr-2.pdf}}
\caption{Boxplots of $D(\wh\bA_1\wh\bX\wh\bP_1,\bL_1\bF\bR_1')$ defined in (\ref{DX}) when $(r_1,r_2)=(2,3)$, $(k_1,k_2)=(1,2)$, and $(\delta_1,\delta_2)=(0,0.9)$ in Example 1. The sample sizes are $300, 500, 1000, 1500, 3000$, respectively. $500$ iterations are used.}\label{fig2}
\end{center}
\end{figure}

Next, we study the accuracy of the estimated loading matrices. The boxplots of $\bar{D}(\mathcal{M}(\wh\bA_1),\mathcal{M}(\bL_1))$ and $\bar{D}(\mathcal{M}(\wh\bP_1),\mathcal{M}(\bR_1))$ are shown in Figure \ref{fig1}(a) and (b), respectively. From Figure~\ref{fig1}, we see that the estimation accuracy of the loading matrix improves as the sample size increases even for moderately large $p_1p_2$, which is in line with our asymptotic theory. Furthermore, we study the estimation accuracy of the estimated factor process by
\begin{equation}\label{DX}
D(\wh\bA_1\wh\bX\wh\bP_1',\bL_1\bF\bR_1')=\frac{1}{n\sqrt{p_1p_2}}\sum_{t=1}^n\|\wh\bA_1\wh\bX_t\wh\bP_1'-\bL_1\bF_t\bR_1\|_2.
\end{equation}
The results are shown in Figure~\ref{fig2}, from which we see that, for fixed $(p_1,p_2)$, the estimation accuracy also improves as the sample size increases. This result is consistent with our Theorem 3 of Section 3.
%%%%%%%%%%%%%%%%%%%%%%%%%%%%%%%%%%%%%%%%%%%%%%%%%%

%\end{document}

%%%%%%%%%%%%%%%%%%%%%%%%%%%%%%%%%
%%%%%%%%%%%%%%%%%%%%%%%%%%%%%%%%%%%%%%%%%%%%%%%%%%

%%%%%%%%%%%%%%%%%%%%%%%%%%%%%%%%%%%%%%%%%%%%%%%%%%

\begin{figure}
\begin{center}
\subfigure[]{\includegraphics[width=0.49\textwidth]{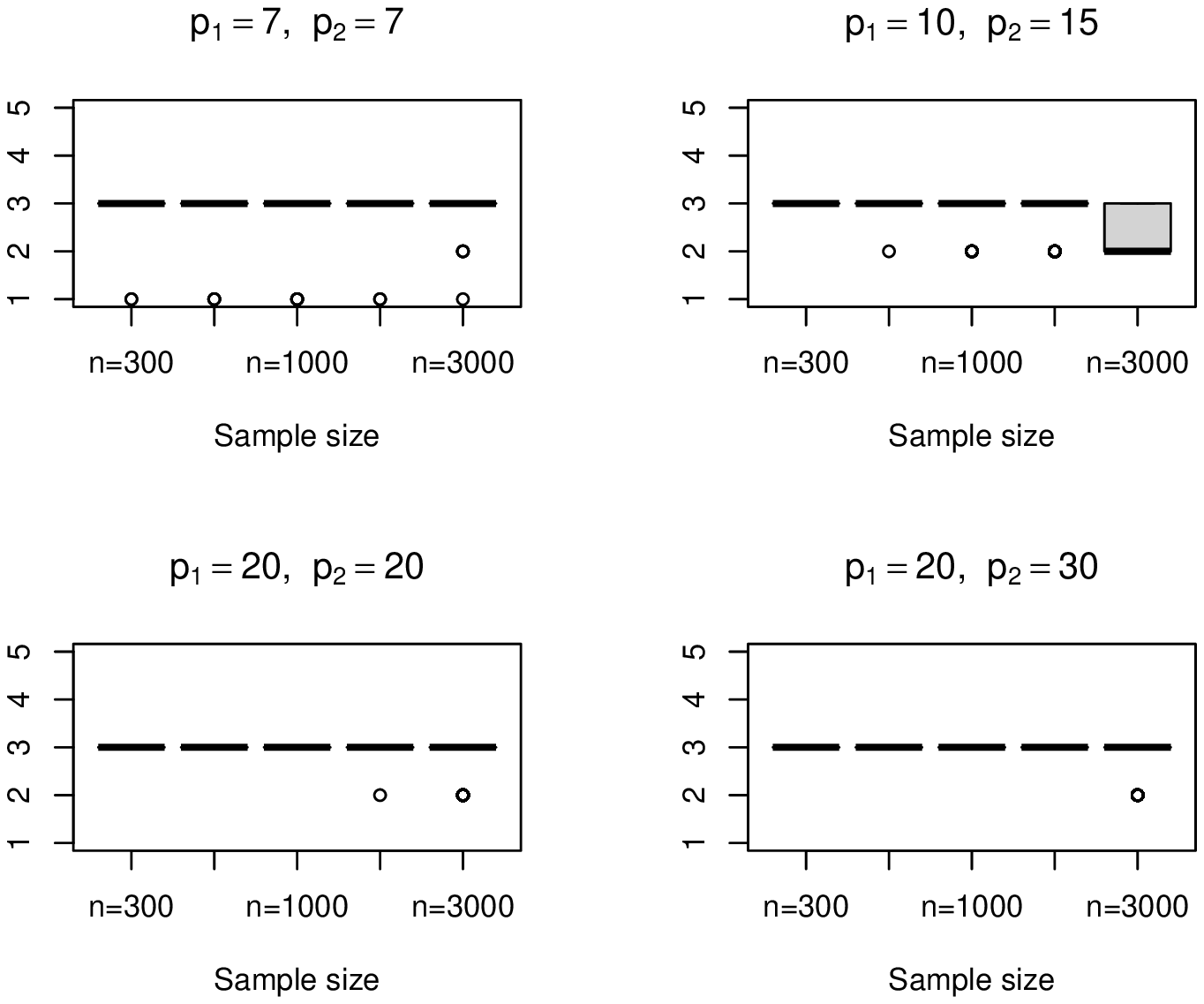}}
\subfigure[]{\includegraphics[width=0.49\textwidth]{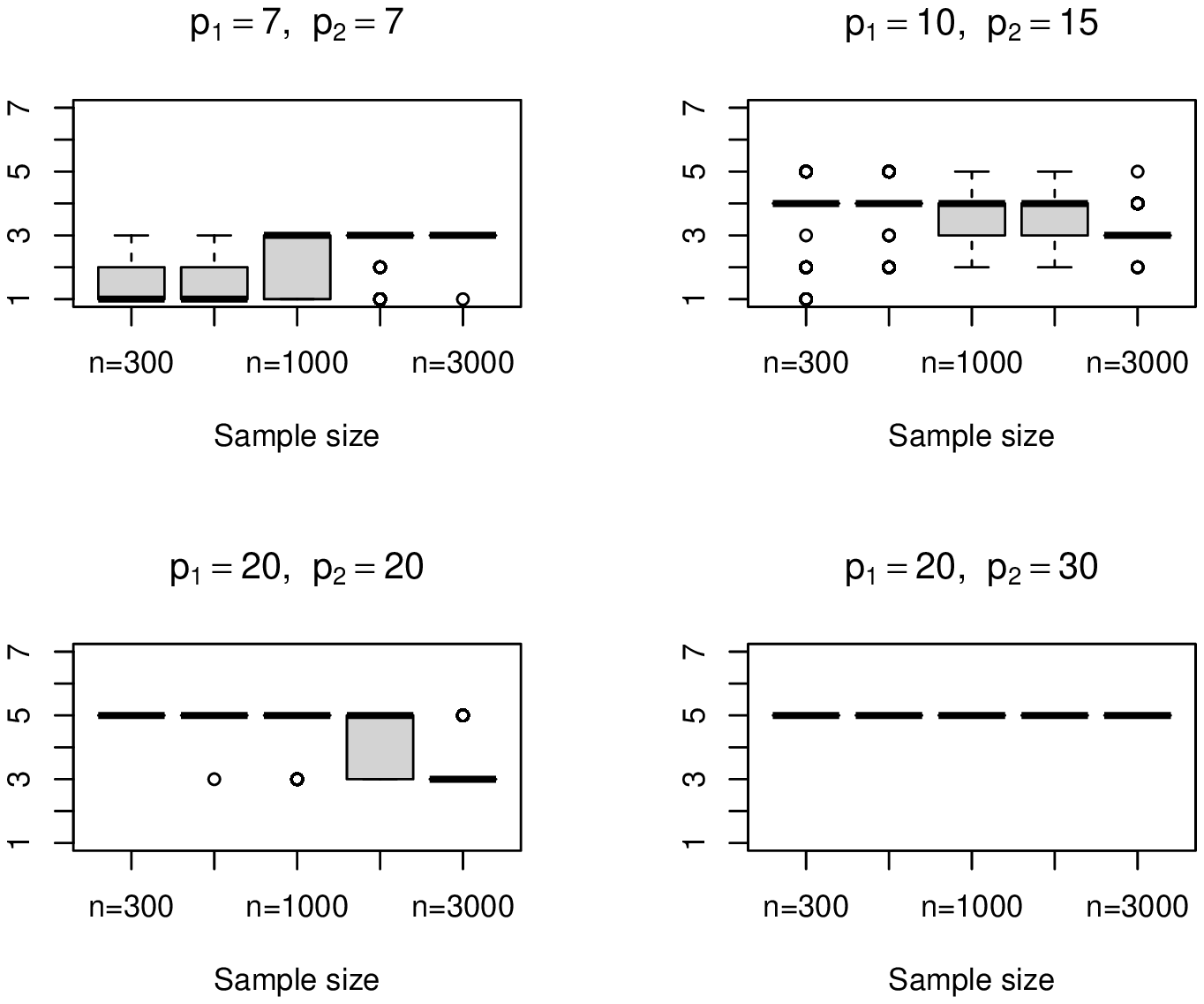}}
\caption{(a) Boxplots of $\wh r_1$ by the ratio-based method of \cite{wang2018}; (b) Boxplots of $\wh r_2$ by the ratio-based method of \cite{wang2018}.  We set $(r_1,r_2)=(2,3)$, $(k_1,k_2)=(1,2)$, and $(\delta_1,\delta_2)=(0.5,0.5)$ in Example 1. The sample sizes are $300, 500, 1000, 1500, 3000$, respectively. $500$ iterations are used.}\label{fig3}
\end{center}
\end{figure}

To see the advantages of the proposed method, we compare it with that of \cite{wang2018} (denoted by WLC) in selecting the order of the matrix-variate factors. For the ratio-based method in WLC, let $\wh\lambda_{i,1},...,\wh\lambda_{i,p_i}$ be the $p_i$ eigenvalues of $\wh\bM_i$ for $i=1,2$, define
\begin{equation}\label{ratio}
\wh r_i=\arg\min_{1\leq j\leq p_i/2}\{\wh\lambda_{i,j+1}/\wh\lambda_{i,j}\},\,\, i=1,2.
\end{equation}
Figure~\ref{fig3}(a)-(b) present the boxplots of $\wh r_1$ and $\wh r_2$, respectively. We see from Figure~\ref{fig3} that the estimated number of factors $\wh r_i$ tend to be the sum of the number of common factors $r_i$ and the number of spiked components of the noises $k_i$ in most of the scenarios. The result indicates that the ratio-based method of \cite{wang2018} may fail to identify the correct dimension of the matrix-variate factor process with dynamic dependence if the covariance of the noise has diverging eigenvalues, while the proposed white noise test continues to work well, as shown in Table~\ref{Table1}.
%%%%%%%%%%%%%%%%%%%%%%%%%%%%%%%%%
%%%%%%%%%%%%%%%%%%%%%%%%%%%%%%%%%%%%%%%%%%%%%%%%%%

Finally, we compare our method with the one of \cite{wang2018} in recovering the common factors since a key difference between the two methods is that we allow some of the eigenvalues of the noise covariance to diverge. We denote our method by GT and the results are reported in Table~\ref{Table2} for $(r_1,r_2)=(2,3)$, $(k_1,k_2)=(1,2)$, and $(\delta_1,\delta_2)=(0.5,0.5)$. From the table, we see that, because the ratio-based method tends to overestimate the dimension of the common factors, the estimation error of our method is much smaller than that obtained by WLC. 
In addition, for a given $(p_1,p_2)$, the estimation error by our method tends to decrease as the sample size increases, which is in agreement with our asymptotic theory. Overall, under the assumption that the noise effect is prominent, the proposed method outperforms the existing one in the literature.
%%%%%%%%%%%%%%%%%%%%%%%%%%%%%%%%%%%%%%%%%%%%%%%%%%%%%%%%
\begin{table}
 \caption{The $D(\wh\bA_1\wh\bX\wh\bP_1',\bL_1\bF\bR_1')$ defined in (\ref{DX}) when $(r_1,r_2)=(2,3)$, $(k_1,k_2)=(1,2)$, and $(\delta_1,\delta_2)=(0.5,0.5)$ in Example 1.  
 The sample sizes used are $n=300, 500, 1000, 1500, 3000$. Standard errors are given in the parentheses and $500$ iterations are used. GT denotes the proposed method and WLC is the one in \cite{wang2018}.}
          \label{Table2}
          
\footnotesize{
\begin{center}
 \setlength{\abovecaptionskip}{0pt}
\setlength{\belowcaptionskip}{3pt}

\begin{tabular}{c|c|ccccc}
\hline
&&\multicolumn{5}{c}{$n$}\\
$(p_1,p_2)$&Method&$300$&$500$&$1000$&$1500$&$3000$\\
\hline
 $(7,7)$&GT& 0.862(0.199)& 0.726(0.494)& 0.460(0.164)& 0.447(0.320)& 0.416(0.096)\\
&WLC&1.178(0.036)&1.178(0.029)&1.182(0.026)&1.183(0.025)&1.179(0.047)\\
\hline
 $(10,15)$&GT& 0.652(0.354)& 0.403(0.185)& 0.290(0.310)& 0.254(0.208)& 0.229(0.132)\\
&WLC&0.891(0.060)&0.886(0.066)&0.862(0.084)&0.783(0.148)&0.549(0.165)\\
\hline
 $(20,20)$&GT& 0.530(0.167)& 0.437(0.108)& 0.301(0.147)& 0.191(0.126)& 0.103(0.043)\\
&WLC&0.696(0.011)&0.695(0.010)&0.686(0.036)&0.648(0.072)&0.485(0.117)\\
\hline
 $(20,30)$&GT& 0.485(0.181)& 0.394(0.102)& 0.278(0.122)& 0.181(0.124)& 0.098(0.069)\\
&WLC&0.662(0.010)&0.663(0.008)&0.662(0.005)&0.662(0.005)&0.651(0.044)\\
\hline
\end{tabular}
          \end{center}}
\end{table}
%%%%%%%%%%%%%%%%%%%%%%%%

\subsection{Real Data Analysis}
{\bf Example 2.} In this example, we use the Fama-French return series to illustrate 
application of the proposed method. The data contain monthly returns of 100 portfolios, structured in 
a $10$ by $10$ matrix according to ten levels of market capitalization (Size, in rows from small to large) and ten levels of investment (Inv, in columns from low to high) both of which are factors for  average stock returns considered in \cite{famafrench2015}. The return series spans from July 1963 to December 2019 and consists of 678 monthly observations for each individual process. Therefore, the series forms a $10\times 10\times 678$ tensor-valued data set. The data and relevant  
information are available at 

\url{http://mba.tuck.dartmouth.edu/pages/faculty/ken.french/data_library.html}.

\begin{figure}
\begin{center}
{\includegraphics[width=0.75\textwidth]{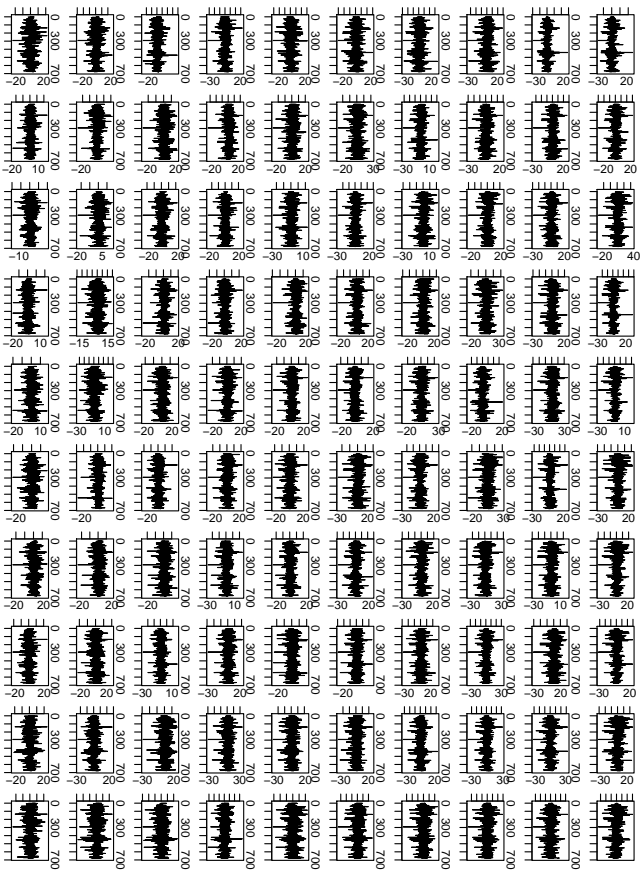}}
%{\includegraphics[width=0.9\textwidth]{p3q6-ycorr-2.pdf}}
\caption{Time series plots of Fama-French 10 by 10 monthly excess return series based on Size and Investment from July 1963 to December 2019.}\label{100series}
\end{center}
\end{figure}

Following \cite{sharpe1964} and \cite{famafrench2015}, we adjust each of  the return series by subtracting the corresponding risk-free asset returns, which are also available from the above website. The missing values were imputed by a simple exponential smoothing method. Time plots of the adjusted $10\times 10$ series are shown in Figure~\ref{100series} with $p_1=p_2=10$ and $n=678$.

%%%%%%%%%%%%%%%%%%%%%%%%%%%%%%%%%
%%%%%%%%%%%%%%%%%%%%%%%%%%%%%%%%%%%%%%%%%%%%%%%%%%

We first applied the method of Section 2.4 with $k_0=2$ and found that  the test statistic $T(m)$ with $m=10$ for testing the null hypothesis $H_0(2)$ defined in Section 2.5 is 4.85, which exceeds the critical value 4.81 based on the limiting Gumbel distribution in \cite{Tsay_2018} with $\alpha=0.05$. But the null hypothesis is not rejected if we increase the order in either the column or row direction. Therefore,  $\wh r_1=2$ and $\wh r_2=2$ implying that a $2\times 2$ matrix-variate latent factor process is detected.  The estimated front and back loading matrices after being multiplied by $30$ are reported in Table~\ref{loading}, which has  several implications.   
First, for Size, it seems that the 10 rows of the portfolios can be divided into two or three groups. 
The one with the smallest size (corresponding to S1) depends on both the first and the second factors 
heavier than the others, the  second smallest size portfolio depends more on the first row factor and less on the second one, and the 3rd to the 10th size portfolios have similar dependence on both the first and the second rows of the matrix-variate factors. Second, for Investment, all the portfolios have similar dependence on the first column of the factor matrix, and the dependence on the second columns seems to have three groups; the lowest investment portfolio (corresponding to Inv1) seems to depend heavily on the second row of the factors, the 5th to the 8th and the 10-th investment portfolios have similar dependence on the first and the second rows of the factors, 
whereas the 2nd to the 4th and the 9th  investment portfolios depend more on the first column of the factors. In addition, the signs of the first coefficients of the size loading and the investment loading are the same, which implies that each return series have a co-movement with respect to the $[1,1]$-factor series. This is understandable since we can treat this common factor as representing the market 
factor of 
the capital asset pricing model (CAPM) of \cite{sharpe1964}. The product of the first coefficients of the size loading and the investment loading can be treated as a market beta, even though 
the starting point of our approach is different from that of CAPM. The usefulness of the 
detected market factor, however, deserves a further investigation.

\begin{figure}
\begin{center}
{\includegraphics[width=0.7\textwidth]{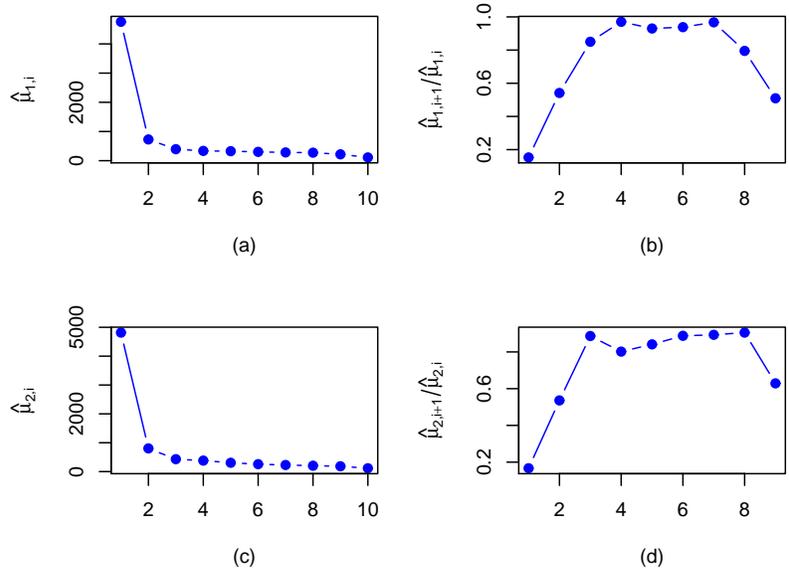}}
%{\includegraphics[width=0.9\textwidth]{p3q6-ycorr-2.pdf}}
\caption{(a) The 10 eigenvalues of $\wh\bS_1$; (b) The plot of ratios of consecutive eigenvalues of $\wh\bS_1$; (c) The 10 eigenvalues of $\wh\bS_2$; (d) The plot of ratios of consecutive 
eigenvalues of $\wh\bS_2$}.\label{fig4}
\end{center}
\end{figure}

\begin{figure}
\begin{center}
\subfigure[]{\includegraphics[width=0.45\textwidth]{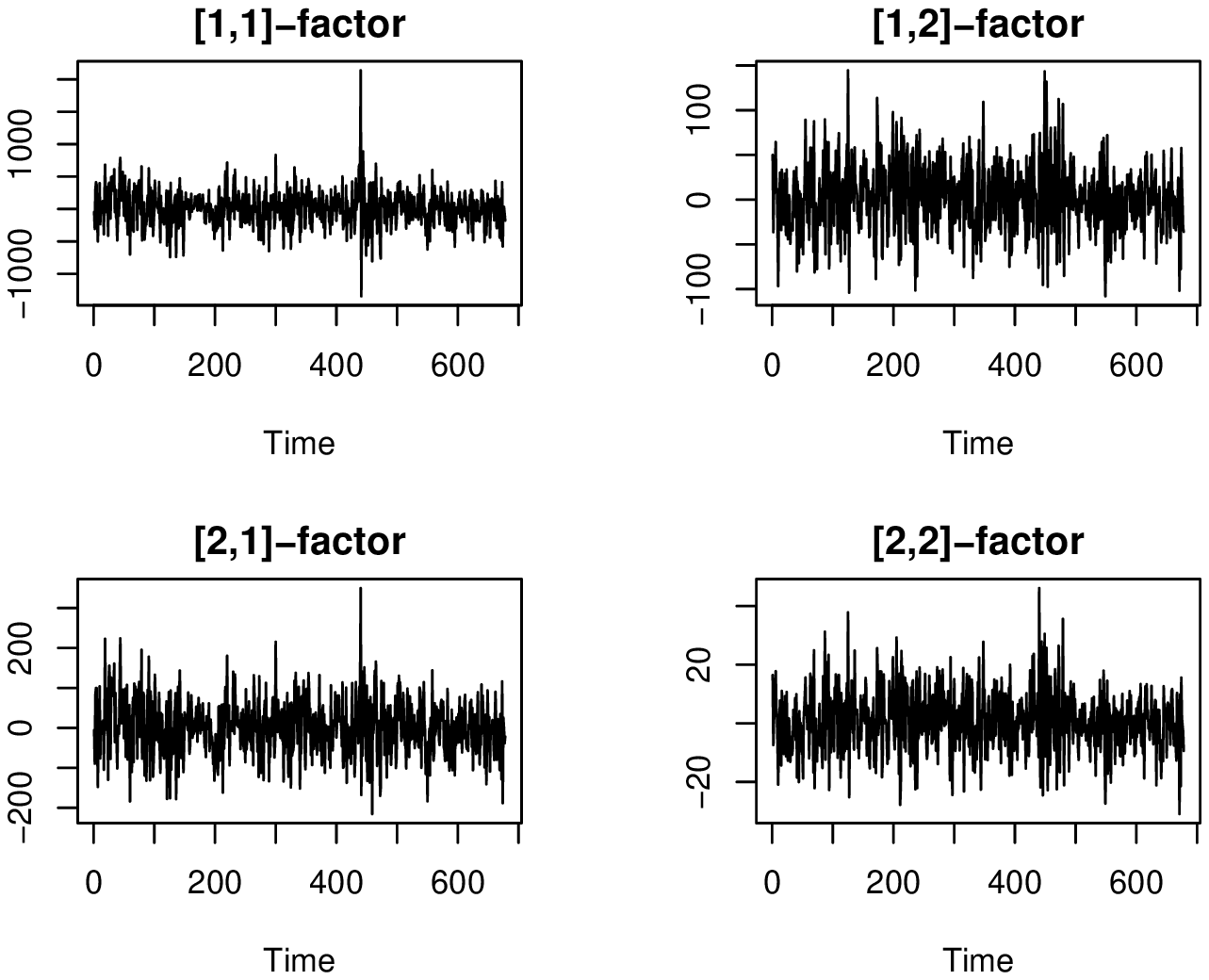}}
\subfigure[]{\includegraphics[width=0.45\textwidth]{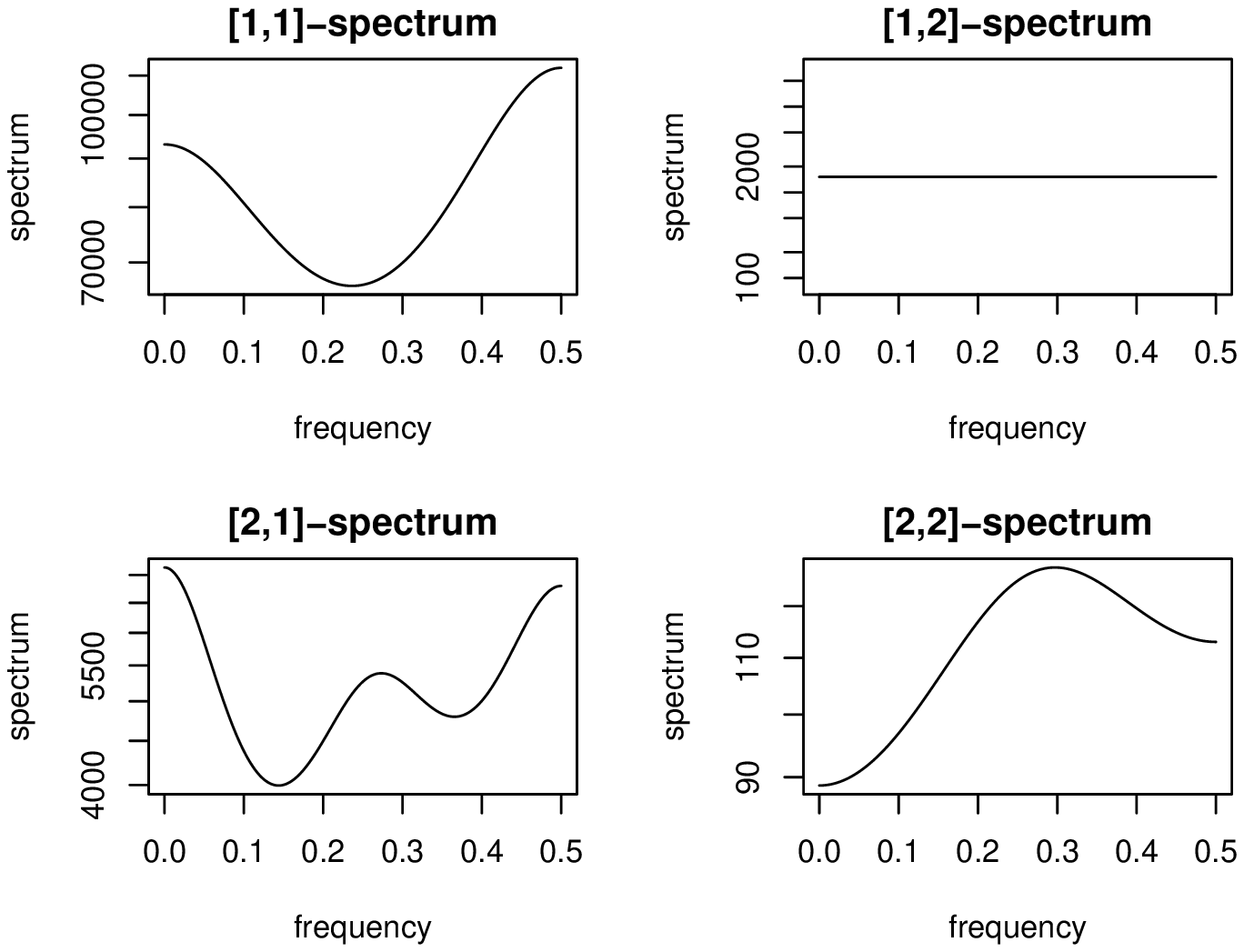}}
\caption{(a) The time series plots of the extracted $2\times 2$ common factors; (b) the corresponding spectrum of the factor processes}.\label{fig5}
\end{center}
\end{figure}

%%%%%%%%%%%%%%%%%%%%%%%%
\begin{table}
\caption{Fama-French series: Size and Investment (Inv) loading matrices after being multiplied  by $30$. The two-dimensional loading vectors are ordered for sizes (S1--S10) and Investment (Inv1--Inv10) from small  to large and from low to high, respectively.}
\label{loading}
\begin{center}
\begin{tabular}{c|cccccccccc}
\hline
Size Factor & S1 & S2 & S3 & S4 & S5 & S6 & S7 & S8 & S9 & S10 \\ \hline
Row 1 & \cellcolor[gray]{0.5}{-18} & \cellcolor[gray]{0.6}{-12} &
\cellcolor[gray]{0.8}{-10} & \cellcolor[gray]{0.8}{-9} &
\cellcolor[gray]{0.8}{-8}
& \cellcolor[gray]{0.8}{-7} &\cellcolor[gray]{0.8}{ -7} & \cellcolor[gray]{0.8}{-6}& \cellcolor[gray]{0.8}{-5}&\cellcolor[gray]{0.8}  {-3} \\
Row 2 & \cellcolor[gray]{0.5}{21} & {1} &
\cellcolor[gray]{0.8}{-6} & \cellcolor[gray]{0.8}{-6} &
\cellcolor[gray]{0.8}{-6}
& \cellcolor[gray]{0.8}{-10} &\cellcolor[gray]{0.8}{ -7} & \cellcolor[gray]{0.8}{-8}& \cellcolor[gray]{0.8}{-8}&  \cellcolor[gray]{0.8}{-7} \\
\hline
\hline
Inv Factor & Inv1 & Inv2 & Inv3 & Inv4 & Inv5 & Inv6 & Inv7 & Inv8 & Inv9 & Inv10 \\ \hline
Column 1 & {\cellcolor[gray]{0.8} -11}  & {\cellcolor[gray]{0.8} -10}
& {\cellcolor[gray]{0.8} -9} & {\cellcolor[gray]{0.8} -8}
& {\cellcolor[gray]{0.8} -8} & {\cellcolor[gray]{0.8} -8} &  {\cellcolor[gray]{0.8}-9 }&  {\cellcolor[gray]{0.8}-9} &  {\cellcolor[gray]{0.8}-10} & {\cellcolor[gray]{0.8} -12 }\\
Column 2 & {\cellcolor[gray]{0.4} 27} &  { -3} &  {-1}  & 1 & { \cellcolor[gray]{0.8}-6}
& {\cellcolor[gray]{0.8}-6} & {\cellcolor[gray]{0.8} -5}
& {\cellcolor[gray]{0.8} -5} & { 1}
& {\cellcolor[gray]{0.8} -7}\\
\hline
\end{tabular}
\end{center}

\end{table}
%%%%%%%%%%%%%%%%%%%%%%%%%%%%%%%%%%%
To obtain the extracted factors, by the two-way projected PCA of Sections 2.3 
and 2.4, we first examine the eigenvalues of the sample covariance matrices $\wh \bS_1$ and $\wh\bS_2$.  From Figure~\ref{fig4}, we see that the first eigenvalues of $\wh\bS_1$ and $\wh\bS_2$ are much larger than the others. Therefore, we choose $\wh k_1=\wh k_2=1$, and the recovered matrix-variate factors are shown in Figure~\ref{fig5}(a) as well as their corresponding spectrum in Figure~\ref{fig5}(b). From Figure~\ref{fig5}, we see that  there are three series with non-trivial spectra, which are all dynamically dependent and they capture most of the dynamic information of the data, and  the $[1,2]$-factor appears to be not serially correlated by itself because its 
sample spectrum is flat. However, this does not imply that the $[1,2]$-factor captures no dynamic information in the detected matrix-variate common factors. 
For example, the lag-1 cross-correlation between the $[1,2]$-factor and the $[2,2]$-factor is 0.08. 
If we test for the zero lag-1 corss correlation between these two series using the long-run covariance matrix calculated by the method in \cite{andrews1991}, the $p$-value is 0.038 implying that 
the two factors are lag-1 cross-correlated. Therefore, the detected 2-by-2 matrix-variate 
common factor process does not violate the assumptions of the proposed model.

%%%%%%%%%%%%%%%%%%%%%%%%%%%%%%%%%
%%%%%%%%%%%%%%%%%%%%%%%%%%%%%%%%%%%%%%%%%%%%%%%%%%

%%%%%%%%%%%%%%%%%%%%%%%%%%%%%%%%%
%%%%%%%%%%%%%%%%%%%%%%%%%%%%%%%%%%%%%%%%%%%%%%%%%%

Next we examine and compare the forecasting performance of the extracted factors via the proposed method (denoted by GT) and those by \cite{wang2018} (denoted by WLC). We estimate the models using the data in the time span $[1,\tau]$ with $\tau=558,...,678-h$ for the $h$-step ahead forecasts, i.e., we use  returns of the last ten years for out-of-sample forecasting. For the method of \cite{wang2018}, the estimated dimension of the matrix-variate factor is $(\wh r_1,\wh r_2)=(1,1)$.
 For simplicity, we employ a simple AR(1) model for each detected common factor to 
 produce forecasts.  
We also fit a scalar AR(1) (denoted by SAR) model to each individual return series as a benchmark approach in out-of-sample forecasting. 
The following two criteria are used to measure the forecast errors:
\begin{equation}\label{fef}
\text{FE}_F(h)=\frac{1}{120-h+1}\sum_{\tau=558}^{678-h}\frac{1}{\sqrt{p_1p_2}}\|\wh\bY_{\tau+h}-\bY_{\tau+h}\|_F,
\end{equation}
and
\begin{equation}\label{fe2}
\text{FE}_2(h)=\frac{1}{120-h+1}\sum_{\tau=558}^{678-h}\frac{1}{\sqrt{p_1p_2}}\|\wh\bY_{\tau+h}-\bY_{\tau+h}\|_2,
\end{equation}
where $p_1=p_2=10$. Table~\ref{Table4} reports the 1-step to 4-step ahead forecast errors of Equations (\ref{fef}) and (\ref{fe2}) for the methods GT,  WLC, and SAR. The smallest forecast error of each step is shown in boldface. From the table, we see that our proposed method is capable of producing accurate forecasts and the associated forecast errors based on the extracted factors by our method are smaller than that based on the factor extracted by WLC or the benchmark approach SAR.  
Although the difference in forecasting errors between the three methods used in 
Table~\ref{Table4} is small, it is generally not easy to produce accurate forecasts in 
asset returns and the improvements by our proposed method could have substantial 
implications to practitioners, especially over the ten-year horizon.

%%%%%%%%%%%%%%%%%
\begin{table}[h]
 \caption{The 1-step to 4-step ahead out-of-sample forecast errors of various methods for Example 2. GT denotes the proposed method, WLC denotes the forecasting errors based on the extracted factor by the method in \cite{wang2018}, and SAR denotes a scalar AR model to each individual return series.
 Boldface numbers denote the smallest error for a given forecast horizon.}
          \label{Table4}
\begin{center}
 \setlength{\abovecaptionskip}{0pt}
\setlength{\belowcaptionskip}{3pt}

\begin{tabular}{c|cccccccc}
\hline
&\multicolumn{3}{c}{FE$_{F}(h)$} &&&\multicolumn{3}{c}{FE$_{2}(h)$}\\
\cline{2-4}\cline{7-9}
Step $h$&GT&WLC&SAR&&&GT&WLC&SAR\\
\cline{1-4}\cline{7-9}
1&{\bf 4.51}&4.61&4.60&&&{\bf 4.04}&4.14&4.14\\
2&{\bf 4.47}&4.51&4.60&&& {\bf 3.98}&4.04&4.14\\
3&{\bf 4.48}&4.51&4.60&&&{\bf 4.00}&4.04&4.14\\
4&{\bf 4.47}&4.49&4.57&&&{\bf 3.98}&4.01&4.11\\
\hline
\end{tabular}
          \end{center}
\end{table}
%%%%%%%%%%%%%%%%%%%%%%%%%%%%%%%%%%%%%%%%%%%%%%%%%%

In conclusion, for the monthly excess return series considered, our method not only produces interpretable factors, but also improves the out-of-sample forecasting. We like to emphasize that the proposed method is different from the traditional factor model analysis, especially those based on 
the conventional principal component analysis. The proposed model explores a different aspect 
of the data via a two-way transformation. Finally, the out-of-sample forecasting can be improved if we adopt some regularization method, but we do not pursue it here. The proposed method is intended as another tool for modeling high-dimensional and possibly highly-correlated matrix-variate time series.

\section{Concluding Remarks}
This paper proposed a new approach to analyze high-dimensional, dynamically dependent matrix-variate data in the presence of prominent noise effect. The proposed approach is an extension 
of that for vector time series in \cite{TiaoTsay_1989} and \cite{gaotsay2018b}. 
The approach not only can reduce the dimensionality of the matrix-variate data, but also preserves the structure of the matrix to mitigate loss in information. The proposed approach is easy to implement for high-dimensional matrix-variate time series data and empirical results show that it can effectively extract the number of common factors from complex data. In addition, the extracted  common factors could be useful in out of sample predictions. 
%Thus, the proposed approach expands the toolkits for analyzing high-dimensional matrix-variate data.

%%%%%%%%%%%%%%%%

\section*{Appendix: Proofs}
\renewcommand{\theequation}{A.\arabic{equation}}
\setcounter{equation}{0}
%{\color{red} What is $sep(\bD_1,{\bf{0}})$? }

 For ease in presentation, we set the blocks $\bZ_{12,t}$ and $\bZ_{21,t}$ to ${\bf 0}$ in model (\ref{m-factor}) since the convergence rates will be dominated by those produced by $\bF_{t}$ and $\bZ_{22,t}$. Thus, for simplicity, we denote $\bZ_t=\bZ_{22,t}$ and $\bE_t=\bE_{22,t}$.
We also use $C$ as a generic constant whose value may change at different places.

\noindent
{\bf Proof of Theorem 1.} We only show it for $\bA_1$ and $\bB_2$ since the cases for $\bP_1$ and $\bQ_2$ are similar. As $p_1$ and $p_2$ are finite, we have the following facts:
\[\|\bSigma_{y,ij}(k)\|_2\leq C,\,\,\text{and}\,\,\sigma_{r_1}(\bSigma_{y,ij}(k))\geq C>0\,\,\text{for}\,\, 1\leq k\leq k_0, 1\leq i,j\leq p_2,\]
and therefore, $\lambda_{1,r_1}=\lambda_{r_1}(\bM_1)\geq C>0$. Furthermore, under Assumptions 1-2, by a similar argument as the proofs of Theorem 1 in \cite{gaotsay2018b}, we have
\[\|\wh\bSigma_{y,ij}(k)-\bSigma_{y,ij}(k)\|_2\leq Cn^{-1/2},\,\,\text{for}\,\, 1\leq k\leq k_0, 1\leq i,j\leq p_2.\]
Thus,
\begin{align}\label{m1:rate}
\|\wh\bM_1-\bM_1\|_2&\leq \sum_{k=1}^{k_0}\sum_{i=1}^{p_2}\sum_{j=1}^{p_2}\left\{\|\wh\bSigma_{y,ij}(k)-\bSigma_{y,ij}(k)\|_2^2+2\|\bSigma_{y,ij}(k)\|_2\|\wh\bSigma_{y,ij}(k)-\bSigma_{y,ij}(k)\|_2\right\}\notag\\
&=O_p(n^{-1/2}).
\end{align}
To bound $\|\wh\bA_1-\bA\|_2$, we make use of the perturbation matrix theory in \cite{golub1996} (Theorem 8.1.10). See also \cite{johnstone-lu2009}, \cite{LamYaoBathia_Biometrika_2011} and \cite{gaotsay2018b}. Note that
\begin{equation}\label{m1:decom}
\left(\begin{array}{c}
\bA_1'\\
\bB_1'
\end{array}\right)\bM_1\left(\bA_1,\bB_1\right)=\left(\begin{array}{cc}
\bD_1&\bf{0}\\
\bf{0}&\bf{0}
\end{array}\right).
\end{equation}
with $sep(\bD_1,{\bf{0}})=\lambda_{r_1}(\bM_1)$, where $sep(\bD_1,\bD_2)$ denotes the minimum distance between the eigenvalues of $\bD_1$ and $\bD_2$. By Lemma 1 in \cite{gaotsay2018b} and a similar argument as the proof of Theorem 1 therein,
\[\|\wh\bA_1-\bA_1\|_2\leq \frac{\|\wh\bM_1-\bM_1\|_2}{\lambda_{1,r_1}(\bM_1)}=O_p(n^{-1/2}).\]
Similarly, we also have $\|\wh\bB_1-\bB_1\|_2=O_p(n^{-1/2})$, $\|\wh\bP_1-\bP_1\|_2=O_p(n^{-1/2})$, and $\|\wh\bQ_1-\bQ_1\|_2=O_p(n^{-1/2})$.

To bound $\|\wh\bB_2-\bB_2\|_2$ and $\|\wh\bQ_2-\bQ_2\|_2$, by a similar argument as above, we only need to bound $\|\wh\bS_1-\bS_1\|_2$ since $\lambda_{v_1}(\bS_1)\geq C>0$. As $p_1$ and $p_2$ are finite,  and $\bB_1$, $\bQ_1$ and their corresponding estimators are all half-orthonormal matrices, by the properties of Kronecker product, 
\[\|\wh\bOmega_{y_i}-\bOmega_{y_i}\|_2=O_p(n^{-1/2})\,\,\text{and}\,\,\|\wh\bQ_1\wh\bQ_1'\otimes\wh\bB_1\wh\bB_1'-\bQ_1\bQ_1'\otimes\bB_1\bB_1'\|_2=O_p(n^{-1/2}).\]
Therefore,
\[\|\wh\bS_1-\bS_1\|_2\leq C\|\wh\bOmega_{y_i}(\wh\bQ_1\wh\bQ_1'\otimes\wh\bB_1\wh\bB_1')\wh\bOmega_{y_i}'-\bOmega_{y_i}(\bQ_1\bQ_1'\otimes\bB_1\bB_1')\bOmega_{y_i}'\|_2=O_p(n^{-1/2}),\]
and 
\[\|\wh\bB_2-\bB_2\|_2\leq\frac{\|\wh\bS_1-\bS_1\|_2}{\lambda_{v_1}(\bS_1)}=O_p(n^{-1/2}).\]
Similarly, we have $\|\wh\bQ_2-\bQ_2\|_2=O_p(n^{-1/2})$.

Furthermore, note that
\begin{align}\label{axp}
\wh\bA_1\wh\bX_t\wh\bP_1'=&\wh\bA_1(\wh\bB_2'\wh\bA_1)^{-1}\wh\bB_2'\bY_t\wh\bQ_2(\wh\bP_1'\wh\bQ_2)^{-1}\wh\bP_1'\notag\\
=&\wh\bA_1(\wh\bB_2'\wh\bA_1)^{-1}\wh\bB_2'\bA_1\bX_t\bP_1'\wh\bQ_2(\wh\bP_1'\wh\bQ_2)^{-1}\wh\bP_1'+\wh\bA_1(\wh\bB_2'\wh\bA_1)^{-1}\wh\bB_2'\bA_2\bE_t\bP_2'\wh\bQ_2(\wh\bP_1'\wh\bQ_2)^{-1}\wh\bP_1'\notag\\
=&\wh\bA_1\bX_t\wh\bP_1'+\wh\bA_1\bX_t(\bP_1-\wh\bP_1)'\wh\bQ_2(\wh\bP_1'\wh\bQ_2)^{-1}\wh\bP_1'\notag\\
&+\wh\bA_1(\wh\bB_2'\wh\bA_1)^{-1}\wh\bB_2'(\bA_1-\wh\bA_1)\bX_t\bP_1'\wh\bQ_2(\wh\bP_1'\wh\bQ_2)^{-1}\wh\bP_1'\notag\\
&+\wh\bA_1(\wh\bB_2'\wh\bA_1)^{-1}\wh\bB_2'\bA_2\bE_t\bP_2'\wh\bQ_2(\wh\bP_1'\wh\bQ_2)^{-1}\wh\bP_1'\notag\\
=:& \Pi_1+\Pi_2+\Pi_3+\Pi_4.
\end{align}
By the upper bounds we have shown above, 
\[\|\Pi_1-\bA_1\bX_t\bP_1'\|_2\leq O_p(\|\wh\bA_1-\bA_1\|_2+\|\wh\bP_1-\bP_1\|_2)=O_p(n^{-1/2}),\]
\[\|\Pi_2\|_2\leq C\|\wh\bP_1-\bP_1\|_2=O_p(n^{-1/2}),\|\Pi_3\|_2\leq C\|\wh\bA_1-\bA_1\|_2=O_p(n^{-1/2})\]
and
\[\|\Pi_4\|_2\leq C\|\wh\bB_2'\bA_2\|_2+C\|\wh\bP_2'\bQ_2\|_2=O_p(n^{-1/2}).\]
This completes the proof. $\Box$
\vskip 0.1cm

To prove the consistency when the dimension is high, we introduce a few useful lemmas first.
%%%%%%%%%%%%%%%%%%%%%%%%%%%%%%%%%%
\begin{lemma}
If Assumptions 1-5 hold, then
\[\sum_{i=1}^{p_2}\sum_{j=1}^{p_2}\|\bSigma_{y,ij}(k)\|_2^2=O_p(p_1^{2-2\delta_1}p_2^{2-2\delta_1})\,\,\text{for}\,\, 1\leq k\leq k_0,\]
and
\[\sum_{k=1}^{k_0}\sum_{i=1}^{p_2}\sum_{j=1}^{p_2}\|\wh\bSigma_{y,ij}(k)-\bSigma_{y,ij}(k)\|_2^2= O_p(p_1^2p_2^2n^{-1}).\]
\end{lemma}
{\bf Proof.} By (\ref{m-factor}), we rewrite (\ref{sigma:k}) as
\[\bSigma_{y,ij}(k)=\bL_1\bSigma_{fr,ij}(k)\bL_1',\]
where $\bSigma_{fr,ij}=\cov(\bF_t\br_{1,i\sbullet}',\bF_{t-k}\br_{1,j\sbullet}')$ and $\bSigma_{fzr,ij}=\cov(\bF_t\br_{1,i\sbullet}',\bZ_{t-k}\br_{2,j\sbullet}')$.

Let $\bff_t=\vc(\bF_t)$, then
\[\bSigma_{y,ij}(k)=\bL_1\cov(\bF_t\br_{1,i\sbullet}',\bF_{t-k}\br_{1,j\sbullet}')\bL_1'=\bL_1(\br_{1,i\sbullet}\otimes\bI_{r_1})\cov(\bff_t,\bff_{t-k})(\br_{1,j\sbullet}'\otimes\bI_{r_1})\bL_1',\]
and
\begin{align}\label{sigy:nm}
\sum_{i=1}^{p_2}\sum_{j=1}^{p_2}\|\bSigma_{y,ij}(k)\|_2^2\leq& C\|\bL_1\|_2^4\sum_{i=1}^{p_2}\sum_{j=1}^{p_2}\|\br_{1,i\sbullet}\otimes\bI_{r_1}\|_2^2\|\br_{1,j\sbullet}'\otimes\bI_{r_1}\|_2^2\notag\\
\leq &C\|\bL_1\|_2^4\|\bR_1\|_F^4=O_p(p_1^{2-2\delta_1}p_2^{2-2\delta_1}).
\end{align}
On the other hand,
\begin{align}\label{sighat}
\wh\bSigma_{y,ij}(k)=&\frac{1}{n}\sum_{t=k+1}^n(\bL_1\bF_t\br_{1,i\sbullet'}+\bL_2\bZ_t\br_{2,i\sbullet}')(\bL_1\bF_{t-k}\br_{1,j\sbullet}'+\bL_2\bZ_{t-k}\br_{2,j\sbullet}')'\notag\\
=&\frac{1}{n}\sum_{t=k+1}^{n}(\bL_1\bF_t\br_{1,i\sbullet}'\br_{1,j\sbullet}\bF_{t-k}'\bL_1')+\frac{1}{n}\sum_{t=k+1}^{n}(\bL_1\bF_t\br_{1,i\sbullet}'\br_{2,j\sbullet}\bZ_{t-k}'\bL_2')\notag\\
&+\frac{1}{n}\sum_{t=k+1}^{n}(\bL_2\bZ_t\br_{2,i\sbullet}'\br_{1,j\sbullet}\bF_{t-k}'\bL_1')+\frac{1}{n}\sum_{t=k+1}^{n}(\bL_2\bZ_t\br_{2,i\sbullet}'\br_{2,j\sbullet}\bZ_{t-k}'\bL_2')\notag\\
=:&\bDelta_{1}^{ijk}+\bDelta_{2}^{ijk}+\bDelta_{3}^{ijk}+\bDelta_{4}^{ijk}.
\end{align}

\begin{equation}\label{delt:1}
\bDelta_{1}^{ijk}-\bSigma_{y,ij}(k)=\bL_1(\br_{1,i\sbullet}\otimes \bI_{r_1})\frac{1}{n}\sum_{t=k+1}^n [\bff_t\bff_{t-k}'-\cov(\bff_t,\bff_{t-k})](\br_{1,j\sbullet}'\otimes \bI_{r_1})\bL_1'.
\end{equation}
By a similar argument as the proof of Theorem 1 in \cite{gaotsay2018b},
\[\|\frac{1}{n}\sum_{t=k+1}^n [\bff_t\bff_{t-k}']-\cov(\bff_t,\bff_{t-k})\|_2=O_p(n^{-1/2}),\]
and hence
\begin{align}\label{delt1:df}
\sum_{i=1}^{p_2}\sum_{j=1}^{p_2}\|\bDelta_{1}^{ijk}-\bSigma_{y,ij}(k)\|_2^2\leq &Cn^{-1}\|\bL_1\|_2^4(\sum_{i=1}^{p_2}\|\br_{1,i\sbullet}\otimes \bI_{r_1}\|_2^2)^2\notag\\
=&O_p(n^{-1}\|\bL_1\|_2^4\|\bR_1\|_2^4)=O_p(p_1^{2-2\delta_1}p_2^{2-2\delta_1}n^{-1}).
\end{align}

Let $\bz_t=\vc(\bZ_t)$. By Assumption 4, we decompose $\bL_2$ and $\bR_2$ as
\begin{equation}\label{l2r2}
\bL_2=\bA_{21}\bD_{21}\bU_{21}'+\bA_{22}\bD_{22}\bU_{22}'\,\,\text{and}\,\,\bR_2=\bP_{21}\bLambda_{21}\bV_{21}'+\bP_{22}\bLambda_{22}\bV_{22}',
\end{equation}
where $\bD_{21}=\diag(d_1,...,d_{k_1})$, $\bD_{22}=\diag(d_{k_{1}+1},...,d_{v_1})$, $\bLambda_{21}=\diag(\gamma_1,...,\gamma_{k_2})$ and $\bLambda_{22}=\diag(\gamma_{k_2+1},...,\gamma_{v_2})$. The dimensions of other matrices can be defined accordingly. Then
\begin{align}\label{delt2}
\Delta_2^{ijk}=&\frac{1}{n}\sum_{t=k+1}^n\bL_1\bF_t\br_{1,i\sbullet}'\br_{2,j\sbullet}\bZ_{t-k}'\bL_2'\notag\\
=&\frac{1}{n}\sum_{t=k+1}^n\bL_1\bF_t\br_{1,i\sbullet}'\bp_{21,j\sbullet}\bLambda_{21}\bV_{21}'\bZ_{t-k}'\bU_{21}\bD_{21}\bA_{21}'\notag\\
&+\frac{1}{n}\sum_{t=k+1}^n\bL_1\bF_t\br_{1,i\sbullet}'\bp_{22,j\sbullet}\bLambda_{22}\bV_{22}'\bZ_{t-k}'\bU_{21}\bD_{21}\bA_{21}'\notag\\
&+\frac{1}{n}\sum_{t=k+1}^n\bL_1\bF_t\br_{1,i\sbullet}'\bp_{21,j\sbullet}\bLambda_{21}\bV_{21}'\bZ_{t-k}'\bU_{22}\bD_{22}\bA_{22}'\notag\\
&+\frac{1}{n}\sum_{t=k+1}^n\bL_1\bF_t\br_{1,i\sbullet}'\bp_{22,j\sbullet}\bLambda_{22}\bV_{22}'\bZ_{t-k}'\bU_{22}\bD_{22}\bA_{22}'\notag\\
=:&\Delta_{2,1}^{ijk}+\Delta_{2,2}^{ijk}+\Delta_{2,3}^{ijk}+\Delta_{2,4}^{ijk}.
\end{align} 
%%%%%%%%%%%%%%%%%%%%%%%%%%%%%%%%
Note that
\begin{align}\label{d21}
\Delta_{2,1}^{ijk}=\bL_1(\br_{1i\sbullet}\otimes \bI_{r_1})\frac{1}{n}\sum_{t=k+1}^n\bff_t\bz_{t-k,11}'(\bLambda_{21}\bp_{21,j\sbullet}'\otimes\bI_{k_1})\bD_{21}\bA_{21}',
\end{align}
where $\bz_{t,i_1i_2}=\bV_{2i_1}'\otimes \bU_{2i_2}'\vc(\bZ_t)$ for $i_1,i_2=1,2$. Then
\begin{equation}\label{dt21:nm}
\sum_{i=1}^{p_2}\sum_{j=1}^{p_2}\|\Delta_{2,1}^{ijk}\|_2^2\leq Cn^{-1}\|\bL_1\|_2^2\|\bD_{21}\|_2^2\|\bR_1\|_F^2\|\bLambda_{21}\|_2^2=O_p(p_1^{2-\delta_1-\delta_2}p_2^{2-\delta_1-\delta_2}n^{-1}).
\end{equation}
For $\Delta_{2,2}^{ijk}$, we note that the covariance of $\bz_{t,i_1i_2}$ is identity, and hence each component of $(\bp_{22,j\sbullet}\bLambda_{22}\otimes \bI_{k_1})\bz_{t,21}\in\mathbb{R}^{k_1}$ has finite $2\gamma$-th moment by Assumption 2. Thus,
\begin{equation}\label{dt22:nm}
\sum_{i=1}^{p_2}\sum_{j=1}^{p_2}\|\Delta_{2,2}^{ijk}\|_2^2\leq Cp_2n^{-1}\|\bL_1\|_2^2\|\bR_1\|_F^2\|\bD_{21}\|_2^2=O_p(p_1^{2-\delta_1-\delta_2}p_2^{2-\delta_1}n^{-1}).
\end{equation}
Similarly, we have
\begin{align}\label{dt23}
\sum_{i=1}^{p_2}\sum_{j=1}^{p_2}\|\Delta_{2,3}^{ijk}\|_2^2\leq& C\|\bL_1\|_2^2\|\bR_1\|_F^2\|\frac{1}{n}\sum_{t=k+1}^n\bff_t\bz_{t-k,12}'\|_2^2\|\bLambda_{21}\|_2^2\notag\\
=&O_p(p_1^{2-\delta_1}p_2^{2-\delta_1-\delta_2}n^{-1}),
\end{align}
and
\begin{align}\label{dt24}
\sum_{i=1}^{p_2}\sum_{j=1}^{p_2}\|\Delta_{2,4}^{ijk}\|_2^2\leq& C\|\bL_1\|_2^2\|\bR_1\|_F^2\sum_{j=1}^{p_2}\|\frac{1}{n}\sum_{t=k+1}^n\bff_t((\bp_{22,j\sbullet}\otimes\bI_{v_1-k_1})\bz_{t,22})'\|_2^2\notag\\
\leq&Cp_1p_2n^{-1}\|\bL_1\|_2^2\|\bR_1\|_F^2=O_p(p_1^{2-\delta_1}p_2^{2-\delta_1}n^{-1}).
\end{align}
Therefore,
\begin{equation}\label{delt2:norm}
\sum_{i=1}^{p_2}\sum_{j=1}^{p_2}\|\Delta_{2}^{ijk}\|_2^2=O_p(p_1^{2-\delta_1}p_2^{2-\delta_1}n^{-1}).
\end{equation}
By symmetry, we also have 
\begin{equation}\label{delt3:norm}
\sum_{i=1}^{p_2}\sum_{j=1}^{p_2}\|\Delta_{3}^{ijk}\|_2^2=O_p(p_1^{2-\delta_1}p_2^{2-\delta_1}n^{-1}).
\end{equation}
Now turn to $\Delta_4^{ijk}$. By (\ref{sighat}) and (\ref{l2r2}),
\begin{align}\label{det4:dm}
\Delta_{4}^{ijk}=&\frac{1}{n}\sum_{t=k+1}^n\left\{\bA_{21}\bD_{21}\bU_{21}'\bZ_t\bV_{21}\bLambda_{21}\bp_{21,i\sbullet}'+\bA_{21}\bD_{21}\bU_{21}'\bZ_t\bV_{22}\bLambda_{22}\bp_{22,i\sbullet}'\right.\notag\\
&\left.+\bA_{22}\bD_{22}\bU_{22}'\bZ_t\bV_{21}\bLambda_{21}\bp_{21,i\sbullet}'+\bA_{22}\bD_{22}\bU_{22}'\bZ_t\bV_{22}\bLambda_{22}\bp_{22,i\sbullet}'\right\}\notag\\
&\times\left\{\bA_{21}\bD_{21}\bU_{21}'\bZ_{t-k}\bV_{21}\bLambda_{21}\bp_{21,j\sbullet}'+\bA_{21}\bD_{21}\bU_{21}'\bZ_{t-k}\bV_{22}\bLambda_{22}\bp_{22,j\sbullet}'\right.\notag\\
&\left.+\bA_{22}\bD_{22}\bU_{22}'\bZ_{t-k}\bV_{21}\bLambda_{21}\bp_{21,j\sbullet}'+\bA_{22}\bD_{22}\bU_{22}'\bZ_{t-k}\bV_{22}\bLambda_{22}\bp_{22,j\sbullet}'\right\}'\notag\\
=:&\Delta_{4,1}^{ijk}+...+\Delta_{4,16}^{ijk},
\end{align}
where there are $16$ cross-product terms and we will specify them in order as follows.
\begin{align}\label{det41}
\Delta_{4,1}^{ijk}=&\frac{1}{n}\sum_{t=k+1}^n[\bA_{21}\bD_{21}\bU_{21}'\bZ_t\bV_{21}\bLambda_{21}\bp_{21,i\sbullet}'][\bA_{21}\bD_{21}\bU_{21}'\bZ_{t-k}\bV_{21}\bLambda_{21}\bp_{21,j\sbullet}']'\notag\\
=&\bA_{21}\bD_{21}(\bp_{21,i\sbullet}\bLambda_{21}\otimes\bI_{k_1})[\frac{1}{n}\sum_{t=k+1}^n\bz_{t,11}\bz_{t,11}'](\bLambda_{21}\bp_{21,j\sbullet}'\otimes\bI_{k_1})\bD_{21}\bA_{21}',
\end{align}
and
\begin{equation}\label{det41:sum}
\sum_{i=1}^{p_2}\sum_{j=1}^{p_2}\|\Delta_{4,1}^{ijk}\|_2^2\leq Cn^{-1}\|\bD_{21}\|_2^2\|\bLambda_{21}\|_2^2\|\bLambda_{21}\|_2^2\|\bD_{21}\|_2^2=O_p(p_1^{2-2\delta_2}p_2^{2-2\delta_2}n^{-1}).
\end{equation}
%%%%%%%%%%%%%%%%%%%%%%%%%%%%%%%%%%%%%%%%%
\begin{align}\label{det42}
\Delta_{4,2}^{ijk}=&\frac{1}{n}\sum_{t=k+1}^n[\bA_{21}\bD_{21}\bU_{21}'\bZ_t\bV_{21}\bLambda_{21}\bp_{21,i\sbullet}'][\bA_{21}\bD_{21}\bU_{21}'\bZ_{t-k}\bV_{22}\bLambda_{22}\bp_{22,j\sbullet}']'\notag\\
=&\bA_{21}\bD_{21}(\bp_{21,i\sbullet}\bLambda_{21}\otimes\bI_{k_1})[\frac{1}{n}\sum_{t=k+1}^n\bz_{t,11}\bz_{t,21}'(\bLambda_{22}\bp_{22,j\sbullet}'\otimes\bI_{k_1})]\bD_{21}\bA_{21}',
\end{align}
and
\begin{equation}\label{det42:sum}
\sum_{i=1}^{p_2}\sum_{j=1}^{p_2}\|\Delta_{4,2}^{ijk}\|_2^2\leq Cp_2n^{-1}\|\bD_{21}\|_2^2\|\bLambda_{21}\|_2^2\|\bD_{21}\|_2^2=O_p(p_1^{2-2\delta_2}p_2^{2-\delta_2}n^{-1}).
\end{equation}
%%%%%%%%%%%%%%%%%%%%%%%

\begin{align}\label{det43}
\Delta_{4,3}^{ijk}=&\frac{1}{n}\sum_{t=k+1}^n[\bA_{21}\bD_{21}\bU_{21}'\bZ_t\bV_{21}\bLambda_{21}\bp_{21,i\sbullet}'][\bA_{22}\bD_{22}\bU_{22}'\bZ_{t-k}\bV_{21}\bLambda_{21}\bp_{21,j\sbullet}']'\notag\\
=&\bA_{21}\bD_{21}(\bp_{21,i\sbullet}\bLambda_{21}\otimes\bI_{k_1})[\frac{1}{n}\sum_{t=k+1}^n\bz_{t,11}\bz_{t,12}'](\bLambda_{21}\bp_{21,j\sbullet}'\otimes\bI_{v_1-k_1})\bD_{22}\bA_{22}',
\end{align}
and
\begin{equation}\label{det43:sum}
\sum_{i=1}^{p_2}\sum_{j=1}^{p_2}\|\Delta_{4,3}^{ijk}\|_2^2\leq Cp_1n^{-1}\|\bD_{21}\|_2^2\|\bLambda_{21}\|_2^2\|\bLambda_{21}\|_2^2=O_p(p_1^{2-\delta_2}p_2^{2-2\delta_2}n^{-1}).
\end{equation}

%%%%%%%%%%%%%%%%%%%%%%%%%%%%%%%%%%%%%%%%%%%%%%%%%%%
\begin{align}\label{det44}
\Delta_{4,4}^{ijk}=&\frac{1}{n}\sum_{t=k+1}^n[\bA_{21}\bD_{21}\bU_{21}'\bZ_t\bV_{21}\bLambda_{21}\bp_{21,i\sbullet}'][\bA_{22}\bD_{22}\bU_{22}'\bZ_{t-k}\bV_{22}\bLambda_{22}\bp_{22,j\sbullet}']'\notag\\
=&\bA_{21}\bD_{21}(\bp_{21,i\sbullet}\bLambda_{21}\otimes\bI_{k_1})[\frac{1}{n}\sum_{t=k+1}^n\bz_{t,11}\bz_{t,22}'(\bLambda_{22}\bp_{22,j\sbullet}'\otimes\bI_{v_1-k_1})]\bD_{22}\bA_{22}',
\end{align}
and
\begin{equation}\label{det44:sum}
\sum_{i=1}^{p_2}\sum_{j=1}^{p_2}\|\Delta_{4,4}^{ijk}\|_2^2\leq Cp_1p_2n^{-1}\|\bD_{21}\|_2^2\|\bLambda_{21}\|_2^2=O_p(p_1^{2-\delta_2}p_2^{2-\delta_2}n^{-1}).
\end{equation}
%%%%%%%%%%%%%%%%%%%%%%%%%%%%%%%
\begin{align}\label{det45}
\Delta_{4,5}^{ijk}=&\frac{1}{n}\sum_{t=k+1}^n[\bA_{21}\bD_{21}\bU_{21}'\bZ_t\bV_{22}\bLambda_{22}\bp_{22,i\sbullet}'][\bA_{21}\bD_{21}\bU_{21}'\bZ_{t-k}\bV_{21}\bLambda_{21}\bp_{21,j\sbullet}']'\notag\\
=&\bA_{21}\bD_{21}[\frac{1}{n}\sum_{t=k+1}^n(\bp_{22,i\sbullet}\bLambda_{22}\otimes\bI_{k_1})\bz_{t,21}\bz_{t,11}'](\bLambda_{21}\bp_{21,j\sbullet}'\otimes\bI_{k_1})\bD_{21}\bA_{21}',
\end{align}
and
\begin{equation}\label{det45:sum}
\sum_{i=1}^{p_2}\sum_{j=1}^{p_2}\|\Delta_{4,5}^{ijk}\|_2^2\leq Cp_2n^{-1}\|\bD_{21}\|_2^2\|\bLambda_{21}\|_2^2\|\bD_{21}\|_2^2=O_p(p_1^{2-2\delta_2}p_2^{2-\delta_2}n^{-1}).
\end{equation}
%%%%%%%%%%%%%%%%%%%%%%%%%%%%%%%

\begin{align}\label{det46}
\Delta_{4,6}^{ijk}=&\frac{1}{n}\sum_{t=k+1}^n[\bA_{21}\bD_{21}\bU_{21}'\bZ_t\bV_{22}\bLambda_{22}\bp_{22,i\sbullet}'][\bA_{21}\bD_{21}\bU_{21}'\bZ_{t-k}\bV_{22}\bLambda_{22}\bp_{22,j\sbullet}']'\notag\\
=&\bA_{21}\bD_{21}[\frac{1}{n}\sum_{t=k+1}^n(\bp_{22,i\sbullet}\bLambda_{22}\otimes\bI_{k_1})\bz_{t,21}\bz_{t,21}'(\bLambda_{21}\bp_{21,j\sbullet}'\otimes\bI_{k_1})]\bD_{21}\bA_{21}',
\end{align}
and
\begin{equation}\label{det46:sum}
\sum_{i=1}^{p_2}\sum_{j=1}^{p_2}\|\Delta_{4,6}^{ijk}\|_2^2\leq Cp_2^2n^{-1}\|\bD_{21}\|_2^2\|\bD_{21}\|_2^2=O_p(p_1^{2-2\delta_2}p_2^{2}n^{-1}).
\end{equation}
%%%%%%%%%%%%%%%%%%%%%%%%%%%%%%%

\begin{align}\label{det47}
\Delta_{4,7}^{ijk}=&\frac{1}{n}\sum_{t=k+1}^n[\bA_{21}\bD_{21}\bU_{21}'\bZ_t\bV_{22}\bLambda_{22}\bp_{22,i\sbullet}'][\bA_{22}\bD_{22}\bU_{22}'\bZ_{t-k}\bV_{21}\bLambda_{21}\bp_{21,j\sbullet}']'\notag\\
=&\bA_{21}\bD_{21}[\frac{1}{n}\sum_{t=k+1}^n(\bp_{22,i\sbullet}\bLambda_{22}\otimes\bI_{k_1})\bz_{t,21}\bz_{t,12}'](\bLambda_{21}\bp_{21,j\sbullet}'\otimes\bI_{v_1-k_1})\bD_{22}\bA_{22}',
\end{align}
and
\begin{equation}\label{det47:sum}
\sum_{i=1}^{p_2}\sum_{j=1}^{p_2}\|\Delta_{4,7}^{ijk}\|_2^2\leq Cp_1p_2n^{-1}\|\bD_{21}\|_2^2\|\bLambda_{21}\|_2^2=O_p(p_1^{2-\delta_2}p_2^{2-\delta_2}n^{-1}).
\end{equation}
%%%%%%%%%%%%%%%%%%%%%%%%%%%%%%%

\begin{align}\label{det48}
\Delta_{4,8}^{ijk}=&\frac{1}{n}\sum_{t=k+1}^n[\bA_{21}\bD_{21}\bU_{21}'\bZ_t\bV_{22}\bLambda_{22}\bp_{22,i\sbullet}'][\bA_{22}\bD_{22}\bU_{22}'\bZ_{t-k}\bV_{22}\bLambda_{22}\bp_{22,j\sbullet}']'\notag\\
=&\bA_{21}\bD_{21}[\frac{1}{n}\sum_{t=k+1}^n(\bp_{22,i\sbullet}\bLambda_{22}\otimes\bI_{k_1})\bz_{t,21}\bz_{t,22}'(\bLambda_{22}\bp_{22,j\sbullet}'\otimes\bI_{v_1-k_1})]\bD_{22}\bA_{22}',
\end{align}
and
\begin{equation}\label{det48:sum}
\sum_{i=1}^{p_2}\sum_{j=1}^{p_2}\|\Delta_{4,8}^{ijk}\|_2^2\leq Cp_1p_2^2n^{-1}\|\bD_{21}\|_2^2=O_p(p_1^{2-\delta_2}p_2^{2}n^{-1}).
\end{equation}
%%%%%%%%%%%%%%%%%%%%%%%%%%%%%%%

\begin{align}\label{det49}
\Delta_{4,9}^{ijk}=&\frac{1}{n}\sum_{t=k+1}^n[\bA_{22}\bD_{22}\bU_{22}'\bZ_t\bV_{21}\bLambda_{21}\bp_{21,i\sbullet}'][\bA_{21}\bD_{21}\bU_{21}'\bZ_{t-k}\bV_{21}\bLambda_{21}\bp_{21,j\sbullet}']'\notag\\
=&\bA_{22}\bD_{22}(\bp_{21,i\sbullet}\bLambda_{21}\otimes\bI_{v_1-k_1})[\frac{1}{n}\sum_{t=k+1}^n\bz_{t,12}\bz_{t,11}'](\bLambda_{21}\bp_{21,j\sbullet}'\otimes\bI_{k_1})\bD_{21}\bA_{21}',
\end{align}
and
\begin{equation}\label{det49:sum}
\sum_{i=1}^{p_2}\sum_{j=1}^{p_2}\|\Delta_{4,9}^{ijk}\|_2^2\leq Cp_1n^{-1}\|\bLambda_{21}\|_2^2\|\bLambda_{21}\|_2^2\|\bD_{21}\|_2^2=O_p(p_1^{2-\delta_2}p_2^{2-2\delta_2}n^{-1}).
\end{equation}
%%%%%%%%%%%%%%%%%%%%%%%%%%%%%%%

\begin{align}\label{det410}
\Delta_{4,10}^{ijk}=&\frac{1}{n}\sum_{t=k+1}^n[\bA_{22}\bD_{22}\bU_{22}'\bZ_t\bV_{21}\bLambda_{21}\bp_{21,i\sbullet}'][\bA_{21}\bD_{21}\bU_{21}'\bZ_{t-k}\bV_{22}\bLambda_{22}\bp_{22,j\sbullet}']'\notag\\
=&\bA_{22}\bD_{22}(\bp_{21,i\sbullet}\bLambda_{21}\otimes\bI_{v_1-k_1})[\frac{1}{n}\sum_{t=k+1}^n\bz_{t,12}\bz_{t,21}'(\bLambda_{22}\bp_{22,j\sbullet}'\otimes\bI_{k_1})]\bD_{21}\bA_{21}',
\end{align}
and
\begin{equation}\label{det410:sum}
\sum_{i=1}^{p_2}\sum_{j=1}^{p_2}\|\Delta_{4,10}^{ijk}\|_2^2\leq Cp_1p_2n^{-1}\|\bLambda_{21}\|_2^2\|\bD_{21}\|_2^2=O_p(p_1^{2-\delta_2}p_2^{2-\delta_2}n^{-1}).
\end{equation}
%%%%%%%%%%%%%%%%%%%%%%%%%%%%%%%

\begin{align}\label{det411}
\Delta_{4,11}^{ijk}=&\frac{1}{n}\sum_{t=k+1}^n[\bA_{22}\bD_{22}\bU_{22}'\bZ_t\bV_{21}\bLambda_{21}\bp_{21,i\sbullet}'][\bA_{22}\bD_{22}\bU_{22}'\bZ_{t-k}\bV_{21}\bLambda_{21}\bp_{21,j\sbullet}']'\notag\\
=&\bA_{22}\bD_{22}(\bp_{21,i\sbullet}\bLambda_{21}\otimes\bI_{v_1-k_1})[\frac{1}{n}\sum_{t=k+1}^n\bz_{t,12}\bz_{t,12}'](\bLambda_{21}\bp_{21,j\sbullet}'\otimes\bI_{v_1-k_1})\bD_{22}\bA_{22}',
\end{align}
and
\begin{equation}\label{det411:sum}
\sum_{i=1}^{p_2}\sum_{j=1}^{p_2}\|\Delta_{4,11}^{ijk}\|_2^2\leq Cp_1^2n^{-1}\|\bLambda_{21}\|_2^2\|\bLambda_{21}\|_2^2=O_p(p_1^{2}p_2^{2-2\delta_2}n^{-1}).
\end{equation}
%%%%%%%%%%%%%%%%%%%%%%%%%%%%%%%

\begin{align}\label{det412}
\Delta_{4,12}^{ijk}=&\frac{1}{n}\sum_{t=k+1}^n[\bA_{22}\bD_{22}\bU_{22}'\bZ_t\bV_{21}\bLambda_{21}\bp_{21,i\sbullet}'][\bA_{22}\bD_{22}\bU_{22}'\bZ_{t-k}\bV_{22}\bLambda_{22}\bp_{22,j\sbullet}']'\notag\\
=&\bA_{22}\bD_{22}(\bp_{21,i\sbullet}\bLambda_{21}\otimes\bI_{v_1-k_1})[\frac{1}{n}\sum_{t=k+1}^n\bz_{t,12}\bz_{t,22}'(\bLambda_{22}\bp_{22,j\sbullet}'\otimes\bI_{v_1-k_1})]\bD_{22}\bA_{22}',
\end{align}
and
\begin{equation}\label{det412:sum}
\sum_{i=1}^{p_2}\sum_{j=1}^{p_2}\|\Delta_{4,12}^{ijk}\|_2^2\leq Cp_1^2p_2n^{-1}\|\bLambda_{21}\|_2^2=O_p(p_1^{2}p_2^{2-\delta_2}n^{-1}).
\end{equation}
%%%%%%%%%%%%%%%%%%%%%%%%%%%%%%%
\begin{align}\label{det413}
\Delta_{4,13}^{ijk}=&\frac{1}{n}\sum_{t=k+1}^n[\bA_{22}\bD_{22}\bU_{22}'\bZ_t\bV_{22}\bLambda_{22}\bp_{22,i\sbullet}'][\bA_{21}\bD_{21}\bU_{21}'\bZ_{t-k}\bV_{21}\bLambda_{21}\bp_{21,j\sbullet}']'\notag\\
=&\bA_{22}\bD_{22}[\frac{1}{n}\sum_{t=k+1}^n(\bp_{22,i\sbullet}\bLambda_{22}\otimes\bI_{v_1-k_1})\bz_{t,22}\bz_{t,11}'](\bLambda_{21}\bp_{21,j\sbullet}'\otimes\bI_{k_1})\bD_{21}\bA_{21}',
\end{align}
and
\begin{equation}\label{det413:sum}
\sum_{i=1}^{p_2}\sum_{j=1}^{p_2}\|\Delta_{4,13}^{ijk}\|_2^2\leq Cp_1p_2n^{-1}\|\bLambda_{21}\|_2^2\|\bD_{21}\|_2^2=O_p(p_1^{2-\delta_2}p_2^{2-\delta_2}n^{-1}).
\end{equation}
%%%%%%%%%%%%%%%%%%%%%%%%%%%%%%%

\begin{align}\label{det414}
\Delta_{4,14}^{ijk}=&\frac{1}{n}\sum_{t=k+1}^n[\bA_{22}\bD_{22}\bU_{22}'\bZ_t\bV_{22}\bLambda_{22}\bp_{22,i\sbullet}'][\bA_{21}\bD_{21}\bU_{21}'\bZ_{t-k}\bV_{22}\bLambda_{22}\bp_{22,j\sbullet}']'\notag\\
=&\bA_{22}\bD_{22}[\frac{1}{n}\sum_{t=k+1}^n(\bp_{22,i\sbullet}\bLambda_{22}\otimes\bI_{v_1-k_1})\bz_{t,22}\bz_{t,21}'(\bLambda_{22}\bp_{22,j\sbullet}'\otimes\bI_{k_1})]\bD_{21}\bA_{21}',
\end{align}
and
\begin{equation}\label{det414:sum}
\sum_{i=1}^{p_2}\sum_{j=1}^{p_2}\|\Delta_{4,14}^{ijk}\|_2^2\leq Cp_1p_2^2n^{-1}\|\bLambda_{21}\|_2^2=O_p(p_1^{2-\delta_2}p_2^{2}n^{-1}).
\end{equation}
%%%%%%%%%%%%%%%%%%%%%%%%%%%%%%%

\begin{align}\label{det415}
\Delta_{4,15}^{ijk}=&\frac{1}{n}\sum_{t=k+1}^n[\bA_{22}\bD_{22}\bU_{22}'\bZ_t\bV_{22}\bLambda_{22}\bp_{22,i\sbullet}'][\bA_{22}\bD_{22}\bU_{22}'\bZ_{t-k}\bV_{21}\bLambda_{21}\bp_{21,j\sbullet}']'\notag\\
=&\bA_{22}\bD_{22}[\frac{1}{n}\sum_{t=k+1}^n(\bp_{22,i\sbullet}\bLambda_{22}\otimes\bI_{v_1-k_1})\bz_{t,22}\bz_{t,12}'](\bLambda_{21}\bp_{21,j\sbullet}'\otimes\bI_{v_1-k_1})\bD_{22}\bA_{22}',
\end{align}
and
\begin{equation}\label{det415:sum}
\sum_{i=1}^{p_2}\sum_{j=1}^{p_2}\|\Delta_{4,15}^{ijk}\|_2^2\leq Cp_1^2p_2n^{-1}\|\bLambda_{21}\|_2^2=O_p(p_1^{2}p_2^{2-\delta_2}n^{-1}).
\end{equation}
%%%%%%%%%%%%%%%%%%%%%%%%%%%%%%%

\begin{align}\label{det416}
\Delta_{4,16}^{ijk}=&\frac{1}{n}\sum_{t=k+1}^n[\bA_{22}\bD_{22}\bU_{22}'\bZ_t\bV_{22}\bLambda_{22}\bp_{22,i\sbullet}'][\bA_{22}\bD_{22}\bU_{22}'\bZ_{t-k}\bV_{22}\bLambda_{22}\bp_{22,j\sbullet}']'\notag\\
=&\bA_{22}\bD_{22}[\frac{1}{n}\sum_{t=k+1}^n(\bp_{22,i\sbullet}\bLambda_{22}\otimes\bI_{v_1-k_1})\bz_{t,22}\bz_{t,12}'(\bLambda_{22}\bp_{22,j\sbullet}'\otimes\bI_{v_1-k_1})]\bD_{22}\bA_{22}',
\end{align}
and
\begin{equation}\label{det416:sum}
\sum_{i=1}^{p_2}\sum_{j=1}^{p_2}\|\Delta_{4,16}^{ijk}\|_2^2\leq Cp_1^2p_2^2n^{-1}=O_p(p_1^{2}p_2^{2}n^{-1}).
\end{equation}
%%%%%%%%%%%%%%%%%%%%%%%%%%%%%%%
We gather the results in Equations (\ref{sighat})-(\ref{det416:sum}) and obtain
\[\sum_{k=1}^{k_0}\sum_{i=1}^{p_2}\sum_{j=1}^{p_2}\|\wh\bSigma_{y,ij}(k)-\bSigma_{y,ij}(k)\|_2^2=O_p(p_1^{2}p_2^2n^{-1}).\]
This completes the proof. $\Box$
\vskip 0.1cm
%%%%%%%%%%%%%%%%%%%%%%%%%%%%%%%
%%%%%%%%%%%%%%%%%%%%%%%%%%%%%%%%%%
\begin{lemma}
Assume Assumptions 1-5 hold. If $p_1^{\delta_1}p_2^{\delta_1}n^{-1/2}=o(1)$, then
\[\sum_{i=1}^{p_2}\|\bOmega_{y_i}(\bQ_1\otimes\bB_1)\|_2^2=O_p(p_1^{2-2\delta_2}p_2^{2-\delta_2}),\]
and
\[\sum_{i=1}^{p_2}\|\wh\bOmega_{y_i}(\wh\bQ_1\otimes\wh\bB_1)-\bOmega_{y_i}(\bQ_1\otimes\bB_1)\|_2^2= O_p(p_1^2p_2^2n^{-1}+p_1^{2-2\delta_2}p_2^{2-\delta_2}(\|\wh\bQ_1-\bQ_1\|_2^2+\|\wh\bB_1-\bB_1\|_2^2)).\]
\end{lemma}
%%%%%%%%%%%%%%%%%%%%%%%%%%%%%%%%%%%%%
{\bf Proof.} By (\ref{m-factor}) and (\ref{omgy}),
\begin{align}\label{sum:omgy}
\sum_{i=1}^{p_2}\|\bOmega_{y_i}(\bQ_1\otimes\bB_1)\|_2^2=&\sum_{i=1}^{p_2}\|\cov(\bL_2\bZ_t\br_{2,i\sbullet}',\vc(\bL_2\bZ_2\bR_2'))(\bQ_1\otimes\bB_1))\|_2^2\notag\\
=&\sum_{i=1}^{p_2}\|\bL_2(\br_{2,i\sbullet}\otimes \bI_{v_1})(\bR_2'\otimes\bL_2')(\bQ_1\otimes\bB_1)\|_2^2\notag\\
\leq& \|\bL_2\|_2^2\|\bR_2\|_2^2\|\bL_2\|_2^2\sum_{i=1}^{p_2}(\|\bp_{21,i\sbullet}\bLambda_{21}\bV_{21}'\|_2+\|\bp_{22,i\sbullet}\bLambda_{22}\bV_{22}'\|_2^2)\notag\\
=&Cp_1^{2-2\delta_2}p_2^{1-\delta_2}(p_2^{1-\delta_2}+p_2)=O_p(p_1^{2-2\delta_2}p_2^{2-\delta_2}).
\end{align}
%%%%%%%%%%%%%%%%%%%%%%%%%%%%%%%%
Furthermore,
\begin{align}\label{omg:hat}
\wh\bOmega_{y_i}(\wh\bQ_1\otimes\wh\bB_1)=&\frac{1}{n}\sum_{t=1}^n\bL_1(\br_{1,i\sbullet}\otimes\bI_{r_1})\bff_t\bff_t'(\bR_1'\otimes\bL_1')(\wh\bQ_1\otimes\wh\bB_1)\notag\\
&+\frac{1}{n}\sum_{t=1}^n\bL_1(\br_{1,i\sbullet}\otimes\bI_{r_1})\bff_t\bz_t'(\bR_2'\otimes\bL_2')(\wh\bQ_1\otimes\wh\bB_1)\notag\\
&+\frac{1}{n}\sum_{t=1}^n\bL_2(\br_{2,i\sbullet}\otimes\bI_{v_1})\bz_t\bff_t'(\bR_1'\otimes\bL_1')(\wh\bQ_1\otimes\wh\bB_1)\notag\\
&+\frac{1}{n}\sum_{t=1}^n\bL_2(\br_{2,i\sbullet}\otimes\bI_{v_1})\bz_t\bz_t'(\bR_2'\otimes\bL_2')(\wh\bQ_1\otimes\wh\bB_1)\notag\\
=:&\bJ_1^{i}+\bJ_2^{i}+\bJ_3^{i}+\bJ_4^{i}.
\end{align}
%%%%%%%%%%%%%%%%%%%%%%%%%%
\begin{align}
\sum_{i=1}^{p_2}\|\bJ_1^i\|_2^2\leq& C\|\bL_1\|_2^2\sum_{i=1}^{p_2}\|\br_{1,i\sbullet}\|_2^2\|\bR_1'\wh\bQ_1\otimes\bL_1'\wh\bB_1\|_2^2\notag\\
\leq &C \|\bR_1\|_2^4\|\bL_1\|_2^4\|\wh\bQ_1-\bQ_1\|_2^2\|\wh\bB_1-\bB_1\|_2^2\notag\\
=&O_p(p_1^{2-2\delta_1}p_2^{2-2\delta_1}\|\wh\bQ_1-\bQ_1\|_2^2\|\wh\bB_1-\bB_1\|_2^2).
\end{align}
We briefly present the rates of the others and the argument is similar as that in Lemma 1.
\begin{align}\label{j2i}
\bJ_2^i=&\frac{1}{n}\sum_{t=1}^n(\br_{1,i\sbullet}\otimes\bL_1)\bff_t\bz_{t,11}(\bLambda_{21}\bP_{21}'\otimes\bD_{21}\bA_{21}')(\wh\bQ_1\otimes\wh\bB_1)\notag\\
&+\frac{1}{n}\sum_{t=1}^n(\br_{1,i\sbullet}\otimes\bL_1)\bff_t\bz_{t,21}'(\bLambda_{22}\bP_{22}'\otimes\bD_{21}\bA_{21}')(\wh\bQ_1\otimes\wh\bB_1)\notag\\
&+\frac{1}{n}\sum_{t=1}^n(\br_{1,i\sbullet}\otimes\bL_1)\bff_t\bz_{t,12}'(\bLambda_{21}\bP_{21}'\otimes\bD_{22}\bA_{22}')(\wh\bQ_1\otimes\wh\bB_1)\notag\\
&+\frac{1}{n}\sum_{t=1}^n(\br_{1,i\sbullet}\otimes\bL_1)\bff_t\bz_{t,22}'(\bLambda_{22}\bP_{22}'\otimes\bD_{22}\bA_{22}')(\wh\bQ_1\otimes\wh\bB_1)\notag\\
=:&\bJ_{2,1}^i+\bJ_{2,2}^i+\bJ_{2,3}^i+\bJ_{2,4}^i.
\end{align}
%%%%%%%%%%%%%%%%%%%
\begin{align}\label{j21:sum}
\sum_{i=1}^{p_2}\|\bJ_{2,1}^i\|_2^2\leq&\sum_{i=1}^{p_2}\|\br_{1,i\sbullet}\otimes\bL_1\|_2^2\|\frac{1}{n}\sum_{t=1}^n\bff_t\bz_{t,11}'\|_2^2\|\bLambda_{21}\bP_{21}'\otimes\bD_{21}\bA_{21}'\|_2^2\notag\\
\leq & Cn^{-1}\|\bL_1\|_2^2\|\bR_1\|_2^2\|\bR_2\|_2^2\|\bL_2\|_2^2\notag\\
=&O_p(p_1^{2-\delta_1-\delta_2}p_2^{2-\delta_1-\delta_2}n^{-1}).
\end{align}
%%%%%%%%%%%%%%%%%%%
\begin{align}\label{j22:sum}
\sum_{i=1}^{p_2}\|\bJ_{2,2}^i\|_2^2\leq&\sum_{i=1}^{p_2}\|\br_{1,i\sbullet}\otimes\bL_1\|_2^2\|\frac{1}{n}\sum_{t=1}^n\bff_t\bz_{t,21}'\|_2^2\|\bLambda_{22}\bP_{22}'\otimes\bD_{21}\bA_{21}'\|_2^2\notag\\
\leq & Cp_2n^{-1}\|\bL_1\|_2^2\|\bR_1\|_2^2\|\bL_2\|_2^2\notag\\
=&O_p(p_1^{2-\delta_1-\delta_2}p_2^{2-\delta_1}n^{-1}).
\end{align}
%%%%%%%%%%%%%%%%%%%%%%%%%%%
\begin{align}\label{j23:sum}
\sum_{i=1}^{p_2}\|\bJ_{2,3}^i\|_2^2\leq&\sum_{i=1}^{p_2}\|\br_{1,i\sbullet}\otimes\bL_1\|_2^2\|\frac{1}{n}\sum_{t=1}^n\bff_t\bz_{t,12}'\|_2^2\|\bLambda_{21}\bP_{21}'\otimes\bD_{22}\bA_{22}'\|_2^2\notag\\
\leq & Cp_1n^{-1}\|\bR_1\|_2^2\|\bL_1\|_2^2\|\bR_2\|_2^2\notag\\
=&O_p(p_1^{2-\delta_1}p_2^{2-\delta_1-\delta_2}n^{-1}).
\end{align}
%%%%%%%%%%%%%%%%%%%%%%%%%%%
\begin{align}\label{j24sum}
\sum_{i=1}^{p_2}\|\bJ_{2,4}^i\|_2^2\leq&\sum_{i=1}^{p_2}\|\br_{1,i\sbullet}\otimes\bL_1\|_2^2\|\frac{1}{n}\sum_{t=1}^n\bff_t\bz_{t,22}'\|_2^2\|\bLambda_{22}\bP_{22}'\otimes\bD_{22}\bA_{22}'\|_2^2\notag\\
\leq & Cp_1p_2n^{-1}\|\bR_1\|_2^2\|\bL_1\|_2^2\notag\\
=&O_p(p_1^{2-\delta_1}p_2^{2-\delta_1}n^{-1}).
\end{align}
%%%%%%%%%%%%%%%%%%%%%%%%%%%
Thus,
\begin{equation}\label{j2:sum}
\sum_{i=1}^{p_2}\|\bJ_2^i\|_2^2=O_p(p_1^{2-\delta_1}p_2^{2-\delta_1}n^{-1}).
\end{equation}
%%%%%%%%%%%%%%%%%%%%%%%%
By a similar argument, we can show that 
\begin{equation}\label{j3:sum}
\sum_{i=1}^{p_2}\|\bJ_3^i\|_2^2=O_p(p_1^{2-\delta_1}p_2^{2-\delta_1}n^{-1}\|\wh\bQ_1-\bQ_1\|_2^2\|\wh\bB_1-\bB_1\|_2^2).
\end{equation}
%%%%%%%%%%%%%%%%%%%%%
We further note that
\begin{align}\label{j4}
\bJ_4^i-\bOmega_{y_i}(\bQ_1\otimes\bB_1)=&\frac{1}{n}\sum_{t=1}^n\left[(\bL_2(\br_{2,i\sbullet}\otimes\bI_{v_1})\bz_t\bz_t'(\bR_2'\otimes\bL_2')-\bOmega_{y_i}\right](\wh\bQ_1\otimes\wh\bB_1)\notag\\
&+\bOmega_{y_i}[\wh\bQ_1\otimes\wh\bB_1-\bQ_1\otimes\bB_1]\notag\\
=:&\bJ_{4,1}^i+\bJ_{4,2}^i.
\end{align}
By a similar argument as that in (\ref{det4:dm}), we have
\begin{equation}\label{j41:sum}
\sum_{i=1}^{p_2}\|\bJ_{4,1}^i\|_2^2=O_p(p_1^2p_2^2n^{-1}),
\end{equation}
and by the first result of Lemma 2, 
\begin{equation}\label{j42:sum}
\sum_{i=1}^{p_2}\|\bJ_{4,2}^i\|_2^2=O_p(p_1^{2-2\delta_2}p_2^{2-\delta_2}(\|\wh\bQ_1-\bQ_1\|_2^2+\|\wh\bB_1-\bB_1\|_2^2)).
\end{equation}
Therefore, 
\begin{equation}\label{j4:sum}
\sum_{i=1}^{p_2}\|\bJ_4^i-\bOmega_{y_i}(\bQ_1\otimes\bB_1)\|_2^2=O_p(p_1^2p_2^2n^{-1}+p_1^{2-2\delta_2}p_2^{2-\delta_2}(\|\wh\bQ_1-\bQ_1\|_2^2+\|\wh\bB_1-\bB_1\|_2^2)).
\end{equation}
By Equations (\ref{omg:hat})--(\ref{j4:sum}), 
\begin{equation*}\label{dif:omg}
\sum_{i=1}^{p_2}\|\wh\bOmega_{y_i}(\wh\bQ_1\otimes\wh\bB_1)-\bOmega_{y_i}(\bQ_1\otimes\bB_1)\|_2^2=O_p(p_1^2p_2^2n^{-1}+p_1^{2-2\delta_2}p_2^{2-\delta_2}(\|\wh\bQ_1-\bQ_1\|_2^2+\|\wh\bB_1-\bB_1\|_2^2)).
\end{equation*}
This completes the proof. $\Box$
%%%%%%%%%%%%%%%%%%%%%%%%
\vskip 0.1cm

%%%%%%%%%%%%%%%%%%%%%%%%%%
\begin{lemma}
Assume Assumptions 1-5 hold. If $p_1^{\delta_1}p_2^{\delta_1}n^{-1/2}=o(1)$, then
\[\|\wh\bM_1-\bM_1\|_2=O_p(p_1^{2-\delta_1}p_2^{2-\delta_1}n^{-1/2}).\]
\end{lemma}

{\bf Proof.} Note that
\begin{align}\label{M:df}
\|\wh\bM_1-\bM_1\|_2\leq & \sum_{k=1}^{k_0}\sum_{i=1}^{p_2}\sum_{j=1}^{p_2}\left\{\|\wh\bSigma_{y,ij}(k)-\bSigma_{y,ij}(k)\|_2^2+2\|\bSigma_{y,ij}(k)\|_2\|\wh\bSigma_{y,ij}(k)-\bSigma_{y,ij}(k)\|_2\right\}\notag\\
\leq&\sum_{k=1}^{k_0}\sum_{i=1}^{p_2}\sum_{j=1}^{p_2}\|\wh\bSigma_{y,ij}(k)-\bSigma_{y,ij}(k)\|_2^2+2\sum_{k=1}^{k_0}\left(\sum_{i=1}^{p_2}\sum_{j=1}^{p_2}\|\bSigma_{y,ij}(k)\|_2^2\right)^{1/2}\notag\\
&\times\left(\sum_{i=1}^{p_2}\sum_{j=1}^{p_2}\|\wh\bSigma_{y,ij}(k)-\bSigma_{y,ij}(k)\|_2^2\right)^{1/2}\notag\\
=&O_p(p_1^{2-\delta_1}p_2^{2-\delta_1}n^{-1/2}).
\end{align}
This completes the proof. $\Box$

%%%%%%%%%%%%%%%%%%%%%%%%%%%%%%%%%
\begin{lemma}
(i) Assume Assumptions 1-5 hold. If $p_1^{\delta_1}p_2^{\delta_1}n^{-1/2}=o(1)$ and $p_1^{\delta_2}p_2^{\delta_2/2}n^{-1/2}=o(1)$, then
\[\|\wh\bS_1-\bS_1\|_2=O_p(p_1^{2-\delta_2}p_2^{2-\delta_2/2}n^{-1/2}+p_1^{2-2\delta_2}p_2^{2-\delta_2}(\|\wh\bQ_1-\bQ_1\|_2+\|\wh\bB_1-\bB_1\|_2)).\]
\end{lemma}
%%%%%%%%%%%%%%%%%%%%%%%%%
{\bf Proof.} Note that
\begin{align}\label{S:df}
\|\wh\bS_1-\bS_1\|_2\leq & \sum_{i=1}^{p_2}\left\{\|\wh\bOmega_{y_i}(\wh\bQ_1\otimes\wh\bB_1)-\bOmega_{y_i}(\bQ_1\otimes\bB_1)\|_2^2+2\|\bOmega_{y_i}(\bQ_1\otimes\bB_1)\|_2\right.\notag\\
&\left.\times\|\wh\bOmega_{y_i}(\wh\bQ_1\otimes\wh\bB_1)-\bOmega_{y_i}(\bQ_1\otimes\bB_1)\|_2\right\}\notag\\
\leq&\sum_{i=1}^{p_2}\|\wh\bOmega_{y_i}(\wh\bQ_1\otimes\wh\bB_1)-\bOmega_{y_i}(\bQ_1\otimes\bB_1)\|_2^2+2\left(\sum_{i=1}^{p_2}\|\bOmega_{y_i}(\bQ_1\otimes\bB_1)\|_2^2\right)^{1/2}\notag\\
&\times\left(\sum_{i=1}^{p_2}\|\wh\bOmega_{y_i}(\wh\bQ_1\otimes\wh\bB_1)-\bOmega_{y_i}(\bQ_1\otimes\bB_1)\|_2^2\right)^{1/2}\notag\\
=&O_p(p_1^{2-\delta_2}p_2^{2-\delta_2/2}n^{-1/2}+p_1^{2-2\delta_2}p_2^{2-\delta_2}(\|\wh\bQ_1-\bQ_1\|_2+\|\wh\bB_1-\bB_1\|_2)),
\end{align}
where the last equality follows from Lemma 2.
This completes the proof. $\Box$

%%%%%%%%%%%%
\begin{lemma}
If Assumptions 1-5 hold, then
\begin{equation}\label{lbdr}
\lambda_{r_1}(\bM_1)\geq Cp_1^{2-2\delta_1}p_2^{2-2\delta_1}.
\end{equation}
\end{lemma}
%%%%%%%%%%%%%%%%%%%%%%%%%%%%
{\bf Proof.} Note that
\begin{align*}
\bSigma_{y,ij}(k)=&\bL_1(\br_{1,i\sbullet}\otimes\bI_{r_1})[E\vc(\bF_t)\vc(\bF_{t-k})](\br_{1,j\sbullet}'\otimes\bI_{r_1})\bL_1'\notag\\
=&\bL_1(\br_{1,i\sbullet}\otimes\bI_{r_1})\bSigma_{f}(k)(\br_{1,j\sbullet}'\otimes\bI_{r_1})\bL_1',
\end{align*}
and $\lambda_{r_1}(\bSigma_{f}(k)\bSigma_{f}(k)')\geq C>0$. The result can be established by a similar argument as the Proof of Lemma 5 in \cite{wang2018}. We omit the details. This competes the proof. $\Box$
%%%%%%%%%%%%%%%%%%%%%%%%%
%%%%%%%%%%%%%%%%%%%%%%%%%%%%%%%%%%
\begin{lemma}
If Assumptions 1-5 hold, then
\[\lambda_{k_1}(\bS_1)\geq Cp_1^{2-2\delta_2}p_2^{2-2\delta_2}.\]
\end{lemma}
%%%%%%%%%%%%%%%%%%%%%%
{\bf Proof.} The proof is similar as Lemma 4. We omit the details here. $\Box$
%%%%%%%%%%%%%%%

\noindent
{\bf Proof of Theorem 2.} By  a similar argument as (\ref{m1:decom}), the matrix perturbation theorem in Lemma 1 of \cite{gaotsay2018b}, and Lemmas 3 and 5 above, we have
\[\|\wh\bA_1-\bA_1\|_2\leq \frac{\|\wh\bM_1-\bM_1\|_2}{\lambda_{r_1}(\bM_1)}=O_p(p_1^{\delta_1}p_2^{\delta_1}n^{-1/2}).\]
Similarly, we can obtain that
\[\|\wh\bB_1-\bB_1\|_2=O_p(p_1^{\delta_1}p_2^{\delta_1}n^{-1/2}),\|\wh\bP_1-\bP_1\|_2=O_p(p_1^{\delta_1}p_2^{\delta_1}n^{-1/2}),\|\wh\bQ_1-\bQ_1\|_2=O_p(p_1^{\delta_1}p_2^{\delta_1}n^{-1/2}).\]
On the other hand, by Lemmas 4 and 6 above, we have
\begin{align*}
\|\wh\bB_2^*-\bB_2^*\|_2\leq&\frac{\|\wh\bS_1-\bS_1\|_2}{\lambda_{k_1}(\bS_1)}=O_p(p_1^{\delta_2}p_2^{3\delta_2/2}n^{-1/2}+p_2^{\delta_2}(\|\wh\bQ_1-\bQ_1\|_2+\|\wh\bB_1-\bB_1\|_2))\notag\\
=&O_p(p_1^{\delta_2}p_2^{3\delta_2/2}n^{-1/2}+p_1^{\delta_1}p_2^{\delta_1+\delta_2}n^{-1/2}).
\end{align*}
Similarly, 
\[\|\wh\bQ_2^*-\bQ_2^*\|_2=O_p(p_1^{3\delta_2/2}p_2^{\delta_2}n^{-1/2}+p_1^{\delta_1+\delta_2}p_2^{\delta_1}n^{-1/2}).\]
This completes the proof. $\Box$
%%%%%%%%%%%%%%%%%%%%%%%%%%%%%%%%%%%%%%%%%%%%%%%%

\noindent
{\bf Proof of Theorem 3.} By the decomposition in Equation (\ref{axp}), we have a similar one as
\[\wh\bA_1\wh\bX_t\wh\bP_1'=\Pi_1^*+\Pi_2^*+\Pi_3^*+\Pi_4^*,\]
where
\[\Pi_1^*=\wh\bA_1\bX_t\wh\bP_1,\,\,\Pi_2^*=\wh\bA_1\bX_t(\bP_1-\wh\bP_1)'\wh\bQ_2\wh\bXi_2(\wh\bP_1'\wh\bQ_2^*\wh\bXi_2)^{-1}\wh\bP_1',\]
\[\Pi_3^*=\wh\bA_1(\wh\bXi_1'\wh\bB_2^*{'}\wh\bA_1)^{-1}\wh\bXi_1'\wh\bB_2^*{'}(\bA_1-\wh\bA_1)\bX_t\bP_1'\wh\bQ_2^*\wh\bXi_2(\wh\bP_1'\wh\bQ_2^*\wh\bXi_2)^{-1}\wh\bP_1',\]
and
\[\Pi_4^*=\wh\bA_1(\wh\bXi_1\wh\bB_2^*{'}\wh\bA_1)^{-1}\wh\bXi_1'\wh\bB_2^*{'}\bA_2\bE_t\bP_2'\wh\bQ_2^*\wh\bXi_2(\wh\bP_1'\wh\bQ_2^*\wh\bXi_2)^{-1}\wh\bP_1'.\]
In addition,
\begin{align}\label{pi1:st}
p_1^{-1/2}p_2^{-1/2}\|\Pi_1^*-\bA_1\bX_t\bP_1'\|_2\leq& p_1^{-1/2}p_2^{-1/2}\|(\wh\bA_1-\bA_1)\bX_t\wh\bP_1'\|_2\notag\\
&+p_1^{-1/2}p_2^{-1/2}\|\bA_1\bX_t(\wh\bP_1-\bP_1)'\|_2\notag\\
\leq& Cp_1^{-\delta_1/2}p_2^{-\delta_1/2}(\|\wh\bA_1-\bA_1\|_2+\|\wh\bP_1-\bP_1\|_2).
\end{align}
%%%%%%%%%%%%%%%%%%%%%%%%%%%%%%%%%%%%
\begin{equation}\label{pi2:st}
p_1^{-1/2}p_2^{-1/2}\|\Pi_2^*\|_2\leq Cp_1^{-\delta_1/2}p_2^{-\delta_1/2}\|\wh\bP_1-\bP_1\|_2.
\end{equation}
%%%%%%%%%%%%%%%%%%%%%%%%%%%%%%%
\begin{equation}\label{pi3:st}
p_1^{-1/2}p_2^{-1/2}\|\Pi_3^*\|_2\leq Cp_1^{-\delta_1/2}p_2^{-\delta_1/2}\|\wh\bA_1-\bA_1\|_2.
\end{equation}
%%%%%%%%%%%%%%%%%%%%%%%%%%%%%%%%%%%%
Note that
\begin{align}\label{aep}
\bA_2\bE_t\bP_2'=&\bA_{21}\bD_{21}\bU_{21}'\bZ_t\bV_{21}\bLambda_{21}\bP_{21}'+\bA_{21}\bD_{21}\bU_{21}'\bZ_t\bV_{22}\bLambda_{22}\bP_{22}\notag\\
&+\bA_{22}\bD_{22}\bU_{22}'\bZ_t\bV_{21}\bLambda_{21}\bP_{21}'+\bA_{22}\bD_{22}\bU_{22}'\bZ_t\bV_{22}\blambda_{22}\bP_{22}'.
\end{align}
Then we can write $\Pi_4^*$ as
\begin{align}\label{pi4:dec}
\Pi_4^*=&\wh\bA_1(\wh\bXi_1\wh\bB_2^*{'}\wh\bA_1)^{-1}\wh\bXi_1'\wh\bB_2^*{'}\bA_{21}\bD_{21}\bU_{21}'\bZ_t\bV_{21}\bLambda_{21}\bP_{21}'\wh\bQ_2^*\wh\bXi_2(\wh\bP_1'\wh\bQ_2^*\wh\bXi_2)^{-1}\wh\bP_1'\notag\\
&+\wh\bA_1(\wh\bXi_1\wh\bB_2^*{'}\wh\bA_1)^{-1}\wh\bXi_1'\wh\bB_2^*{'}\bA_{21}\bD_{21}\bU_{21}'\bZ_t\bV_{22}\bLambda_{22}\bP_{22}\wh\bQ_2^*\wh\bXi_2(\wh\bP_1'\wh\bQ_2^*\wh\bXi_2)^{-1}\wh\bP_1'\notag\\
&+\wh\bA_1(\wh\bXi_1\wh\bB_2^*{'}\wh\bA_1)^{-1}\wh\bXi_1'\wh\bB_2^*{'}\bA_{22}\bD_{22}\bU_{22}'\bZ_t\bV_{21}\bLambda_{21}\bP_{21}'\wh\bQ_2^*\wh\bXi_2(\wh\bP_1'\wh\bQ_2^*\wh\bXi_2)^{-1}\wh\bP_1'\notag\\
&+\wh\bA_1(\wh\bXi_1\wh\bB_2^*{'}\wh\bA_1)^{-1}\wh\bXi_1'\wh\bB_2^*{'}\bA_{22}\bD_{22}\bU_{22}'\bZ_t\bV_{22}\bLambda_{22}\bP_{22}'\wh\bQ_2^*\wh\bXi_2(\wh\bP_1'\wh\bQ_2^*\wh\bXi_2)^{-1}\wh\bP_1'\notag\\
=:&\Pi_{4,1}^*+\Pi_{4,2}^*+\Pi_{4,3}^*+\Pi_{4,4}^*.
\end{align}
 %%%%%%%%%%%%%%%%%%%%%%%%%%%%
 \begin{equation}\label{pi41:nm}
 p_1^{-1/2}p_2^{-1/2}\|\Pi_{4,1}^*\|_2\leq Cp_1^{-\delta_2/2}p_2^{-\delta_2/2}\|\wh\bB_2^*-\bB_2^*\|_2 \|\wh\bQ_2^*-\bQ_2^*\|_2.
 \end{equation}
  \begin{equation}\label{pi42:nm}
 p_1^{-1/2}p_2^{-1/2}\|\Pi_{4,2}^*\|_2\leq Cp_1^{-\delta_2/2}\|\wh\bB_2^*-\bB_2^*\|_2.
 \end{equation}
  \begin{equation}\label{pi43:nm}
 p_1^{-1/2}p_2^{-1/2}\|\Pi_{4,3}^*\|_2\leq Cp_2^{-\delta_2/2}\|\wh\bQ_2^*-\bQ_2^*\|_2.
 \end{equation}
Let $\wh\bXi_1=(\bxi_{1,1},...,\bxi_{1,k_1})$ and $\wh\bXi_2=(\bxi_{2,1},...,\bxi_{2,k_2})$ and they are two half-orthogonal matrices. Denote $\bZ_t^*=\wh\bB_2^*{'}\bA_{22}\bD_{22}\bU_{22}'\bZ_t\bV_{22}\bLambda_{22}\bP_{22}'\wh\bQ_2^*\wh\bXi_2$. Then the variance of the $(i,j)$-th element of $\wh\bXi_1'\bZ_t^*\wh\bXi_2^*$ is 
\[\Var(\bxi_{1,i}'\bZ_t^*\bxi_{2,j})=\Var\{(\bxi_{2,j}'\otimes\bxi_{1,i}')\vc(\bZ_t^*)\}\]
and $\|\Cov(\vc(\bZ_t^*))\|_2\leq C$. Thus,
\[|\Var(\bxi_{1,i}'\bZ_t^*\bxi_{2,j})|\leq C(\bxi_{2,j}'\bxi_{2,j}\otimes\bxi_{1,i}'\bxi_{1,i})\leq C, \]
which implies that $|\bxi_{1,i}'\bZ_t^*\bxi_{2,j}|=O_p(1)$ and
\[p_1^{-1/2}p_2^{-1/2}\|\Pi_{4,4}^*\|_2=O_p(p_1^{-1/2}p_2^{-1/2}).\]
Therefore,
\begin{equation}\label{pi4:st}
p_1^{-1/2}p_2^{-1/2}\|\Pi_{4}^*\|_2\leq C(p_1^{-\delta_2/2}\|\wh\bB_2^*-\bB_2^*\|_2+p_2^{-\delta_2/2}\|\wh\bQ_2^*-\bQ_2^*\|_2+p_1^{-1/2}p_2^{-1/2}).
\end{equation}
Finally, Theorem 3 follows from the upper bounds in (\ref{pi1:st})-(\ref{pi4:st}). This completes the proof. $\Box$

%%%%%%%%%%%%%%%%%%%%%%%%%%%%%%%%%%%%%%%%%%%%%%%%%%%%%%%%%
\noindent
{\bf Proof of Theorem 4.} (i) The proof follows the arguments in the proof of Theorem 6 in \cite{gaotsay2018b}. Let $ u_{ij,t}=\bb_{1,i}'\bY_t\bq_{1,j}$ be the $(i,j)$-th element of $\bB_1'\bY_t\bQ_1$, where $\bb_{1,i}$ and $\bq_{1,j}$ are the $i$-th and $j$-th columns of $\bB_1$ and $\bQ_1$, respectively. By the proof of Theorem 3 in \cite{changyaozhou2017}, we only need to show  that
\begin{equation}\label{dif:uhat}
\frac{1}{n}\sum_{t=1}^n (\wh u_{ij,t}-u_{ij,t})^2=o_p(1),\quad 1\leq i\leq p_1-r_1,\,\,1\leq j\leq p_2-r_2,
\end{equation}
where $\wh u_{ij,t}=\wh\bb_{1,i}'\bY_t\wh\bq_{1,j}$ and $\wh\bb_{1,i}$ and $\wh\bq_{1,j}$ are the $i$-th and $j$-th column of $\wh\bB_1$ and $\wh\bQ_1$ if we ignore some orthogonal rotations. Note that
\[\wh\bb_{1,i}'\bY_t\wh\bq_{1,j}=\wh\bb_{1,i}'\bA_1\bX_t\bP_1'\wh\bq_{1,j}+\wh\bb_{1,i}'\bA_2\bE_t\bP_2'\wh\bq_{1,j},\,\,\text{and}\,\,\bb_{1,i}'\bY_t\bq_{1,j}=\bb_{1,i}'\bA_2\bE_t\bP_2'\bq_{1,j}.\]
Then it suffices to guarantee
\begin{equation}\label{gt1}
\frac{1}{n}\sum_{t=1}^n(\wh\bb_{1,i}'\bA_1\bX_t\bP_1'\wh\bq_{1,j})^2=o_p(1),
\end{equation}
and
\begin{equation}\label{gt2}
\frac{1}{n}\sum_{t=1}^n(\wh\bb_{1,i}'\bA_2\bE_t\bP_2'\wh\bq_{1,j}-\bb_{1,i}'\bA_2\bE_t\bP_2'\bq_{1,j})^2=o_p(1).
\end{equation}
We recall that $\bL_1=\bA_1\bW_1$ and $\bR_1=\bP_1\bG_1$ with $\|\bW_1\|_2\asymp p_1^{(1-\delta_1)/2}$ and $\|\bG_1\|_2\asymp p_2^{(1-\delta_1)/2}$ and since
\begin{equation*}
(\wh\bb_{1,i}'\bA_1\bX_t\bP_1'\wh\bq_{1,j})^2=(\wh\bq_{1,j}'\bP_1\bG_1\otimes\wh\bb_{1,i}'\bA_1\bW_1)\bff_t\bff_t'(\bG_1'\bP_1'\wh\bq_{1,j}\otimes\bW_1'\bA_1'\wh\bb_{1,i}),
\end{equation*}
then
\begin{align}\label{byp:hat}
\left|\frac{1}{n}\sum_{t=1}^n(\wh\bb_{1,i}'\bA_1\bX_t\bP_1'\wh\bq_{1,j})^2\right|\leq C\|\wh\bQ_1-\bQ_1\|_2^2\|\wh\bB_1-\bB_1\|_2^2\|\bW_1\|_2^2\|\bG_1\|_2^2=O_p(p_1^{1+3\delta_1}p_2^{1+3\delta_1}n^{-2}).
\end{align}
On the other hand,
\[\wh\bb_{1,i}'\bA_2\bE_t\bP_2'\wh\bq_{1,j}-\bb_{1,i}'\bA_2\bE_t\bP_2'\bq_{1,j}=(\wh\bq_{1,j}'\otimes\wh\bb_{1,i}'-\bq_{1,j}'\otimes\bb_{1,i}')(\bR_2\otimes\bL_2)\bz_t,\]
and then
\begin{align}\label{bypd:hat}
\left|\frac{1}{n}\sum_{t=1}^n(\wh\bb_{1,i}'\bA_2\bE_t\bP_2'\wh\bq_{1,j}-\bb_{1,i}'\bA_2\bE_t\bP_2'\bq_{1,j})^2\right|\leq& (\|\wh\bb_{1,i}-\bb_{1,i}\|_2+\|\wh\bq_{1,j}-\bq_{1,j}\|_2)^2\notag\\
&\times\|\bR_2\|_2^2\|\bL_2\|_2^2\|\frac{1}{n}\sum_{t=1}^n\bz_t\bz_t'\|_2\notag\\
\leq & C(\|\wh\bB_1-\bB_1\|_2^2+\|\wh\bQ_1-\bQ_1\|_2^2)\|\bL_2\|_2^2\|\bR_2\|_2^2\notag\\
=& O_p(p_1^{1+2\delta_1-\delta_2}p_2^{1+2\delta_1-\delta_2}n^{-1}),
\end{align}
where we use the property that $\|\frac{1}{n}\sum_{t=1}^n\bz_t\bz_t'\|_2$ is bounded since $\bz_t$ is sub-Gaussian; see the proof of Lemma 4 in \cite{gaotsay2018b} or Theorem 4.3.5 of \cite{vershynin2018}. Therefore we require $p_1^{1+3\delta_1}p_2^{1+3\delta_1}n^{-2}=o(1)$ and $p_1^{1+2\delta_1-\delta_2}p_2^{1+2\delta_1-\delta_2}n^{-1}=o(1)$. \\
(ii) By (\ref{m-factor}),
\begin{align}\label{byq-d}
\wh\bb_{1,i}'\bY_{t}\wh\bq_{1,j}=&\wh\bb_{1,i}'\bL_1\bF_t\bR_1'\wh\bq_{1,j}+(\wh\bb_{1,i}-\bb_{1,i})'\bL_2\bZ_t\bR_2'\wh\bq_{1,j}'+\bb_{1,i}'\bL_2\bZ_t\bR_2'(\wh\bq_{1,j}-\bq_{1j})+\bb_{1i}'\bL_2\bZ_t\bR_2\bq_{1,j}\notag\\
=&:\alpha_1+\alpha_2+\alpha_3+\alpha_4.
\end{align}
By Assumptions 3, 4, and 6, we can show that for any unit vector $\bv_1\in\mathbb{R}^{r_1r_2}$ and $\bv_2\in\mathbb{R}^{v_1v_2}$ such that 
\[P(\max_{1\leq t\leq n}|\bv_1'\vc(\bF_t)|>x)\leq Cn\exp(-Cx^2),\]
\[P(\max_{1\leq t\leq n}|\bv_2'\vc(\bZ_t)|>x)\leq Cn\exp(-Cx^2),\]
and thus
\[P(\max_{1\leq t\leq n}\|(\bV_2'\otimes\bU_{2}')\vc(\bZ_t)\|_2>x)\leq Cp_1p_2n\exp(-Cx^2/(p_1p_2)).\]
It follows that $\max_{1\leq t\leq n}|\bv_1'\vc(\bF_t)|=O_p(\sqrt{\log (n)})$ and $\max_{1\leq t\leq n}\|(\bV_2'\otimes\bU_{2}')\vc(\bZ_t)\|_2=O_p(p_1^{1/2}p_2^{1/2}\sqrt{\log(np_1p_2)})$. Therefore,
\begin{align*}
\max_{1\leq i\leq v_1,1\leq j\leq v_2}\max_{1\leq t\leq n}\|\alpha_1\|_2\leq& C(\|\wh\bB_1-\bB_1\|_2+\|\wh\bQ_1-\bQ_1\|_2)\max_{1\leq t\leq n}\|\bL_1\bF_t\bR_1'\|_2\notag\\
\leq& Cp_1^{(1+\delta_1)/2}p_2^{(1+\delta_1)/2}n^{-1/2}\sqrt{\log(n)},
\end{align*}
\begin{align*}
\max_{1\leq i\leq v_1,1\leq j\leq v_2}\max_{1\leq t\leq n}\|\alpha_2\|_2\leq& C(\|\wh\bB_1-\bB_1\|_2)\max_{1\leq t\leq n}\|\bL_2\bZ_t\bR_2'\|_2\notag\\
\leq& Cp_1^{1+\delta_1-\delta_2/2}p_2^{1+\delta_1-\delta_2/2}n^{-1/2}\sqrt{\log(np_1p_2)},
\end{align*}
and 
\[\max_{1\leq i\leq v_1,1\leq j\leq v_2}\max_{1\leq t\leq n}\|\alpha_3\|_2\leq Cp_1^{1+\delta_1-\delta_2/2}p_2^{1+\delta_1-\delta_2/2}n^{-1/2}\sqrt{\log(np_1p_2)}.\]
We only need to show that the effect of the estimators on the noise term $\alpha_2$ is asymptotically negligible, which is sufficient to show (ii). Therefore, 
\[p_1^{1+\delta_1-\delta_2/2}p_2^{1+\delta_1-\delta_2/2}n^{-1/2}\sqrt{\log(np_1p_2)}=o_p(1),\]
which is the one specified in Theorem 4(ii). This completes the proof. $\Box$

%%%%%%%%%%%%%%%%%%%%%%%%%%%%%%%%%%%%%


\begin{thebibliography}{99}


%\bibitem[Bai, 2003]{bai2003}
%Bai, J. (2003). \newblock Inferential theory for factor models of large dimensions. 
%\newblock {\em Econometrica}, {\bf 71(1)}, 135--171.
%\bibitem[{Arenas-Garc\'{i}a and Camps-Valls (2008)}]{arenas2008}
%Arenas-Garc\'{i}a, J. and Camps-Valls, G. (2008). 
%\newblock Efficient kernel orthonormalized PLS for remote sensing applications. 
%\newblock {\em IEEE Transactions on Geoscience and Remote Sensing}, {\bf 46(10)}, 2872--2881.


\bibitem[{Andrews(1991)}]{andrews1991}
Andrews, D. W. (1991). Heteroskedasticity and autocorrelation consistent covariance matrix estimation. 
\newblock {\em Econometrica}, {\bf 59}, 817--858.




\bibitem[{Bai (2003)}]{Bai_Econometrica_2003}
Bai J. (2003) Inferential theory for factor models of large dimensions. \textsl{Econometrica}, {\bf 71(1)}, 135--171.



\bibitem[{Bai and Ng (2002)}]{BaiNg_Econometrica_2002}
Bai, J. and Ng, S. (2002). Determining the number of factors in approximate factor models. \textsl{Econometrica}, {\bf 70}, 191--221.


%\bibitem[{Bernstein (2009)}]{bernstein2009}
%Bernstein, D. S. (2009). 
%\newblock {\em Matrix Mathematics: Theory, Facts, and Formulas}. 
%\newblock Princeton Univ. Press.

\bibitem[{Black(1986)}]{black1986}
Black, F. (1986). 
\newblock Noise. \newblock {\em The Journal of Finance}, {\bf 41(3)}, 528--543.

%\bibitem[Bickel and Levina, 2008]{bickel2008}
%Bickel, P. J. and Levina, E. (2008).
%\newblock  Covariance regularization by thresholding.
%\newblock {\em The Annals of Statistics}, {\bf 36(6)}, 2577--2604.


\bibitem[{Box and Tiao(1977)}]{BoxTiao_1977}
Box, G. E. P. and Tiao, G. C. (1977). A canonical analysis of multiple time series. {\sl Biometrika}, {\bf 64}, 355--365.

%\bibitem[Chang~et~al., 2015]{changguoyao2015}
%Chang, J., Guo, B. and Yao, Q. (2015).
%\newblock High dimensional stochastic regression with latent factors, endogeneity and nonlinearity. \newblock {\em Journal of Econometrics}, %{\bf 189(2)}, 297--312.


\bibitem[Chang~et~al., 2017]{changyaozhou2017}
Chang, J., Yao, Q. and Zhou, W. (2017). 
\newblock Testing for high-dimensional white noise using maximum cross-correlations. 
\newblock {\em Biometrika}, {\bf 104(1)}, 111--127.

\bibitem[Chen~et~al., 2020]{chentsaychen2018}
Chen, E.Y., Tsay, R.S., and Chen, R. (2020). 
\newblock Constrained factor models for high-dimensional matrix-variate time series. 
\newblock {\em Journal of the American Statistical Association}, {\bf 115(530)}, 775--793.


\bibitem[Chen~et~al., 2020]{chenxiaoyang2020}
Chen, R., Xiao, H., and Yang, D. (2020). Autoregressive models for matrix-valued time series. {\em Journal of Econometrics} (forthcoming).

\bibitem[{Davis { et al.}(2012)}]{Davis2012}
Davis, R. A., Zang, P. and Zheng, T. (2012). Sparse vector autoregressive modelling. Available at {\sl arXiv:1207.0520}.

%\bibitem[Ding and Cook, 2017]{dingcook2017}
%Ding, S. and  Cook, R. D. (2017). \newblock Matrix variate regressions and envelope models.
%\newblock {\em Journal of the Royal Statistical Society: Series B (Statistical Methodology)}, {To appear}.


%\bibitem[{Diebold and Mariano(1995)}]{diebold-1995}
%Diebold, F. X., and Mariano, R. S. (1995). Comparing predictive accuracy.
%\newblock {\em Journal of Business \& Economic Statistics}, {\bf 13(3)}, 253--263.


\bibitem[{Fama and French(2015)}]{famafrench2015}
Fama, E. F. and French, K. R. (2015). A five-factor asset pricing model. 
\newblock {\em Journal of Financial Economics}, {\bf 116(1)}, 1--22.


\bibitem[{Forni { et al.}(2000)}]{forni2000}
Forni, M., Hallin, M., Lippi, M. and Reichlin, L. (2000).
\newblock Reference cycles: the NBER methodology revisited (No. 2400). Centre for Economic Policy Research.

\bibitem[{Forni { et al.}(2005)}]{forni2005}
Forni, M., Hallin, M., Lippi, M. and Reichlin, L. (2005). 
\newblock The generalized dynamic factor model: one-sided estimation and forecasting. 
\newblock {\em Journal of the American Statistical Association}, {\bf 100(471)}, 830--840.


\bibitem[{Gao{ et al.}(2019)}]{gaoetal2017}
Gao, Z., Ma, Y., Wang, H. and Yao, Q. (2019). 
\newblock Banded spatio-temporal autoregressions. 
\newblock {\em Journal of Econometrics,} {\bf 208(1)}, 211--230.


\bibitem[{Gao and Tsay(2019)}]{gaotsay2018a}
Gao, Z. and Tsay, R. S. (2019). 
\newblock A structural-factor approach for modeling high-dimensional time series and space-time data. 
\newblock {\em Journal of Time Series Analysis}, {\bf 40}, 343--362.

\bibitem[{Gao and Tsay(2020a)}]{gaotsay2020a}
Gao, Z. and Tsay, R. S. (2020a). 
\newblock Modeling high-dimensional unit-root time series.
\newblock {\em International Journal of Forecasting} (forthcoming).

\bibitem[{Gao and Tsay(2020b)}]{gaotsay2018b}
Gao, Z. and Tsay, R. S. (2020b). 
\newblock Modeling high-dimensional time series: a factor model with
dynamically dependent factors and diverging eigenvalues.
\newblock Available at {\em arXiv:1808.07932}.
%\bibitem[{Govaerts and Pryce (1989)}]{govaerts1989}
%Govaerts, W. and Pryce, J. D. (1989). 
%\newblock A singular value inequality for block matrices. 
%\newblock {\em Linear Algebra and its Applications}, {\bf 125}, 141--148.

\bibitem[Golub and Van Loan, 1996]{golub1996}
Golub, G. H. and Van Loan, C. F. (1996).
\newblock Matrix computations.
\newblock Johns Hopkins University Press.

%\bibitem[Gorski~et~al., 2007]{gorski2007}
%Gorski, J., Pfeuffer, F., and Klamroth, K. (2007). \newblock Biconvex sets and optimization with biconvex functions: a survey and extensions. \newblock {\em Mathematical methods of operations research}, {\bf 66(3)}, 373--407.

%\bibitem[Gregory and Head, 1999]{GregoryHead1999}
%Gregory, A. W. and Head, A. C. (1999). 
%\newblock Common and country-specific fluctuations in productivity, investment, and the current account. 
%\newblock {\em Journal of Monetary Economics}, {\bf 44(3)}, 423--451.

%\bibitem[{Guo { et al.}(2016)}]{GuoWangYao_2014}
%Guo, S., Wang, Y. and Yao, Q. (2016). 
%\newblock High dimensional and banded vector autoregression. 
%\newblock {\em Biometrika,} {\bf 103}(4),  889--903.




%\bibitem[Gupta and Nagar, 2000]{gupta2000}
%Gupta, A. K. and Nagar, D. K. (2000). \newblock {\em Matrix Variate Distributions}. Chapman \& Hall/CRC,
%Boca Raton, FL.

%\bibitem[Hoff, 2015]{hoff2015}
%Hoff, P. D. (2015). \newblock Multilinear tensor regression for longitudinal relational data. 
%\newblock {\em The Annals of Applied Statistics}, {\bf 9(3)}, 1169--1193.

\bibitem[Han and Tsay(2020)]{HanTsay2020}
Han, Y. and Tsay, R. S. (2020). 
\newblock High-dimensional linear regression for dependent data with applications to
now-casting.
\newblock {\em Statistica Sinica}, {\bf 30}, 1797--1827. 

\bibitem[Hosking, 1980]{hosking1980}
Hosking, J. R. (1980).
\newblock The multivariate portmanteau statistic. 
\newblock {\em Journal of the American Statistical Association}, {\bf 75(371)}, 602--608.


%\bibitem[Huang and Wang, 2012]{huagwang2012}
%Hung, H. and Wang, C. C. (2012). 
%\newblock Matrix variate logistic regression model with application to EEG data. 
%\newblock {\em Biostatistics}, {\bf 14(1)}, 189--202.


%\bibitem[{Huang and Tsay (2014)}]{Huang_2014}
%Huang, D. and Tsay, R. S. (2014). A refined scalar component approach to multivariate time series modeling. {\sl Manuscript}.

\bibitem[{Johnstone and Lu(2009)}]{johnstone-lu2009}
Johnstone, I.M. and Lu, A.Y. (2009). 
\newblock On consistency and sparsity for principal components analysis in high dimensions. 
\newblock {\em Journal of the American Statistical Association}, {\bf 104(486)}, 682--693.


\bibitem[Lam and Yao(2012)]{lamyao2012}
Lam, C. and Yao, Q. (2012).
\newblock Factor modeling for high-dimensional time series: inference for the number of factors.
\newblock {\em The Annals of Statistics}, {\bf 40(2)}, 694--726.




\bibitem[{Lam {et al.}(2011)}]{LamYaoBathia_Biometrika_2011}
Lam, C., Yao, Q. and Bathia, N. (2011). Estimation of latent factors for high-dimensional time series. \BKA, {\bf98}, 901--918.

%\bibitem[Leng and Tang, 2012]{lengtang2012}
%Leng, C. and Tang, C. Y. (2012). Sparse matrix graphical models. 
%\newblock {\em Journal of the American
%Statistical Association}, {\bf107(499)} 1187-–1200.

%\bibitem[Li~et~al., 2016]{li2016}
%Li, B., Kim, M. K. and Altman, N. (2010). \newblock
%On dimension folding of matrix-or array-valued statistical objects. 
%\newblock {\em The Annals of Statistics}, 1094--1121.


%\bibitem[L\"{u}tkepohl, 2006]{lutkepohl2006}
%L\"{u}tkepohl, H. (2006). 
%\newblock {\em New Introduction to Multiple Time Series Analysis,} 
%\newblock Springer, Berlin.



\bibitem[Onatski, 2010]{onatski2010}
Onatski, A. (2010). \newblock Determining the number of factors from empirical distribution of eigenvalues. 
\newblock {\em The Review of Economics and Statistics}, {\bf 92(4)}, 1004--1016.

\bibitem[Pan and Yao, 2008]{panyao2008}
Pan, J. and Yao, Q. (2008).
\newblock Modelling multiple time series via common factors.
\newblock {\em Biometrika}, {\bf 95(2)}, 365--379.

\bibitem[Rogers {et al.}(2013)]{rogers2013}
Rogers, M., Li, L., and Russell, S. (2013).
\newblock Multilinear dynamic systems for tensor time series. 
\newblock Conference for Neural Information Processing Systems,
https://papers.nips.cc/paper/5117-multilinear-dynamical-systems-for-tensor-time-series.pdf.

%\bibitem[Rockafellar and Wets, 2009]{rockafellarwets2009}
%Rockafellar, R. T., ad Wets, R. J. B. (2009). \newblock {\em Variational Analysis}. Grundlehren Math. Wiss. 317, Springer-Verlag, New York.

%\bibitem[Shang and Hyndman, 2011]{shanghyndman2011}
%Shang, H. L., and Hyndman, R. J. (2011), FDS: Functional Data Sets Package
%in R, Vienna, Austria: R Development Core Team.


\bibitem[Sharpe, 1964]{sharpe1964}
Sharpe, W. F. (1964). Capital asset prices: A theory of market equilibrium under conditions of risk. 
\newblock {\em The Journal of Finance}, {\bf 19(3)}, 425--442.

\bibitem[{Shen {et al.} (2016)}]{shenetal2016}
Shen, D., Shen, H. and Marron, J. S. (2016). 
\newblock A general framework for consistency of principal component analysis.
\newblock {\em Journal of Machine Learning Research}, {\bf 17(150)}, 1--34.


\bibitem[{Shojaie and Michailidis(2010)}]{ShojaieMichailidis_2010}
Shojaie, A. and Michailidis, G. (2010). Discovering graphical Granger causality using the truncated lasso penalty. \textsl{Bioinformatics}, {\bf 26}, 517--523.



\bibitem[{Song and Bickel(2011)}]{SongBickel_2011}
Song, S. and Bickel, P. J. (2011). Large vector auto regressions. Available at {\sl arXiv:1106.3519}.


%\bibitem[{Stock and Watson (1989)}]{StockWatson_1989}
%Stock, J. H. and Watson, M. W. (1989).
%\newblock  New indexes of coincident and leading economic indicators. 
%\newblock {\em NBER Macroeconomics Annual}, {\bf 4}, 351--394.

%\bibitem[{Stock and Watson (1998)}]{StockWatson_1998}
%Stock, J. H. and Watson, M. W. (1998).
%\newblock  Diffusion indexes. 
%\newblock NBER Working Paper 6702.

\bibitem[{Stock and Watson(2002)}]{StockWatson_2002}
Stock, J. H. and Watson, M. W. (2002). Forecasting using principal components from a large number of predictors. \JASA, {\bf 97}, 1167--1179.


\bibitem[{Stock and Watson(2005)}]{StockWatson_2005}
Stock, J. H. and Watson, M. W. (2005). Implications of dynamic factor models for VAR analysis. Available at {\sl www.nber.org/papers/w11467}.

\bibitem[Surana {et al.}(2016)]{surana2016}
Surana, A. Patterson, G., and Rajapakse, I. (2016). 
\newblock Dynamic tensor time series modeling and analysis.  2016 IEEE 55th Conference on Decision and Control (CDC), Las Vegas, NV, 2016, pp. 1637-1642, doi: 10.1109/CDC.2016.7798500.



\bibitem[{Tiao and Tsay(1989)}]{TiaoTsay_1989}
Tiao, G. C. and Tsay, R. S. (1989). Model specification in multivariate time series (with discussion). 
\newblock {\em \JRSSB}{\bf51}, 157--213.


 \bibitem[{Tsay(2014)}]{Tsay_2014}
 Tsay, R. S. (2014). {\sl Multivariate Time Series Analysis}. Wiley, Hoboken, NJ.

 \bibitem[{Tsay(2020)}]{Tsay_2018}
Tsay, R. S. (2020). 
\newblock Testing for serial correlations in high-dimensional time series via extreme value
theory. \newblock {\em Journal of Econometrics}, {\bf 216}, 106–-117.

 \bibitem[{Vershynin(2018)}]{vershynin2018}
Vershynin, R. (2018).
{\em High-Dimensional Probability: An Introduction with Applications in Data Science.}
\newblock Cambridge University Press.

 \bibitem[Walden and Serroukh, 2002]{walden2002}
Walden, A. and Serroukh, A. (2002). \newblock Wavelet analysis of matrix-valued time series. 
\newblock {\em Proceedings:
Mathematical, Physical and Engineering Sciences}, {\bf 458(2017)}, 157–-179.

 \bibitem[{Wang~et~al.(2019)}]{wang2018}
Wang, D., Liu, X. and Chen, R. (2019). 
\newblock Factor Models for Matrix-Valued High-Dimensional Time Series. 
\newblock {\em Journal of Econometrics}, {\bf 208(1)}, 231--248.


\bibitem[Werner~et~al., 2008]{werner2008}
Werner, K., Jansson, M., and Stoica, P. (2008).
\newblock On estimation of covariance matrices with Kronecker product structure.
\newblock {\em IEEE Transactions on Signal Processing}, {\bf 56(2)}, 478--491.


%\bibitem[Xu and Yin, 2013]{xuyin2013}
%Xu, Y., and Yin, W. (2013). \newblock A block coordinate descent method for regularized multiconvex optimization with applications to nonnegative tensor factorization and completion. 
%\newblock {\em SIAM Journal on Imaging Sciences}, {\bf 6(3)}, 1758--1789.

%\bibitem[Xue and Yin, 2014]{xueandyin2014}
%Xue, Y. and Yin, X. (2014). 
%\newblock Sufficient dimension folding for regression mean function. 
%\newblock {\em Journal of Computational and Graphical Statistics}, {\bf 23(4)}, 1028--1043.

%\bibitem[Yin and Li, 2012]{yinli2012}
%Yin, J. and Li, H. (2012). \newblock Model selection and estimation in the matrix normal graphical model.
%\newblock {\em Journal of Multivariate Analysis}, {\bf 107(0)}, 119-–140.

%\bibitem[Zhao and Leng, 2014]{zhaoleng2014}
%Zhao, J. and Leng, C. (2014). Structured lasso for regression with matrix covariates. 
%\newblock {\em Statistica
%Sinica}, {\bf 24}, 799-–814.

%\bibitem[Zhou and Li, 2014]{zhouli2014}
%Zhou, H. and Li, L. (2014). \newblock Regularized matrix regression.
%\newblock {\em Journal of the Royal Statistical Society}, {\bf B76(2)}, 463--483.

%\bibitem[Zhou~et~al., 2013]{zhou2013}
%Zhou, H., Li, L. and Zhu, H. (2013). 
%\newblock Tensor regression with applications in neuroimaging data analysis. 
%\newblock {\em Journal of the American Statistical Association}, {\bf 108(502)}, 540--552.

%\bibitem[Zhou, 2014]{zhou2014}
%Zhou, S. (2014). \newblock Gemini: Graph estimation with matrix variate normal instances. 
%\newblock {\em The Annals of Statistics}, {\bf 42(2)}, 532-–562.
%%%%%%%%%%%%%%%%%%%%%%%%%%%%%%%%%%%%%%%%%%%%%%%%%%%%%%%%%%%%%%%%%%%%%%%%%%%%%%%%%%%%%%%%%%%%%%%%%%%%%%%%%%%%%%%%%%%%%%%%%%%%
\end{thebibliography}
\end{document}